\documentclass[11pt, a4paper]{article}
\pdfoutput=1
\usepackage{jcappub}

\usepackage{amssymb,amsmath,latexsym,mathrsfs}
\usepackage{graphicx} 
\usepackage{bm}
\usepackage{url}
\usepackage{epsfig,graphicx,verbatim,xspace,multirow,mathtools}
\usepackage{array}
\usepackage{enumitem}
\usepackage{booktabs}

\begin{document}

\title{Precision Early Universe Thermodynamics made simple: \\$N_{\rm eff}$ and Neutrino Decoupling in the Standard Model and beyond}

\hspace*{124mm}{\tt KCL-2019-85}

\author{Miguel Escudero Abenza}\note{ORCID: \url{http://orcid.org/0000-0002-4487-8742}}
\emailAdd{miguel.escudero@kcl.ac.uk}

\affiliation{Theoretical Particle Physics and Cosmology Group \\
Department of Physics, King's College London, Strand, London WC2R 2LS, UK}

\abstract{Precision measurements of the number of effective relativistic neutrino species and the primordial element abundances require accurate theoretical predictions for early Universe observables in the Standard Model and beyond. Given the complexity of accurately modelling the thermal history of the early Universe, in this work, we extend a previous method presented by the author in~\cite{Escudero:2018mvt} to obtain simple, fast and accurate early Universe thermodynamics. The method is based upon the approximation that all relevant species can be described by thermal equilibrium distribution functions characterized by a temperature~and a chemical potential. We apply the method to neutrino decoupling in the Standard Model and find $N_{\rm eff}^{\rm SM} = 3.045$ -- a result in excellent agreement with previous state-of-the-art calculations. We apply the method to study the thermal history of the Universe in the presence of a very light ($1\,\text{eV}<m_\phi < 1\,\text{MeV}$) and weakly coupled ($\lambda \lesssim 10^{-9}$) neutrinophilic scalar. We find our results to be in excellent agreement with the solution to the exact Liouville equation. Finally, we release a code:~\href{https://github.com/MiguelEA/nudec_BSM}{NUDEC\_BSM} (available in both \texttt{Mathematica} and \texttt{Python} formats), with which neutrino decoupling can be accurately and efficiently solved in the Standard Model and beyond:~\url{https://github.com/MiguelEA/nudec_BSM}.  
}

\maketitle
\newpage

\section{Introduction}
\vspace{0.15cm}

The Planck collaboration~\cite{Akrami:2018vks} reports unprecedented precision measurements of the number of effective relativistic neutrino species, $N_{\rm eff}$. Within the framework of $\Lambda$CDM, Planck legacy data analyses yield: $N_{\rm eff} = 2.99\pm0.34$ at 95\% CL~\cite{planck}. Additionally, upcoming CMB experiments like SPT-3G~\cite{Benson:2014qhw} and the Simons Observatory~\cite{Ade:2018sbj} will soon improve upon Planck's precision on $N_{\rm eff}$, and future experiments such as CMB-S4~\cite{Abazajian:2019eic}, PICO~\cite{Hanany:2019lle}, CORE~\cite{DiValentino:2016foa} or CMB-HD~\cite{Sehgal:2019ewc} are expected to deliver a 1\% precision determination of $N_{\rm eff}$. Likewise, the primordial helium and deuterium abundances are now measured with 1\% precision~\cite{pdg}; and in the future, the primordial deuterium abundance could be measured with 0.1\% precision~\cite{Cooke:2016rky,Grohs:2019cae}. 

\vspace{0.1cm}

Precision determinations of the primordial element abundances and $N_{\rm eff}$ simultaneously serve as confirmation of the thermal history in the Standard Model (SM) and a strong constraint on many scenarios beyond, see e.g.~\cite{Pospelov:2010hj,Iocco:2008va,Sarkar:1995dd,Dolgov:2000jw,Kawasaki:2017bqm,Hasegawa:2019jsa,Cadamuro:2010cz,Cicoli:2012aq,Brust:2013xpv,Vogel:2013raa,Boehm:2013jpa,Wilkinson:2016gsy,Weinberg:2013kea,deSalas:2015glj,Gariazzo:2019gyi}. However, precision measurements also represent a challenge in finding accurate theoretical predictions for early Universe observables. In fact, obtaining an accurate prediction for $N_{\rm eff}$ in the Standard Model is already a non-trivial task~\cite{deSalas:2016ztq,Mangano:2005cc} (see also~\cite{Dolgov:2002wy,Dicus:1982bz,Hannestad:1995rs,Dodelson:1992km,Dolgov:1997mb,Esposito:2000hi,Mangano:2001iu,Birrell:2014uka,Grohs:2015tfy,Froustey:2019owm}). 

\vspace{0.1cm}

The challenge in obtaining accurate predictions for $N_{\rm eff}$ arises mainly as a result of the fact that neutrinos decouple from electrons and positrons in the early Universe at temperatures $T_\nu^{\rm dec}\sim 2\,\text{MeV}$~\cite{Dolgov:2002wy}. Since this temperature is rather close to the electron mass, residual out-of-equilibrium electron-positron annihilations to neutrinos heat the neutrino fluid leading to $N_{\rm eff}^{\rm SM} > 3$. However, this is not the only relevant effect since finite temperature corrections alter $N_{\rm eff}$ at the level of $\Delta N_{\rm eff}^{\rm SM} \sim 0.01$~\cite{Heckler:1994tv,Fornengo:1997wa,Bennett:2019ewm}, and neutrinos start to oscillate prior to neutrino decoupling~\cite{Hannestad:2001iy,Dolgov:2002ab}. Thus, state-of-the-art treatments of neutrino decoupling account for non-thermal neutrino spectral distortions, finite temperature corrections and neutrino oscillations, leading to the Standard Model prediction of $N_{\rm eff}^{\rm SM} = 3.045$~\cite{deSalas:2016ztq,Mangano:2005cc}\footnote{Next to leading order finite temperature corrections are expected to shift this number by -0.001~\cite{Bennett:2019ewm}.}. To obtain such an accurate number, state-of-the-art calculations resort to the density matrix formalism and solve a system of hundreds of \textit{stiff} integro-differential equations for the neutrino distributions -- something which is computationally challenging. Although this approach is very accurate, its complexity represents a drawback in studying generalized scenarios Beyond the Standard Model (BSM).

\vspace{0.1cm}

In this work -- building upon~\cite{Escudero:2018mvt} -- we propose a simplified, accurate and fast method to calculate early Universe BSM thermodynamics. The method is based on the approximation that any species can be described by thermal equilibrium distribution functions with evolving temperature and chemical potential. Within this approximation, simple ordinary differential equations for the time evolution of the temperature and chemical potential of any given species can be derived. These equations account for all relevant interactions, are easy to solve, and track accurately the thermodynamics.

\vspace{0.1cm}

This study extends the approach of~\cite{Escudero:2018mvt} by allowing for non-negligible chemical potentials. This is important since, unlike in the SM, chemical potentials cannot be neglected in many BSM theories. In addition, we include the next-to-leading order finite temperature corrections to the electromagnetic plasma from~\cite{Bennett:2019ewm}, and spin-statistics and the electron mass in the SM reaction rates. Including these effects is required in order to obtain $N_{\rm eff}$ in the Standard Model with an accuracy of $0.001$. 
\newpage
When applying the method to neutrino decoupling in the Standard Model, and by accounting for finite temperature corrections, and the electron mass and spin-statistics in the $\nu$-$\nu$ and $\nu$-$e$ interaction rates, we find $N_{\rm eff}^{\rm SM} =3.045$. A result that is in excellent agreement with previous accurate calculations in the literature. We also find very good agreement with previous literature (better than 0.1\%) for the neutrino number density, the entropy density and the helium and deuterium primordial element abundances. 

In order to illustrate the implications of the method for BSM physics, we solve for the early Universe thermodynamics in the presence of a very light ($1\,\text{eV} <m_\phi < 1\,\text{MeV}$) and weakly coupled $\lambda \sim \mathcal{O}(10^{-12})$ neutrinophilic scalar. The phenomenology and cosmology of this scenario was first highlighted in~\cite{Chacko:2003dt}. Recently, using the methods developed in this study and by analyzing Planck legacy data~\cite{Akrami:2018vks,planck}, strong constraints on the parameter space were derived in~\cite{Escudero:2019gvw}. In this work, we compute all relevant thermodynamic quantities and cosmological parameters and compare them to the results obtained by solving the exact Liouville equation for the neutrino-scalar system. We find excellent agreement between the two computations and we therefore claim that the proposed method can be used to find fast and accurate thermodynamic quantities in BSM scenarios. We note that the method can be applied to other frameworks not necessarily restricted to neutrino decoupling.

\vspace{0.1cm}

Finally, to facilitate the implementation of the method to the interested reader, we publicly release a \texttt{Mathematica} and \texttt{Python} code to solve for neutrino decoupling: \href{https://github.com/MiguelEA/nudec_BSM}{NUDEC\_BSM}. The typical execution time of the code in an average computer is $\mathcal{O}(10)\,\text{s}$. Thanks to its simplicity, speed and accuracy we believe it can represent a useful tool to study BSM early Universe thermodynamics. With \texttt{NUDEC\_BSM}, neutrino decoupling can be solved in the Standard Model, in the presence of dark radiation, with MeV-scale species in thermal equilibrium during neutrino decoupling, and in the presence of a light and weakly coupled neutrinophilic scalar. Given these examples, other scenarios should be easy to implement.

\vspace{0.2cm}
\noindent \textbf{Structure of this work:}\vspace{0.1cm}

This work is organized as follows: In Section~\ref{sec:BSM_neutrinodec}, we present the main method to solve for the thermodynamic evolution of any species in the early Universe. We discuss the approximations upon which the method is based and provide a first-principles derivation of the relevant evolution equations for the temperature and chemical potential describing the thermodynamics. In Section~\ref{sec:SM_neutrinodec}, we apply the method to neutrino decoupling in the Standard Model and compare with previous treatments. In Section~\ref{sec:boson_1to2}, we solve for the thermal history of the Universe in the presence of a very light and weakly coupled neutrinophilic scalar. We present a detailed comparison between the solution using our method and that of solving the exact Liouville equation. In Section~\ref{sec:discussion}, we discuss the results, and argue on theoretical grounds the reason the proposed method is very accurate in many scenarios. We also discuss the limitations of our approach. In Section~\ref{sec:BSM_recipe}, we outline a recipe to model the early Universe thermodynamics in generic extensions of the Standard Model. We conclude in Section~\ref{sec:summary}.

The practitioner is also referred to the Appendices~\ref{sec:app_TOTAL}. There we outline various details, comparison between the proposed method against solutions to the Liouville equation, and useful thermodynamic formulae.

\newpage 

\section{Fast and Precise Thermodynamics in the Early Universe}\label{sec:BSM_neutrinodec}
In this section we develop a simplified method to track the thermodynamic (in and out of equilibrium) evolution of any species in the Standard Model and beyond. We first start by reviewing the Liouville equation that governs the time evolution of the distribution function of a given species. Then, we assume that such distribution function is well-described by a thermal equilibrium distribution. Starting from the Liouville equation, we find ordinary differential equations for the time evolution of the temperature and a chemical potential describing the thermodynamics. Finally, we provide useful formulae for the number and energy density transfer rates for decays, annihilations and scatterings.

\subsection{The Liouville equation}\label{}

The distribution function $f$ determines the thermodynamics of any given species in the early Universe. In a fully homogeneous and isotropic Universe, the time evolution of the distribution function $f$ is governed by the Liouville equation~\cite{Bernstein:1988bw,Kolb:1990vq,Dolgov:2002wy}:
\begin{align}\label{eq:Liouville}
\frac{\partial f }{\partial t}  - H p  \frac{\partial f }{\partial p}  = \mathcal{C}[f]\, ,
\end{align} 
where $p$ is the momentum, $H  =(8 \pi\rho_{\rm T}/(3m_{\rm Pl}^2))^{1/2}$ is the Hubble rate, $\rho_{\rm T}$ is the total energy density of the Universe, $m_{\rm Pl} = 1.22\times 10^{19}\,\text{GeV}$ is the Planck mass, and $\mathcal{C}[f]$ is the collision term. In full generality, the collision term for a particle $\psi$ is defined as~\cite{Dolgov:2002wy}:
\begin{align}\label{eq:Collisionoperator}
& \mathcal{C}[f_\psi] \equiv - \frac{1}{2 E_\psi} \sum_{X,Y} \int  \prod_i d\Pi_{X_i} \prod_j d\Pi_{Y_j}  (2\pi)^4 \delta^4 (p_\psi + p_X - p_Y)\times  \\
&\left[|\mathcal{M}|^2_{\psi+X\to Y} f_\psi \prod_i f_{X_i} \prod_j \left[1 \pm f_{Y_j} \right] - |\mathcal{M}|^2_{Y\to \psi+X} \prod_j  f_{Y_j}   \left[ 1 \pm f_\psi \right] \prod_i \left[ 1 \pm f_{X_i} \right] \right]\,. \nonumber
\end{align}
The collision term accounts for any process $\psi + X \to Y$ where $X \equiv \sum_i X_i$ and $Y \equiv \sum_j Y_j$ represent the initial and final states accompanying $\psi$, and the indices $i,\,j$ label particle number. Here, $\mathcal{M}_{\psi + X \to Y}$ is the probability amplitude for the process $\psi + X \to Y$. The sum is taken over all possible initial $\psi + X$ and final states $Y$, $d\Pi_{X_i} \equiv \frac{1}{(2 \pi)^3} \frac{d^3 p_{X_i}}{2 E_{X_i} }$, and the $+$ signs correspond to bosons, while the $-$ signs correspond to fermions.

\subsection{Approximations}\label{sec:approximations}

The actual solution to the Liouville equation~\eqref{eq:Liouville} strongly depends upon the processes that the given species is undergoing in the early Universe. It is well-known (see e.g.~\cite{Kolb:1990vq}), that when interactions between a particle $\psi$ and the rest of the plasma are efficient, the distribution function of the $\psi$ species is well described by equilibrium distribution functions. Namely, fermions and bosons follow the usual Fermi-Dirac (FD) and Bose-Einstein (BE) distribution functions:
\begin{align}
f_{\rm FD}(E) = \frac{1}{1+e^{(E-\mu)/T }}\,,\qquad f_{\rm BE}(E) = \frac{1}{-1+e^{(E-\mu)/T }} \,,
\end{align}
respectively. Here, $T$ is the temperature, $\mu$ is the chemical potential, and $E = \sqrt{p^2 + m^2}$ is the energy with $m$ being the mass of the particle. Note that for bosons: $\mu <0$~\cite{Landau:1980mil}.

If scattering/annihilation/decay processes are not fully efficient, the distribution function of a given species may not exactly be described by a Fermi-Dirac or Bose-Einstein formula. However, in this work, in order to hugely simplify the Liouville equation~\eqref{eq:Liouville} we shall assume that any species is described by a thermal equilibrium distribution, namely by a Fermi-Dirac or Bose-Einstein function. Thanks to this approximation, we can find simple differential equations for the time evolution of the temperature and chemical potential that fully describe the thermodynamics of a given system. As we shall see, this approach -- although only rendering an approximate solution to the Liouville equation -- allows one to track very accurately the relevant thermodynamic quantities.

\subsection{Temperature and chemical potential evolution for a generic species}
Upon integrating Equation~\eqref{eq:Liouville} with the measures $g {d^3 p} /(2\pi)^3$ and $g E {d^3 p} /(2\pi)^3$, for a particle with $g$ internal degrees of freedom, we find:
\begin{align}
  \frac{dn}{dt} + 3 H n &= \frac{\delta n}{\delta t}  =  \int g  \, \frac{d^3p}{(2\pi)^3} \, \mathcal{C}[f]  \label{eq:n_general}  \, , \\
 \frac{d\rho}{dt} + 3 H (\rho + p) &=\frac{\delta \rho}{\delta t} = \int g \,  E \, \frac{d^3p}{(2\pi)^3} \, \mathcal{C}[f] \label{eq:rho_general}  \, ,
\end{align}
where $n$, $\rho$ and $p$ are the number, energy and pressure densities the species, respectively. $ \delta n/\delta t$ and $ \delta \rho/\delta t$ represent the number and energy transfer rate between a given particle and the rest of the plasma. 

 By summing Equation~\eqref{eq:rho_general} over all species in the Universe, one finds the usual continuity equation:
\begin{align}\label{eq:rho_continuity}
 \frac{d\rho_{\rm T}}{dt}  &=-  3 H (\rho_{\rm T} + p_{\rm T}) \,.
\end{align}
From a microphysics perspective, this continuity equation arises as a result of energy conservation in each process in the plasma; while from the fluid perspective, it simply results from the conservation of the total stress-energy tensor $\nabla_\mu T^{\mu \nu} = 0$.

By applying the chain rule to~\eqref{eq:n_general} and~\eqref{eq:rho_general}, we find:
\begin{subequations}\label{eq:dTdmu_generic}
 \begin{align}
\frac{dT}{dt} &=  -\left( \frac{dn}{dt} \frac{\partial \rho}{\partial \mu} - \frac{d\rho}{dt}\frac{\partial n}{\partial \mu}  \right)\bigg/\left( \frac{\partial n}{\partial \mu} \frac{\partial \rho}{\partial T}-\frac{\partial n}{\partial T} \frac{\partial \rho}{\partial \mu}  \right)\, ,\\
\frac{d\mu}{dt} &= -\left(\frac{d\rho}{dt}\frac{\partial n}{\partial T} -\frac{dn}{dt} \frac{\partial \rho}{\partial T} \right)\bigg/ \left(\frac{\partial n}{\partial \mu} \frac{\partial \rho}{\partial T}-\frac{\partial n}{\partial T} \frac{\partial \rho}{\partial \mu} \right)\, .\label{eq:aux_23}
\end{align}
\end{subequations}
This set of equations can be considerably simplified if chemical potentials are neglected. This typically occurs as a result of some efficient interactions. In the Standard Model, for example, efficient $e^+e^- \leftrightarrow \gamma \gamma$ and $e^+e^- \leftrightarrow \gamma \gamma \gamma$ interactions allow one to set $\mu_e(t) = \mu_\gamma(t) = 0$. If, $d\mu/dt = 0$ then $\frac{d n}{d t} = \frac{d \rho}{d t}   \frac{\partial n}{\partial T} / \frac{\partial \rho}{\partial T}$ and the previous equations~\eqref{eq:dTdmu_generic} are simplified to:
\begin{align}\label{eq:dTdt_nochem}
\frac{dT}{dt} &= \frac{d \rho}{dt} \bigg/  \frac{\partial \rho}{\partial T} = \left[- 3 H (\rho + p) + \frac{\delta \rho}{\delta t}\right]\bigg/  \frac{\partial \rho}{\partial T}\, .
\end{align}
Therefore, Equation~\eqref{eq:dTdt_nochem} can be used to track the thermodynamics of a species when chemical potentials are negligible (which occurs in many scenarios). 

The set of Equations \eqref{eq:dTdmu_generic} can be rewritten in terms of the Hubble parameter and the energy and number density transfer rates. Explicitly, they read:
\begin{subequations}\label{eq:dTdmu_generic_simple}
\begin{align}
\frac{dT}{dt} &=\frac{1}{\frac{\partial n}{\partial \mu} \frac{\partial \rho}{\partial T}-\frac{\partial n}{\partial T} \frac{\partial \rho}{\partial \mu} }\left[ -3 \,H \, \left((p+\rho)\frac{\partial n}{\partial \mu}-n \frac{\partial \rho}{\partial \mu} \right)+ \frac{\partial n}{\partial \mu} \, \frac{\delta \rho}{\delta t} - \frac{\partial \rho}{\partial \mu} \, \frac{\delta n}{\delta t} \right]  
 \, ,\label{eq:dT_dt_simple} \\
\frac{d\mu}{dt} &=\frac{-1}{\frac{\partial n}{\partial \mu} \frac{\partial \rho}{\partial T}-\frac{\partial n}{\partial T} \frac{\partial \rho}{\partial \mu} } \left[ -3 \,H \, \left((p+\rho)\frac{\partial n}{\partial T}-n \frac{\partial \rho}{\partial T} \right)+ \frac{\partial n}{\partial T} \, \frac{\delta \rho}{\delta t} - \frac{\partial \rho}{\partial T} \, \frac{\delta n}{\delta t} \right]   \, . \label{eq:dmu_dt_simple}
\end{align}
\end{subequations}
In general, $n$, $\rho$, $p$ and their derivatives cannot be written analytically unless the given species follows Maxwell-Boltzmann (MB) statistics and $m=0$, or unless $\mu = m=0$ for FD, BE and MB statistics. Hence, in general, one requires numerical integration in order to obtain thermodynamic quantities. In Appendix~\ref{app:Thermo_formulae} we list all the relevant thermodynamic formulae while in Appendix~\ref{app:MB_limit} we particularize the evolution equations for the case of MB statistics.

\vspace{0.2cm}
\subsection{Energy and number density transfer rates}\label{sec:interaction_rates}

In order to solve for the early Universe thermodynamics of a given system, the last ingredient needed is to calculate the expressions for $\delta n/\delta t$ and $\delta \rho/\delta t$ that enter Equations~\eqref{eq:dTdmu_generic_simple} and~\eqref{eq:dTdt_nochem}. This step is not generic and has to be carried out for each particular scenario. Here, in order to facilitate the procedure, we write down relevant formulae for decays, annihilations and scattering processes that can be useful in many cases. We write down all the expressions in the Maxwell-Boltzmann approximation. Namely, by considering $f_{\rm MB} = e^{-(E-\mu)/T}$. The spin-statistics correction to the rates typically represents a $\lesssim 20\%$ correction to the Maxwell-Boltzmann rate, but the advantage of the rates in the Maxwell-Boltzmann approximation is that they can be written analytically for $1\leftrightarrow 2$ processes and as a one-dimensional integral for $2\leftrightarrow2$ annihilation processes. We write down the main formulae here and the reader is referred to Appendix~\ref{app:rates_MB} for detailed calculations and for spin-statistics corrections to the rates in relevant cases.

\subsubsection*{Decay rates}\label{sec:BSM_recipe_decay}
The collision term for $1\leftrightarrow 2$ decay processes including quantum statistics can be written analytically~\cite{Kawasaki:1992kg,Starkman:1993ik}. Here, we outline the results for the most relevant decay process in which a particle $a$, characterized by $T_a$ and $\mu_a$, decays into two identical and massless states characterized by $T'$ and $\mu'$. In the Maxwell-Boltzmann approximation, the rates read as follows (see Appendix~\ref{app:rates_decay}):
\begin{align}
\frac{\delta n_a}{\delta t} &= \frac{\Gamma_a  g_a m_a^2 }{2 \pi ^2}\left[T' e^{\frac{2 \mu' }{T'}} K_1\left(\frac{m_a}{T'}\right)-T_a e^{\frac{\mu_a}{T_a}} K_1\left(\frac{m_a}{T_a}\right)\right]\,,\label{eq:dndt_decays}\\ 
\frac{\delta \rho_a}{\delta t} &= \frac{\Gamma_a  g_a m_a^3 }{2 \pi ^2}\left[T' e^{\frac{2 \mu' }{T'}} K_2\left(\frac{m_a}{T'}\right)-T_a e^{\frac{\mu_a}{T_a}} K_2\left(\frac{m_a}{T_a}\right)\right]\,,\label{eq:drhodt_decays}
\end{align} 
where $\Gamma_a$ is the decay rate at rest, $g_a$ is the number of internal degrees of freedom of the $a$ particle, $m_a$ is its mass, and $K_j$ are modified Bessel functions of the $j$ kind.

\vspace{1cm}
\subsubsection*{Annihilation rates}\label{sec:BSM_recipe_ann}
For the process $1+2\leftrightarrow 3+4$ where $T_1 = T_2 = T$, $\mu_1 = \mu_2 = \mu$, $T_3 = T_4 = T'$, and $\mu_3 = \mu_4 = \mu'$, the number and energy density transfer rate for the $1$ particle read (see Appendix~\ref{app:rates_ann}):
\begin{align}\label{eq:drho_dt_ann}
\frac{\delta n_1}{\delta t} &= \frac{g_1g_2}{8 \pi ^4} \int_{s_{\rm min}}^{\infty} d s  \,p_{12}^2\,\sqrt{s}\, \sigma(s)\, \left[T'\,e^{\frac{2\mu'}{T'}}\, K_1\left(\frac{\sqrt{s}}{T'}\right)-T \,e^{\frac{2\mu}{T}}\,K_1\left(\frac{\sqrt{s}}{T}\right)\right]\,,\\
\frac{\delta \rho_1}{\delta t} &=  \frac{ g_1g_2 }{16 \pi^4 }\, \int_{s_{\rm min}}^{\infty} d s  \,p_{12}^2 
  \left(s +m_2^2 -m_1^2\right) \! \sigma(s) \! \left[ T' \, e^{\frac{2\mu'}{T'}}\, K_2\left(\frac{\sqrt{s}}{T'}\right) -  T \, e^{\frac{2\mu}{T}}\, K_2\left(\frac{\sqrt{s}}{T}\right)  \right] \, ,\label{eq:deltan_dt_ann}
\end{align}
where $s$ is the Mandelstam variable, $s_{\rm min} = \text{min}[(m_1+m_2)^2, \, (m_3+m_4)^2]$,  $p_{12} = [s-(m_1+m_2)^2]^{1/2} [s-(m_1-m_2)^2]^{1/2} /(2\sqrt{s})$, and $\sigma(s)$ is the usual cross section for the process (namely, the cross section summed over final spin states and averaged over initial spins). 

These equations can be further simplified if all species are massless. Yielding:
\begin{align}\label{eq:drho_dt_ann_mless}
\frac{\delta n_1}{\delta t} &= \frac{g_1g_2}{32 \pi ^4} \int_{0}^{\infty} d s  \,s^{3/2}\, \sigma(s) \left[T'\,e^{\frac{2\mu'}{T'}}\, K_1\left(\frac{\sqrt{s}}{T'}\right)-T \,e^{\frac{2\mu}{T}}\,K_1\left(\frac{\sqrt{s}}{T}\right)\right]\,,\\
\frac{\delta \rho_1}{\delta t} &=  \frac{ g_1g_2 }{64 \pi^4 }\, \int_{0}^{\infty} d s  \,\,s^2\,\, \sigma(s)\,  \left[ T' \, e^{\frac{2\mu'}{T'}}\, K_2\left(\frac{\sqrt{s}}{T'}\right) -  T \, e^{\frac{2\mu}{T}}\, K_2\left(\frac{\sqrt{s}}{T}\right)  \right] \, .\label{eq:deltan_dt_ann_mless}
\end{align}
One must be careful with the spin factors. For example, for the energy transfer rate of a single neutrino species and for the process $\nu_\alpha \bar{\nu}_\alpha \leftrightarrow e^+ e^-$, $g_1 = g_2 = 1$. On the other hand, if one considers the energy transfer rate of an electron and the same process, then $g_1 = g_2 = 2$. Of course, the spin averaged/summed cross section takes into account this factor of 4 so that $\delta \rho_e/\delta t = - \delta \rho_\nu/\delta t$. 

\noindent We exemplify the use of Equations~\eqref{eq:drho_dt_ann_mless} and \eqref{eq:deltan_dt_ann_mless} for typical annihilation  cross sections:
\begin{align}\label{eq:drho_dt_ann_mless_examples}
\qquad \sigma(s)  &= A/s\,\,\,\,\,\,\, \,\longrightarrow \,\,\, \frac{\delta n_1}{\delta t} =A \frac{g_1 g_2}{8\pi^4} \, T'^4 e^{\frac{2\mu'}{T'}}\,,\,\,\,\,\,\,\,\,\,\,\,\,\qquad \frac{\delta \rho_1}{\delta t} =A \frac{g_1 g_2}{4\pi^4} \, T'^5 e^{\frac{2\mu'}{T'}} \,, \\
\sigma(s) &= A \,\frac{s}{M_\star^4} \,\,\,\longrightarrow \,\,\, \frac{\delta n_1}{\delta t}=\frac{A}{M_\star^4} \frac{24 g_1 g_2}{\pi^4} \, T'^8 e^{\frac{2\mu'}{T'}} \,,\qquad \frac{\delta \rho_1}{\delta t} = \frac{A}{M_\star^4} \frac{96 g_1 g_2}{\pi^4} \, T'^9 e^{\frac{2\mu'}{T'}}\,,
\end{align}
where we only show the $3+4\to 1+2$ contribution to $\delta n_1/\delta t$ and $\delta \rho_1/\delta t$ for concreteness.
\vspace{-0.1cm}
\subsubsection*{Scattering rates}\label{sec:BSM_recipe_scatt}
Scattering processes by definition yield $\delta n/\delta t= 0$ and if annihilation or decay interactions are efficient the energy exchange from scatterings is typically subdominant. For example, in the Standard Model, electron-positron annihilations to neutrinos are a factor of 5 times more efficient transferring energy than electron-neutrino scatterings~\cite{Escudero:2018mvt}. However, even if scattering interactions can typically be neglected in the energy transfer rates, their effect is very important since efficient scatterings distribute energy within the relevant species and lead to distribution functions that resemble equilibrium ones -- and this is the key approximation that leads to a simple evolution of the thermodynamics, see Section~\ref{sec:approximations}. 

Scattering rates are substantially more complicated to work out than decay or annihilation rates and simple expressions for them can only be found in very few cases. One such case is the scattering between two massless particles that interact via a four-point interaction. We perform the exact phase space integration and report the rates for such process in Appendix~\ref{app:MB_scatterings} in the Maxwell-Boltzmann approximation. We also provide spin-statistics corrections for them. Finally, we point the reader to~\cite{Bringmann:2006mu,Bringmann:2009vf,Binder:2016pnr} where, in the context of WIMP dark matter, scatterings of non-relativistic particles against massless ones have been investigated in detail.

\newpage 
\section{Neutrino Decoupling in the Standard Model}\label{sec:SM_neutrinodec}
In this section we solve for neutrino decoupling in the Standard Model. To this end, by using Equation~\eqref{eq:dTdt_nochem} we write down differential equations describing the time evolution of the temperature of the electromagnetic plasma $T_\gamma$ and the neutrino fluid $T_\nu$. This calculation is therefore fully analogous to the one performed by the author in~\cite{Escudero:2018mvt}, but here we update it by accounting for next-to-leading order finite temperature corrections~\cite{Bennett:2019ewm}, and spin-statistics and the electron mass in the $\nu$-$\nu$ and $\nu$-$e$ interaction rates. 

\subsection{Approximations}\label{sec:SM_neutrinodec_approx}
We solve for neutrino decoupling in the Standard Model by using the following well-justified approximations:

\begin{enumerate}
\item \textit{Neutrinos follow perfect Fermi-Dirac distributions}. Non-thermal corrections to an instantaneously decoupled neutrino distribution function are present in the Standard Model, but they encode less than $1\%$ of the total energy density carried by the neutrinos~\cite{deSalas:2016ztq,Mangano:2005cc}. Here we will show (see Section~\ref{sec:comparison}) that, in fact, thermal distribution functions can account very accurately for the energy density carried out by neutrinos. 
\item \textit{Neglect neutrino oscillations}. Neutrino oscillations are active at the time of neutrino decoupling~\cite{Hannestad:2001iy,Dolgov:2002ab,deSalas:2016ztq,Mangano:2005cc}. However, we neglect them on the grounds that their impact on $N_{\rm eff}$ in the Standard Model is small, $\Delta N_{\rm eff}^{\rm SM} = 0.0007$~\cite{deSalas:2016ztq}. As to approximately mimic the effect of neutrino oscillations we describe the neutrino fluid with a single temperature, $T_\nu = T_{\nu_e} =T_{\nu_\mu} =T_{\nu_\tau}$.
\item \textit{Neglect chemical potentials}. The number of photons is not conserved in the early Universe and the electron chemical potential is negligible given the very small baryon-to-photon ratio, $\mu_e/T_\gamma \sim 10^{-9}$. This therefore justifies setting $\mu_e = \mu_\gamma = 0$. In addition, in the Standard Model, neutrino chemical potentials can be neglected since interactions $\bar{\nu}\nu \leftrightarrow e^+e^- \leftrightarrow \gamma \gamma \gamma$ are highly efficient in the relevant temperature range. Nonetheless, we have explicitly checked that allowing for non-negligible neutrino chemical potentials does not render any significant change on any of the results presented in this Section, see Appendix~\ref{app:SM_chemical} for details.
\end{enumerate} 

\vspace{0.1cm}
\subsection{Temperature evolution equations}\label{sec:SM_Tevol}

At the time of neutrino decoupling, electrons and positrons are tightly coupled to photons via efficient annihilations and scatterings. Therefore we can model the electromagnetic sector by the photon temperature $T_\gamma = T_e$. Since chemical potentials are negligible, by using Equation~\eqref{eq:dTdt_nochem} we can write the time evolution of the photon temperature in the SM as:
\begin{align}\label{eq:T_gamma}
\frac{dT_{\gamma}}{dt}  =- \frac{  4 H \rho_{\gamma} + 3 H \left( \rho_{e} + p_{e}\right) +  \frac{\delta \rho_{\nu_e}}{\delta t} +2 \frac{\delta \rho_{\nu_\mu}}{\delta t} }{ \frac{\partial \rho_{\gamma}}{\partial T_\gamma} + \frac{\partial \rho_e}{\partial T_\gamma}  } \,.
\end{align}
Here, $\delta \rho_\nu/\delta t$ accounts for the energy transfer between one neutrino species (including the antineutrino) and the rest of the electromagnetic plasma.  $\delta \rho_\nu/\delta t$ enters this equation because energy conservation ensures $\delta \rho_e/\delta t = -  \delta \rho_{\nu_e}/\delta t - 2 \rho_{\nu_\mu}/\delta t $. 
\vspace{0.1cm}

We account for the next-to-leading order (NLO) Quantum Electrodynamics (QED) finite temperature correction to the electromagnetic energy and pressure densities from~\cite{Bennett:2019ewm}. Across the entire paper, however, when comparing with~\cite{deSalas:2016ztq,Mangano:2005cc} we shall use the leading order (LO) corrections from~\cite{Heckler:1994tv,Fornengo:1997wa} as used in~\cite{deSalas:2016ztq,Mangano:2005cc} to render a fair comparison. We denote the finite temperature corrections to the electromagnetic pressure and energy densities as: $P_{\rm int}$ and $\rho_{\rm int} = -P_{\rm int}+T_\gamma \frac{dP_{\rm int}}{dT_\gamma}$, respectively. Where, in the notation of~\cite{Bennett:2019ewm}, ${dP_{\rm int}}/{dT_\gamma}\equiv (2/3)\, T_\gamma^3 \, [G_2  (\frac{m_e}{T_\gamma})-\frac{m_e}{T_\gamma} \,G_1(\frac{m_e}{T_\gamma})]$ and ${d^2P_{\rm int}}/{dT_\gamma^2}   \equiv  2 \,T_\gamma^2\, G_2(\frac{m_e}{T_\gamma})$. Expressions for $P_{\rm int}$ and $G_{1,2}$ can be found in~\cite{Bennett:2019ewm}. In the code we simply use a tabulated version of $P_{\rm int}$ and its derivatives. 

By using Equation~\eqref{eq:dTdt_nochem}, we find the photon temperature evolution including finite temperature corrections to be:
\vspace{-0.08cm}
\begin{align}\label{eq:T_gamma_QED}
\frac{dT_{\gamma}}{dt}  =- \frac{  4 H \rho_{\gamma} + 3 H \left( \rho_{e} + p_{e}\right) + 3 H T_\gamma  \frac{d P_{\rm int}}{d T_\gamma}+  \frac{\delta \rho_{\nu_e}}{\delta t} +2 \frac{\delta \rho_{\nu_\mu}}{\delta t}  }{ \frac{\partial \rho_{\gamma}}{\partial T_\gamma} + \frac{\partial \rho_e}{\partial T_\gamma} + T_\gamma \frac{d^2 P_{\rm int}}{d T_\gamma^2} } \,.
\end{align}
\vspace{-0.08cm}
The neutrino temperature evolution can also be obtained from~\eqref{eq:dTdt_nochem} and simply reads:
\vspace{-0.05cm}
\begin{align}\label{eq:dTdt_nu_SM}
 \frac{dT_{\nu_\alpha}}{dt} = - H \, T_{\nu_\alpha} +   \frac{\delta \rho_{\nu_\alpha}}{\delta t}\bigg/\frac{\partial \rho_{\nu_\alpha}}{\partial T_{\nu_\alpha} }\,.
\end{align}
Note that the differential equations we have written for the neutrino temperature can be applied to each neutrino flavor $\alpha$ or to the collective neutrino fluid. Applying them to the entire neutrino fluid is a good way of taking into account the fact that neutrino oscillations become active prior to neutrino decoupling. 
\vspace{-0.05cm}
\subsection{Neutrino-neutrino and electron-neutrino interaction rates}\label{sec:SM_interactions}
\vspace{-0.05cm}
In order to solve the relevant system of differential equations, we need to calculate the energy transfer rates $\delta \rho_{\nu_\alpha}/\delta t$. We take into account all relevant processes: $e^+ e^- \leftrightarrow \bar{\nu}_\alpha \nu_\alpha $, $e^\pm \nu_\alpha \leftrightarrow e^\pm \nu_\alpha$, $e^\pm \bar{\nu}_\alpha \leftrightarrow e^\pm \bar{\nu}_\alpha$, $\nu_\alpha \nu_\beta \leftrightarrow \nu_\alpha \nu_\beta $,  $\nu_\alpha \bar{\nu}_\beta \leftrightarrow \nu_\alpha \bar{\nu}_\beta $, and $\bar{\nu}_\alpha \nu_\alpha \leftrightarrow \bar{\nu}_\beta {\nu}_\beta $. We consider the matrix elements reported in Table 2 of~\cite{Dolgov:2002wy}, and we follow the integration method developed in~\cite{Hannestad:1995rs} in order to reduce the collision term to just two dimensions\footnote{The reader is referred to the appendices of~\cite{Fradette:2018hhl} and~\cite{Kreisch:2019yzn} for recent and detailed descriptions of the method. The exact phase space reduced matrix elements are available from the author upon request.}. We write the energy transfer rate as a function of $T_\gamma$ and $T_\nu$ by making corrections to the analytical Maxwell-Boltzmann rates (see Appendix~\ref{app:rates_SM} for details):
\begin{subequations}\label{eq:energyrates_nu_SM}
\begin{align}
\left. \frac{\delta \rho_{\nu_e}}{\delta t} \right|_{\rm SM}^{\rm FD} &= \frac{G_F^2}{\pi^5}\left[4\left(g_{eL}^2+g_{eR}^2\right) \, F(T_\gamma,T_{\nu_e}) + 2 \, F(T_{\nu_\mu},T_{\nu_e}) \right] \, ,\\
\left. \frac{\delta \rho_{\nu_\mu}}{\delta t} \right|_{\rm SM}^{\rm FD} &= \frac{G_F^2}{\pi^5}\left[ 4\left(g_{\mu L}^2+g_{\mu R}^2\right) \, F(T_\gamma,T_{\nu_\mu}) -  F(T_{\nu_\mu},T_{\nu_e}) \right] \, ,
\end{align}
\end{subequations}
where, $G_F$ is Fermi's constant, and as relevant for the energies of interest ($E\ll M_Z$)~\cite{pdg,Erler:2013xha}\footnote{In the first version of the paper we used the tree level values for $g_L$ and $g_R$ with $s_W^2 = 1-M_W^2/M_Z^2= 0.223$. Current numbers account for radiative corrections. $\nu\!-\!e$ rates in this version are $1.8\%$ larger. }:
\begin{align}\label{eq:LowEnergyCouplings}
g_{eL} =0.727 \,,\qquad g_{eR} = 0.233 \,,\qquad g_{\mu L} = -0.273 \,,\qquad g_{\mu R} = 0.233 \,,\qquad 
\end{align} 
and where
\begin{align}\label{eq:G_FermiDirac}
F(T_1,T_2) &=  32 \, f_a^{\rm FD} \,  (T_1^9-T_2^9) + 56 \, f_s^{\rm FD}  \,T_1^4\,T_2^4 \, (T_1-T_2) \, .
\end{align} 
with $f_a^{\rm FD} = 0.884$, $f_s^{\rm FD} = 0.829$, and although the impact of the electron mass on the rates is small (see Appendix~\ref{app:rates_SM}), we incorporate it by interpolating over the exact and numerically precomputed rates including $m_e$. 

\subsection{Results}\label{sec:SM_results}
Here we present the results of solving the system of Equations~\eqref{eq:T_gamma_QED} and~\eqref{eq:dTdt_nu_SM}. We solve them starting from a sufficiently high temperature so that neutrino-electron interactions are highly efficient. In particular, we start the integration from $T_\gamma =T_\nu = 10\,\text{MeV}$, corresponding to $t_0 =1/(2H)|_{T = 10\,\text{MeV}}$. We track the system until $T_\gamma \sim 0.01\,\text{MeV}$, at which point neutrinos have fully decoupled from the plasma and electron-positron annihilation has taken place. We use $10^{-8}$ as the relative and absolute accuracies for the integrator which yields a high numerical accuracy. Given these settings, the typical CPU time used to integrate these equations in \texttt{Mathematica} is $\sim 20\,\text{s}$ and in \texttt{Python} $\lesssim 10\,\text{s}$. We have explicitly checked that the continuity equation is fulfilled at each integration step to a level of $10^{-5}$ or better.

\begin{figure}[t]
\centering
\hspace{0.8cm}\includegraphics[width=0.9\textwidth]{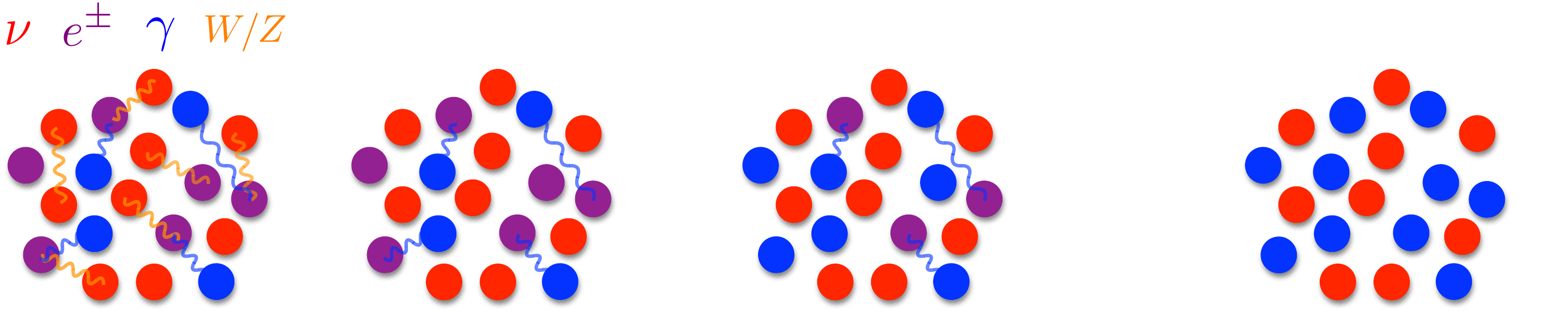}
\includegraphics[width=1.0\textwidth]{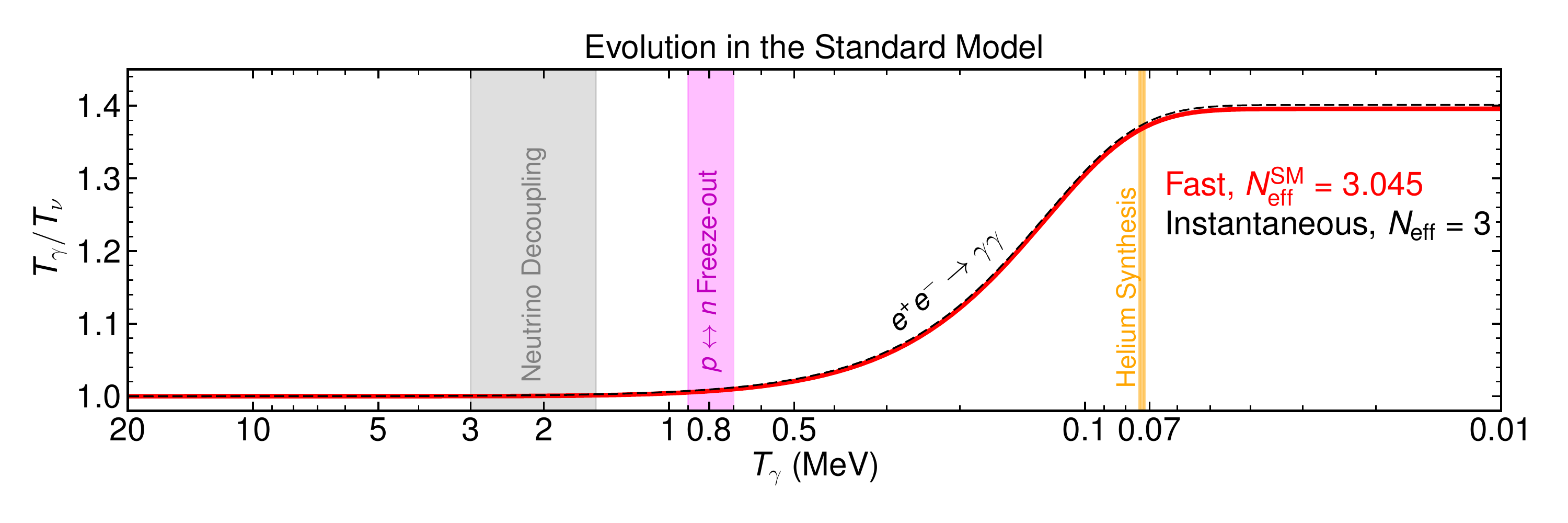}
\vspace{-0.7cm}
\caption{Neutrino temperature evolution in the Standard Model. Key cosmological events are highlighted: neutrino decoupling, proton-to-neutron freeze-out, and helium synthesis. The red line is the result obtained in this work that leads to $N_{\rm eff} = 3.045$ and the black dashed line corresponds to the case of instantaneous decoupling, leading to $N_{\rm eff} = 3$. The upper cartoons represent snapshots of the relative number density of neutrinos (red), electrons (purple) and photons (blue). A data file containing the thermodynamic evolution in the Standard Model can be found on~\href{https://arxiv.org/src/2001.04466/anc/SM_Thermo.dat}{ArXiv}.}\label{fig:SM_temperature}
\end{figure}

In Figure~\ref{fig:SM_temperature} we show the neutrino temperature evolution as a function of $T_\gamma$. We highlight some key cosmological events and also compare with instantaneous neutrino decoupling.

The central parameter to this study is the number of effective relativistic neutrino species as relevant for CMB observations: $N_{\rm eff}$. It is explicitly defined as:
\begin{align}\label{eq:Neff_def}
N_{\rm eff} \equiv  \frac{8}{7} \left(\frac{11}{4}\right)^{4/3} \left(  \frac{\rho_{\rm rad} - \rho_\gamma}{\rho_\gamma} \right) = 3  \left(\frac{11}{4}\right)^{4/3} \,  \left(\frac{T_{\nu}}{T_\gamma} \! \right)^4 \,,
\end{align} 
where in the last step we have assumed that the neutrinos follow perfect Fermi-Dirac distribution functions with negligible chemical potentials.  

In Table~\ref{tab:Neff_SM}, we report the resulting values of $N_{\rm eff}$ and $T_\gamma/T_\nu$ for $T_\gamma \ll m_e$ in the SM under various approximations and considering both a common neutrino temperature ($T_\nu = T_{\nu_e} =T_{\nu_\mu} =T_{\nu_\tau}  $) and by evolving separately $T_{\nu_e}$ and $T_{\nu_{\mu,\,\tau}}=T_{\nu_\mu} =T_{\nu_\tau}  $. \\ One of our main results is that -- by considering spin-statistics and $m_e$ in the $e$-$\nu$ and $\nu$-$\nu$ energy transfer rates -- we find $N_{\rm eff}^{\rm SM} = 3.045$. This result is in excellent agreement with state-of-the-art calculations of $N_{\rm eff}$ in the SM~\cite{deSalas:2016ztq,Mangano:2005cc} that account for non-thermal corrections to the neutrino distribution functions and neutrino oscillations. The reader is referred to Appendix~\ref{app:SM_comparison_indetail} for a comparison in terms of the neutrino distribution functions at $T\! \ll \! m_e$.

Thus, we have found that although describing the neutrino populations by a Fermi-Dirac distribution functions is just an approximation to the actual scenario, it suffices for the purpose of computing $N_{\rm eff}^{\rm SM}$ with a remarkable accuracy.

\begin{table}[t]
\begin{center}
\begin{tabular}{l|cc|ccc}
\hline\hline
Neutrino Decoupling in the SM & \multicolumn{2}{c|}{$T_{\nu_e} = T_{\nu_{\mu,\,\tau}}$} & \multicolumn{3}{c}{$T_{\nu_e} \neq T_{\nu_{\mu,\,\tau}}$} \\\hline
Scenario  	     		         &$T_\gamma/ T_{\nu}$	& $N_\text{eff}$ &$T_\gamma/ T_{\nu_e}$	&$T_\gamma/ T_{\nu_\mu}$	& $N_\text{eff}$ \\ \hline
Instantaneous decoupling 		   	&	1.4010 &	3.000  & 1.4010 & 1.4010 &	3.000	 \\
Instantaneous decoupling + LO-QED &	1.3997 &	3.011 & 1.3997  & 1.3997 &	3.011	 \\
Instantaneous decoupling + NLO-QED&	1.3998 &	3.010 & 1.3998  & 1.3998 &	3.010	 \\
MB collision term 			    	&	1.3961 &	3.043  & 1.3946   &1.3970  & 3.042	 \\
MB collision term + NLO-QED  		  &	1.3950 &	3.052  & 1.3935  & 1.3959 & 3.051 	 	 \\
FD collision term  	  		   	 &      1.3965 &	3.039  &  1.3951 & 1.3973 & 3.038	 	 \\
FD collision term  + NLO-QED	       	 &	1.3954 & 3.049	  & 1.3941  & 1.3962 &  3.048	 	 \\
FD+$m_e$ collision term  		   	 &	1.3969 & 3.036	  & 1.3957  & 1.3976 &  3.035	 \\\hline
FD+$m_e$ collision term + LO-QED   &	1.39568 & 3.046 & 1.3945  & 1.3964 &  3.045 	 	 \\
\textbf{FD+$m_e$ collision term + NLO-QED}   &	1.39578 & \textbf{3.045} & 1.3946  & 1.3965 &  \textbf{3.044} 	 	 \\
  \hline \hline
\end{tabular}
\end{center}\vspace{-0.3cm}
\caption{$N_{\rm eff}$ and $T_\gamma/T_\nu$ as relevant for CMB observations in the SM by taking different approximations and neglecting neutrino chemical potentials. The last row shows the case in which we account for both quantum statistics and the electron mass in the relevant collision terms, for which we find $N_{\rm eff}^{\rm SM} = 3.044-3.045$. Our results are in excellent agreement with state-of-the-art calculations that account for non-thermal neutrino distribution functions and neutrino oscillations~\cite{deSalas:2016ztq,Mangano:2005cc}.}\label{tab:Neff_SM}
\end{table}

\subsection{Comparison with previous calculations in the SM }\label{sec:comparison}

We compare our results for early and late Universe observables from our calculation of neutrino decoupling with state-of-the-art calculations in the Standard Model~\cite{deSalas:2016ztq,Mangano:2005cc,Pisanti:2007hk,Consiglio:2017pot}. These studies account for non-thermal neutrino distribution functions, finite temperature corrections, and neutrino oscillations in the primordial Universe. We compare our results in terms of $N_{\rm eff}$, the energy density of degenerate non-relativistic neutrinos ($\Omega_{\nu}h^2$), the effective number of species contributing to entropy density ($g_{\star s}$), and the primordial abundances of helium ($Y_{\rm P}$) and deuterium ($\text{D/H}_{\rm P}$). To obtain the relative differences in terms of $Y_{\rm P}$ and $\text{D/H}_{\rm P}$ we have modified the BBN code~\texttt{PArthENoPE}~\cite{Pisanti:2007hk,Consiglio:2017pot}. We refer the reader to Appendix~\ref{app:SM_comparison_indetail} for details.

In Table~\ref{tab:SM_summary} we outline our main results and comparison with previous state-of-the-art literature. We find an agreement of better than 0.1\% for any cosmological parameter. The accuracy of our approach in the Standard Model is well within the sensitivity of future CMB experiments to $N_{\rm eff}$~\cite{Benson:2014qhw,Ade:2018sbj, Abazajian:2019eic,Hanany:2019lle,DiValentino:2016foa,Sehgal:2019ewc} and of future measurements of the light element abundances~\cite{Cooke:2016rky,Grohs:2019cae}. 

Finally, in addition to accuracy, we stress that the two other key features of our approach are simplicity and speed. One needs to solve for a handful of ordinary differential equations and the typical execution time of \texttt{NUDEC\_BSM} is $\sim 20\,\text{s}$ in~\texttt{Mathematica} and $\sim 10\,\text{s}$ in \texttt{Python}. Thus, we believe this approach has all necessary features to be used to model early Universe BSM thermodynamics. This is the subject of study of the next sections. 

\begin{table}[t]
\begin{center}
\begin{tabular}{l|ccccc}
\hline\hline
\multicolumn{6}{c}{Neutrino Decoupling in the Standard Model: Key Parameters and Observables}  \\ \hline
Parameter      	   & $N_{\rm eff}$ &  $ Y_{\rm P} $   &$ \text{D/H}|_{\rm P} $ & $g_{\rm \star s}$ & $\sum m_\nu/\Omega_\nu h^2$   \\ \hline
This work   	  &  3.045        &   - & -     & 3.931  &	 93.05 eV	 \\
Difference w.r.t. instantaneous $\nu$-dec    &  1.5 \%    &   0.1 \% &  0.4 \%  & 0.6 \%  &	 1.2 \%        	 \\
Difference w.r.t.~\cite{deSalas:2016ztq,Mangano:2005cc,Pisanti:2007hk,Consiglio:2017pot}    &  0.03 \%          &   0.008 \% &  0.08 \% & 0.05 \%  &	 0.09 \%  	 \\
\hline
Current precision~\cite{planck,pdg}   &  5-6 \%       &   1.2 \% &  1.1 \%   &-  &	 -  	 \\
Future precision~\cite{Abazajian:2019eic,Ade:2018sbj,Cooke:2016rky,Grohs:2019cae}  &  1-2 \%  &   $<1$ \% &  0.1? \%     & -  &	 -      	 \\
\hline\hline
\end{tabular}
\end{center}\vspace{-0.5cm}
\caption{Standard Model results as obtained in this work (\href{https://github.com/MiguelEA/nudec_BSM}{NUDEC\_BSM}) for relevant cosmological parameters by solving for the time evolution of $T_\nu$ and $T_\gamma$ neglecting chemical potentials. The values of $Y_{\rm P} $ and $\text{D/H}_{\rm P} $ have been calculated using a modified version of~\texttt{PArthENoPE}~\cite{Pisanti:2007hk,Consiglio:2017pot}. We compare our results in terms of $N_{\rm eff}$, $g_{\rm \star s}$ and $\sum m_\nu/\Omega_\nu h^2$ against state-of-the-art calculations in the SM~\cite{deSalas:2016ztq,Mangano:2005cc}. We compare the results in terms of $Y_{\rm P} $ and $\text{D/H}|_{\rm P} $ against the default mode of~\texttt{PArthENoPE}~\cite{Pisanti:2007hk,Consiglio:2017pot} that includes non-instantaneous neutrino decoupling as obtained in~\cite{Mangano:2005cc}. Current precision on $N_{\rm eff}$ is from Planck~\cite{planck} and on $Y_{\rm P} $ and $\text{D/H}|_{\rm P} $ is from the PDG~\cite{pdg}. Expected future precision on $N_{\rm eff}$ is from the Simons Observatory~\cite{Ade:2018sbj} and CMB-S4~\cite{Abazajian:2019eic}. Expected precision on $Y_{\rm P} $ and $\text{D/H}|_{\rm P} $ is from~\cite{Cooke:2016rky,Grohs:2019cae}. }\label{tab:SM_summary}
\end{table}
\section{A Very Light and Weakly Coupled Neutrinophilic Boson}\label{sec:boson_1to2}

In this section we study the thermal history of the Universe in the presence of a very light ($1\,\text{eV} < m_\phi < 1\,\text{MeV}$) and weakly coupled ($\lambda < 10^{-9}$) neutrinophilic scalar: $\phi$. This is prototypically the case of majorons~\cite{Chikashige:1980ui,Gelmini:1980re,Schechter:1981cv,Georgi:1981pg} where the very small coupling strengths are associated to the small neutrino masses, and where sub-MeV $\phi$ masses are consistent with the explicit breaking of global symmetries by gravity. We will assume that the neutrinophilic scalar posses the same couplings to all neutrino flavors. Furthermore, we note that even if the scalar does not couple with the same strength to each neutrino flavor, neutrino oscillations in the early Universe~\cite{Hannestad:2001iy,Dolgov:2002ab} will render $T_{\nu_e} \simeq T_{\nu_\mu} \simeq T_{\nu_\tau}$ regardless. 

We begin by explicitly defining the model we consider. We follow by posing and solving the relevant system of differential equations for the temperature and chemical potentials for the neutrinos and the $\phi$ scalar. We solve this system of equations and briefly comment on the phenomenology in the region of parameter space where neutrino-scalar interactions are highly efficient in the early Universe. We finally write down the Liouville equation for the neutrino-scalar system. We solve it and compare with the results as output by the method proposed in this work (\href{https://github.com/MiguelEA/nudec_BSM}{NUDEC\_BSM}). We find an excellent agreement between the two approaches.

\subsection{The Model}\label{sec:model_nu}

The interaction Lagrangian that describes this model is
\begin{align}\label{eq:L_majoron}
\mathcal{L}_{\rm int} = \frac{\lambda}{2}\, \phi\, \sum_i \,\bar{\nu}_i \gamma_5 \nu_i\,,
\end{align}
where $\lambda$ is a coupling constant, $\nu_i$ corresponds to a neutrino mass eigenstate, we have assumed that neutrinos are Majorana particles, and we will restrict ourselves to $1\,\text{eV} < m_\phi < 1\,\text{MeV}$. In addition, for concreteness, we shall consider that there is no primordial abundance of $\phi$ particles and we shall focus on the regime in which the neutrino-$\phi$ interaction strength is very small $\lambda < 10^{-8}$.  In this regime, $2\leftrightarrow 2$ interactions are completely negligible. 

\begin{figure}[t]
\centering
\begin{tabular}{cc}
 \hspace{0.5cm} \includegraphics[width=0.525\textwidth]{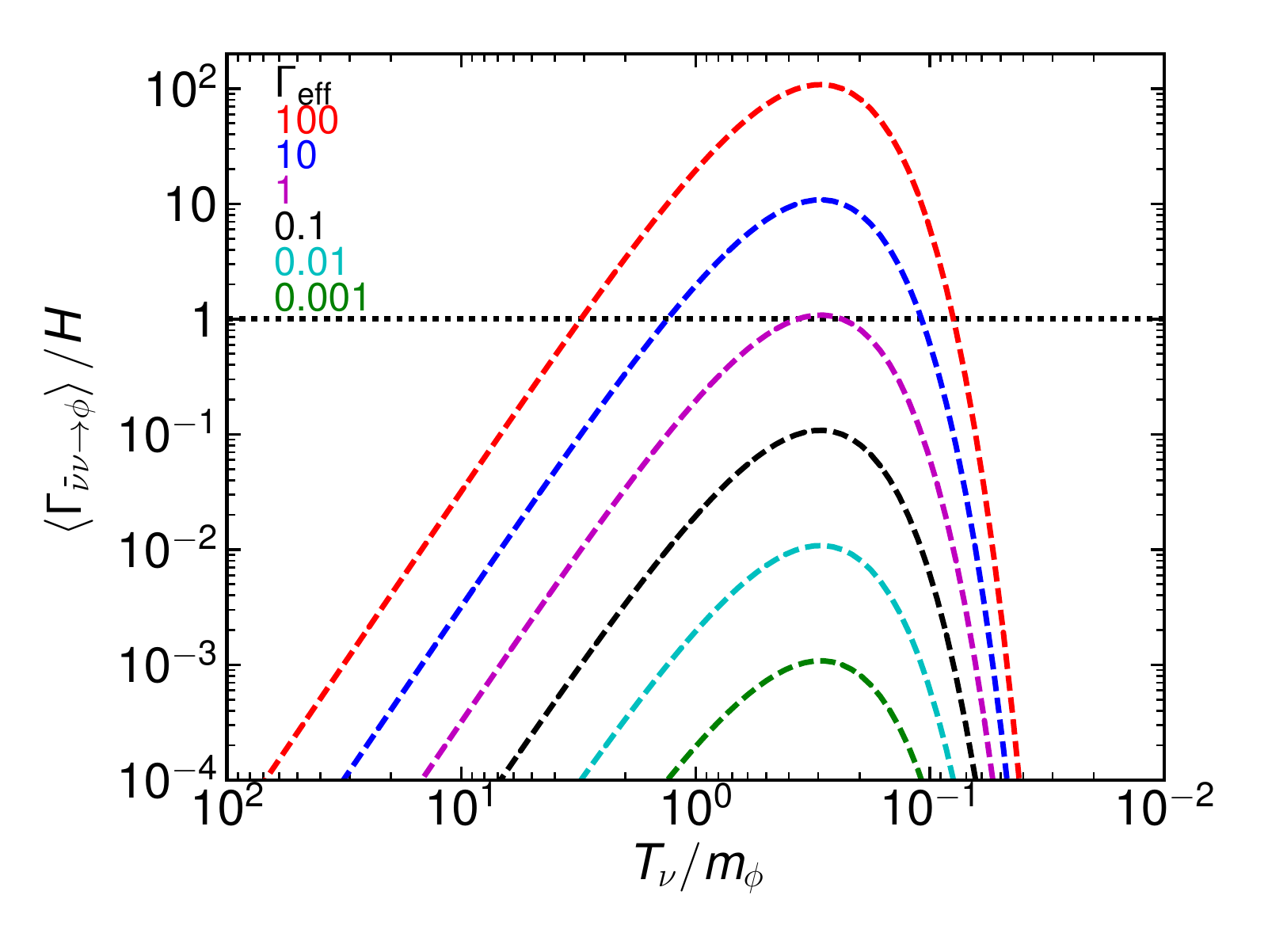} & \hspace{-1.7cm}  \includegraphics[width=0.525\textwidth]{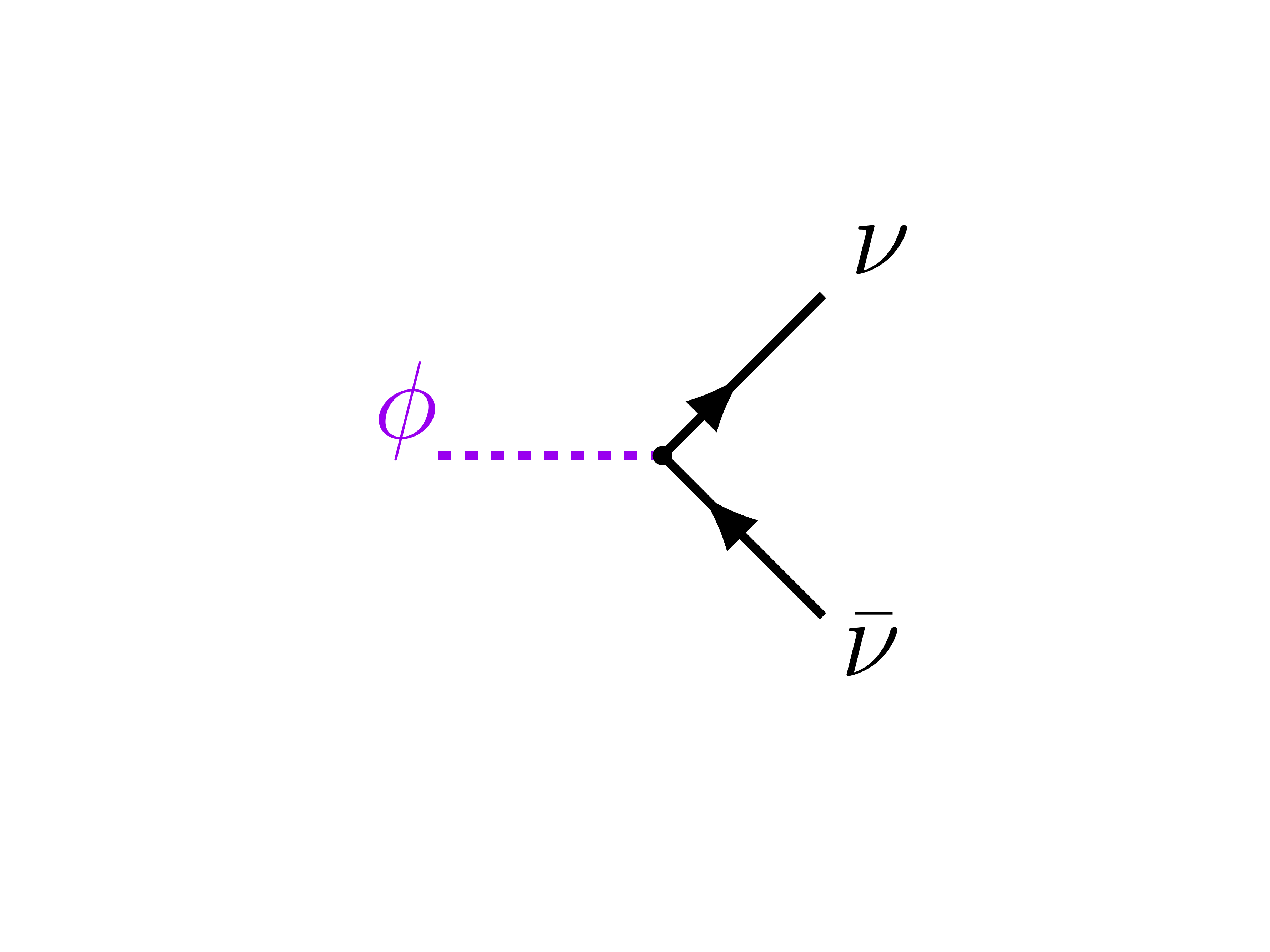}    \\
 \end{tabular}
\vspace{-0.4cm}
\caption{\textit{Left:} Ratio of the thermally averaged neutrino inverse decay rate ($\bar{\nu}\nu\to\phi$) to the Hubble parameter as a function of temperature for different neutrino-$\phi$ interaction strengths, $\Gamma_{\rm eff}$~\eqref{eq:Gammaeff_def}. \textit{Right:} $\phi\to \bar{\nu}\nu$ decay diagram.}\label{fig:Majoron_Rate_Neff}
\end{figure}

Therefore, in this scenario, the only cosmologically relevant processes are decays of $\phi$ into neutrinos and neutrino inverse decays -- as depicted in Figure~\ref{fig:Majoron_Rate_Neff}. The decay width at rest of $\phi$ reads:
\begin{align}
\Gamma_{\phi} &= \frac{\lambda^2}{16\pi} m_\phi \,\sum_i \sqrt{1-\frac{4m_{\nu_i}^2}{m_\phi^2}}  \simeq \frac{3\lambda^2}{16\pi} m_\phi   \, ,
\end{align}
where in the last step we have neglected neutrino masses.

Given the decay width at rest, the number and energy density transfer rates as a result of $\phi \leftrightarrow \bar{\nu}\nu$ interactions, in the Maxwell-Boltzmann approximation, are given by Equations~\eqref{eq:dndt_decays} and~\eqref{eq:drhodt_decays} respectively so that:
\begin{subequations}\label{eq:dndrho_dt_maj}
\begin{align}
\frac{\delta n_\phi}{\delta t} &= \frac{\Gamma_\phi  m_\phi^2 }{2 \pi ^2}\left[T_\nu e^{\frac{2 \mu_\nu }{T_\nu}} K_1\left(\frac{m_\phi}{T_\nu}\right)-T_\phi e^{\frac{\mu_\phi}{T_\phi}} K_1\left(\frac{m_\phi}{T_\phi}\right)\right]\,,\\ 
\frac{\delta \rho_\phi}{\delta t} &= \frac{\Gamma_\phi   m_\phi^3 }{2 \pi ^2}\left[T_\nu e^{\frac{2 \mu_\nu }{T_\nu}} K_2\left(\frac{m_\phi}{T_\nu}\right)-T_\phi e^{\frac{\mu_\phi}{T_\phi}} K_2\left(\frac{m_\phi}{T_\phi}\right)\right]\,.
\end{align}
\end{subequations}
Since the relevant process is $1\leftrightarrow2$, the collective neutrino fluid interaction rates fulfil: $\delta \rho_\nu/\delta t = - \delta \rho_\phi/\delta t $ and $\delta n_\nu/\delta t = - 2\delta n_\phi/\delta t $.

In order to efficiently explore the parameter space and to understand the extent to which departures from thermal equilibrium occur, it is convenient to introduce the effective interaction strength rate:
\begin{align}\label{eq:Gammaeff_def}
\Gamma_{\rm eff} &\equiv \left(\frac{\lambda}{4\times 10^{-12}}\right)^2 \left(\frac{\text{keV}}{m_\phi}\right)   \simeq \frac{\left< \Gamma_{\bar{\nu}\nu \to \phi}\right>}{H} = \frac{1}{H(T_\nu = m_\phi/3)}  \frac{1}{n_\nu} \frac{\delta n_\nu}{\delta t}\bigg|_{\bar{\nu}\nu \to \phi}  \,.
\end{align}
Figure~\ref{fig:Majoron_Rate_Neff} shows the temperature dependence of the thermally averaged interaction rate $\bar{\nu}\nu \to \phi$, and we can clearly appreciate that if $\Gamma_{\rm eff} > 1$ then thermal equilibrium will have been reached between $\phi$ and $\nu$ species.

\vspace{1cm}
\subsection{Evolution equations}\label{sec:evol_boson_1to2}

In order to describe the early Universe thermodynamics of the scenario described above we follow the method outlined in Section~\ref{sec:BSM_neutrinodec}. We can simply use two sets of Equations~\eqref{eq:dTdmu_generic_simple} tracking the scalar and neutrino temperature and chemical potentials ($T_\phi$, $\mu_\phi$, $T_\nu$, $\mu_\nu$) sourced with the number and energy density transfer rates induced by $\phi \leftrightarrow \bar{\nu}\nu$ interactions~\eqref{eq:dndrho_dt_maj}. For concreteness, we shall assume that $\phi \leftrightarrow \bar{\nu}\nu$ interactions become relevant after electron-positron annihilation (although including the effect of $e^+e^-$ annihilations is straightforward). 

Explicitly, the evolution equations describing the thermodynamics read:
\begin{subequations}\label{eq:full_system_maj}
\begin{align}
\frac{dT_\phi}{dt} &=\frac{1}{\frac{\partial n_\phi}{\partial \mu_\phi} \frac{\partial \rho_\phi}{\partial T_\phi}-\frac{\partial n_\phi}{\partial T_\phi} \frac{\partial \rho_\phi}{\partial \mu_\phi} }\left[ -3 H  \left((p_\phi+\rho_\phi)\frac{\partial n_\phi}{\partial \mu_\phi}-n_\phi \frac{\partial \rho_\phi}{\partial \mu_\phi} \right)+ \frac{\partial n_\phi}{\partial \mu_\phi}  \frac{\delta \rho_\phi}{\delta t} - \frac{\partial \rho_\phi}{\partial \mu_\phi}  \frac{\delta n_\phi}{\delta t} \right] , 
 \\
\frac{d\mu_\phi}{dt} &=\frac{-1}{\frac{\partial n_\phi}{\partial \mu_\phi} \frac{\partial \rho_\phi}{\partial T_\phi}-\frac{\partial n_\phi}{\partial T_\phi} \frac{\partial \rho_\phi}{\partial \mu_\phi} } \left[ -3 H \left((p_\phi+\rho_\phi)\frac{\partial n_\phi}{\partial T_\phi}-n_\phi \frac{\partial \rho_\phi}{\partial T_\phi} \right)+ \frac{\partial n_\phi}{\partial T_\phi}  \frac{\delta \rho_\phi}{\delta t} - \frac{\partial \rho_\phi}{\partial T_\phi} \frac{\delta n_\phi}{\delta t} \right]   , \\
\frac{dT_\nu}{dt} &=\frac{1}{\frac{\partial n_\nu}{\partial \mu_\nu} \frac{\partial \rho_\nu}{\partial T_\nu}-\frac{\partial n_\nu}{\partial T_\nu} \frac{\partial \rho_\nu}{\partial \mu_\nu} }\left[ -3 H  \left((p_\nu+\rho_\nu)\frac{\partial n_\nu}{\partial \mu_\nu}-n_\nu \frac{\partial \rho_\nu}{\partial \mu_\nu} \right)-\! \frac{\partial n_\nu}{\partial \mu_\nu}  \frac{\delta \rho_\phi}{\delta t} + 2 \frac{\partial \rho_\nu}{\partial \mu_\nu}  \frac{\delta n_\phi}{\delta t} \right] , 
 \\
\frac{d\mu_\nu}{dt} &=\frac{-1}{\frac{\partial n_\nu}{\partial \mu_\nu} \frac{\partial \rho_\nu}{\partial T_\nu}-\frac{\partial n_\nu}{\partial T_\nu} \frac{\partial \rho_\nu}{\partial \mu_\nu} } \left[ -3 H \left((p_\nu+\rho_\nu)\frac{\partial n_\nu}{\partial T_\nu}-n_\nu \frac{\partial \rho_\nu}{\partial T_\nu} \right)-\! \frac{\partial n_\nu}{\partial T_\nu}  \frac{\delta \rho_\phi}{\delta t} +2 \frac{\partial \rho_\nu}{\partial T_\nu} \frac{\delta n_\phi}{\delta t} \right]  ,  \\
\frac{dT_\gamma}{dt} &= -H \, T_\gamma \,,
\end{align}
\end{subequations}
where $\delta n_\phi/\delta t $ and $\delta \rho_\phi/\delta t$ are given by~\eqref{eq:dndrho_dt_maj} and the thermodynamic quantities of the neutrinos correspond to all neutrinos and antineutrinos, namely, to a Fermi-Dirac fluid with $g_\nu = 6$. 

\vspace{0.2cm}
\noindent \textit{Initial Conditions:}\vspace{0.1cm}

We solve this set of equations with initial conditions: 
\begin{align}\label{eq_initial_con}
T_\nu/m_\phi= 100 \,, \,\,T_\gamma/T_\nu = 1.39578\,, \,\,\mu_\nu /T_\nu =  -10^{-4}\,, \,\,T_\phi/T_\nu = 10^{-3}\,, \,\,\mu_\phi/T_\nu = -10^{-5} \,,
\end{align}
and $t_0 = 1/(2H)$. The initial condition for $T_\gamma/T_\nu$ corresponds to the temperature ratio as obtained after $e^+e^-$ annihilation in the Standard Model. The initial conditions for $T_\phi$ and $\mu_\phi$ ensure that the integration starts with a tiny abundance of $\phi$ particles, $\rho_\phi/\rho_\nu|_{t_0} < 10^{-12}$. In practice, when solving the system of Equations~\eqref{eq:full_system_maj} we split the integration in two. We integrate from $T_\nu/m_\phi= 100 $ until $T_\nu/m_\phi= 20$ setting $m_\phi = 0$ in the thermodynamic formulae since that allows for a great simplification and it is clearly a good approximation in that regime. Then, we ran from $T_\nu/m_\phi= 20$ using thermodynamic formulae with $m_\phi \neq 0$. 

\vspace{0.2cm}
\noindent \textit{Parameter Space:}\vspace{0.1cm}

We solve the set of  Equations~\eqref{eq:full_system_maj} for the range $10^{-6}<\Gamma_{\rm eff} < 10^4$. Since we consider $1\,\text{eV}<m_\phi <1\,\text{MeV}$, this corresponds to coupling strengths $10^{-17}<\lambda <10^{-8}$. Note that the results presented here will only depend upon $\Gamma_{\rm eff}$ and apply to the entire range $1\,\text{eV}<m_\phi <1\,\text{MeV}$. The reason is as follows: for such a mass range, $\phi$ particles become cosmologically relevant in a radiation dominated Universe and the evolution is only dependent upon ratios of temperature/chemical potential to $m_\phi$. Thus, for one single value of $m_\phi$, the results can be mapped into the entire range of $\phi$ masses for a given $\Gamma_{\rm eff}$. For $m_\phi < 1\,\text{eV}$ one would simply need to include the energy densities of dark matter and baryons, which is straightforward. 

\subsection{Thermodynamics in the strongly interacting regime}
If $\Gamma_{\rm eff} \gg 1$, then $\phi \leftrightarrow \bar{\nu}\nu$ interactions are highly efficient in the early Universe and thermal equilibrium is reached between the $\nu$ and $\phi$ populations. In this regime, one can find the thermodynamic evolution of the joint $\nu$-$\phi$ system independently of $\Gamma_{\rm eff}$ and without the actual need to solve for Equations~\eqref{eq:full_system_maj}\footnote{This is analogous to what happens in the SM with electron-positron annihilation. The net heating of the photon bath does not depend upon the interaction strength of electromagnetic interactions but only upon the number of internal degrees of freedom of the electrons and photons and their fermionic and bosonic nature.}.

 When $\Gamma_{\rm eff}\gg 1$, the $\phi$ scalars thermalize with neutrinos while relativistic, $T \gg m_\phi$. Since at such temperatures all particles can be regarded as massless, entropy is conserved and a population of $\phi$ particles is produced at the expense of neutrinos. In this regime, energy and number density conservation can be used to find an expression for the temperature and chemical potential of the joint $\nu$-$\phi$ system after equilibration occurs. By denoting $T_\nu$ as the temperature before equilibration, and $T_{\rm eq}$ and $\mu_{\rm eq}$ the temperature and chemical potential of the neutrinos after equilibration, we can express the previous conditions as:
\begin{align}
\rho_\nu(T_\nu, 0) &= \rho_\nu(T_{\rm eq}, \mu_{\rm eq}) + \rho_\phi(T_{\rm eq}, 2\mu_{\rm eq})\,, \\
n_\nu(T_\nu, 0) &= n_\nu(T_{\rm eq}, \mu_{\rm eq}) + 2 n_\phi(T_{\rm eq}, 2\mu_{\rm eq}) \, ,
\end{align}
where $\mu_{\rm eq}^{\phi}  = 2\mu_{\rm eq}^{\nu }$ as required from the fact that $\phi \leftrightarrow \bar{\nu}\nu$ processes are highly efficient. We can easily solve these equations to find:
 \begin{align}
T_{\rm eq} = 1.120818 \, T_\nu, \qquad \mu_{\rm eq} = -0.64199 \, T_\nu \, .
\end{align}
This result bounds the maximum energy density of the $\phi$ species to be:
 \begin{align}\label{eq:thermal_rhophi}
\frac{\rho_\phi(T_{\rm eq}, 2\mu_{\rm eq})}{\rho_\nu(T_{\rm eq}, \mu_{\rm eq})+\rho_\phi(T_{\rm eq}, 2\mu_{\rm eq})} \simeq 0.09 \,\qquad \text{and equivalently} \qquad \frac{\rho_\phi}{T_\gamma^4} < 0.041 \,.
\end{align}
Therefore, the energy density in $\phi$'s represents less than $10\%$ of the total $\nu$-$\phi$ system. 

Once thermal equilibrium is reached the thermodynamic evolution of the system is fixed. This is a result of the fact that in thermal equilibrium entropy is conserved in the joint $\nu$-$\phi$ system, and we know that for every $\phi$ that decays a pair of $\bar{\nu} \nu$ are produced. This allows us to write down two conservation equations for the entropy and number densities: 
\begin{subequations}\label{eq:continuity_nuphi}
\begin{align}
\left[s_{\nu}(T',\mu')+s_\phi(T',2\mu')\right] a'^3  &= \left[s_{\nu}(T,\mu)+s_\phi(T,2\mu)\right] a^3 \,, \\
\left[n_{\nu}(T',\mu')+2n_{\phi}(T',2\mu')\right]a'^3  & =\left[n_{\nu}(T,\mu)+2n_{\phi}(T,2\mu)\right]a^3  \,.
\end{align}
\end{subequations}
And therefore, this set of equations -- when solved simultaneously -- provide the evolution of $T'$ and $\mu'$ as a function of the scale factor $a'$. In this case: $T_\gamma' a'= T_\gamma a $, and we can find the evolution as a function of $T_\gamma$. We can solve the system starting from the equilibrium conditions when $T \gg m_\phi$ until $T \ll m_\phi$, so that the $\phi$ population has disappeared from the plasma. By following this procedure we find that after the $\phi$ particles decay, the temperature and neutrino chemical potentials are:
\begin{align}\label{eq:full_thermalization}
T_\gamma/T_\nu = 1.33632\,,\qquad T_\nu/\mu_\nu = -7.01941\,,\qquad (T \ll m_\phi,\,\Gamma_{\rm eff} \gg 1)\,,
\end{align}
which implies 
\begin{align}\label{eq:DNeff_majoron_thermal}
\Delta N_{\rm eff} = N_{\rm eff}-N_{\rm eff}^{\rm SM} = 3.163 -3.045 =  0.118 \simeq 0.12 \,.
\end{align}
This is in excellent agreement with the results of~\cite{Chacko:2003dt}. Note that so long as $\Gamma_{\rm eff} \gg 1$, these expressions are independent of the actual coupling strength.

\begin{figure}[t]
\centering
\hspace{0.8cm}\includegraphics[width=0.9\textwidth]{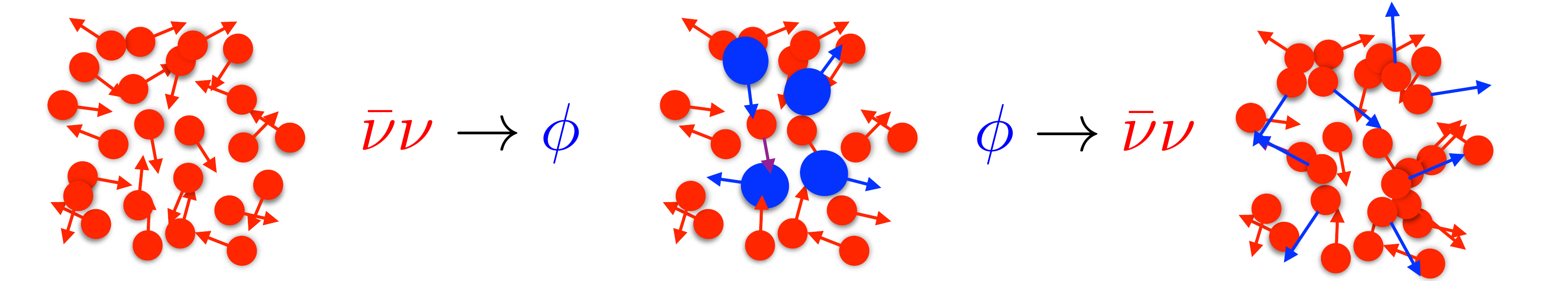}
\includegraphics[width=1.0\textwidth]{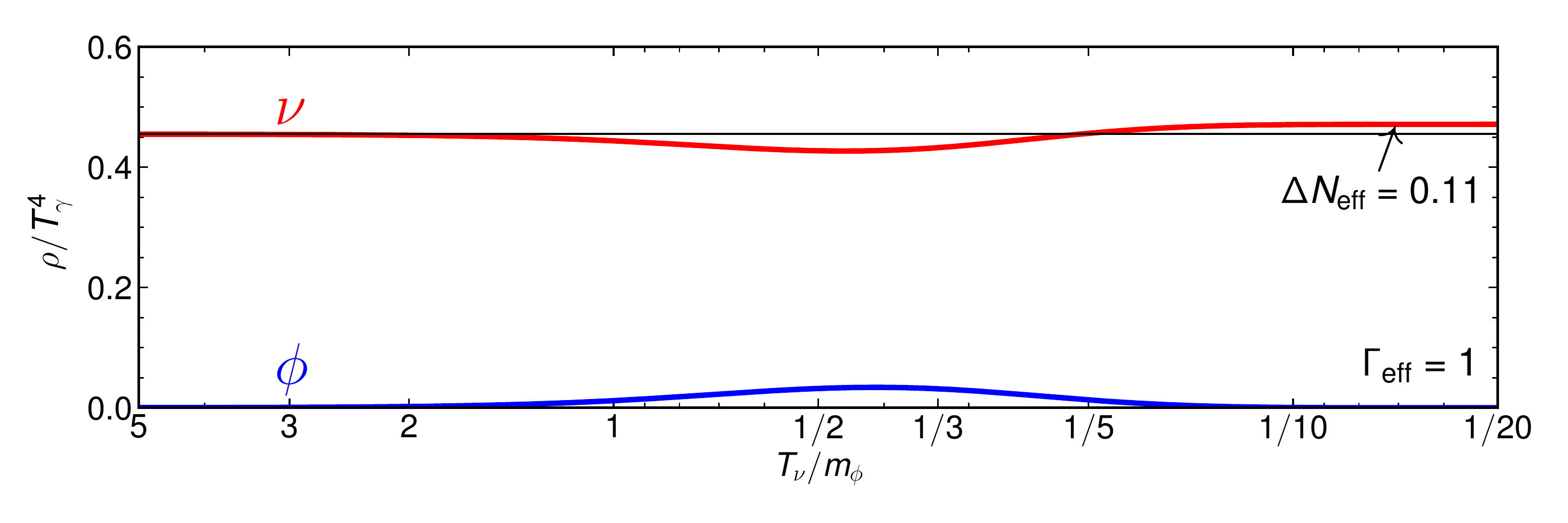}
\vspace{-0.7cm}
\caption{Energy density evolution of a neutrinophilic scalar $\phi$ (blue) coupled to neutrinos (red) in a region of parameter space in which $\phi$-decays to neutrinos and neutrino inverse decays are the only relevant processes. The interaction strength has been chosen to be $\Gamma_{\rm eff} = 1$, which corresponds to a neutrino-$\phi$ coupling strength of $\lambda = 4\times 10^{-12}\,\sqrt{m_\phi/\text{keV}}$. The sketch illustrates an approximate snapshot in comoving space of the joint thermodynamic system. At very high temperatures, the $\bar{\nu}\nu\to\phi$ process is thermally suppressed as a result of the high boost of neutrinos which suppresses the phase space for producing very light $\phi$ particles. When $T_\nu \sim m_\phi$, $\bar{\nu}\nu\to \phi$ processes start to be efficient. As soon as $T_\nu \sim m_\phi/3$ the $\phi$ particles feel they are non-relativistic and decay to neutrinos. However, by then, the $\phi$ population is non-relativistic and the decay products from $\phi\to \bar{\nu}\nu$ are more energetic than the typical neutrino in the plasma, thereby yielding $\Delta N_{\rm eff} = 0.11$ as relevant for CMB observations for $1\,\text{eV} \lesssim m_\phi \lesssim \text{MeV}$. The reader is referred to~\cite{Chacko:2003dt} and~\cite{Escudero:2019gvw} for the cosmological implications of this scenario.  }\label{fig:Majoron_evolution_nu}
\end{figure}

\subsection{Results}\label{sec:results_1to2}

In this section we present the solution to the thermodynamic evolution of the early Universe in the presence of a light and weakly coupled neutrinophilic scalar. 

First, in Figure~\ref{fig:Majoron_evolution_nu} we show the evolution of the energy density in neutrinos and $\phi$ species together with an sketch of the joint thermodynamic system. We can appreciate that the main effect of efficient $\phi \leftrightarrow \bar{\nu}\nu$ interactions $(\Gamma_{\rm eff} > 1)$ is to render a siezable population of $\phi$ particles at $T\gtrsim m_\phi$. As a result of the fact that these particles have a mass, when the temperature of the Universe is $T \lesssim m_\phi$, they do not redshift as pure radiation and upon decay they render a more energetic neutrino population (relative to the photons) than the one at $T\gg m_\phi$. 

\begin{figure}[t]
\centering
 \includegraphics[width=0.53\textwidth]{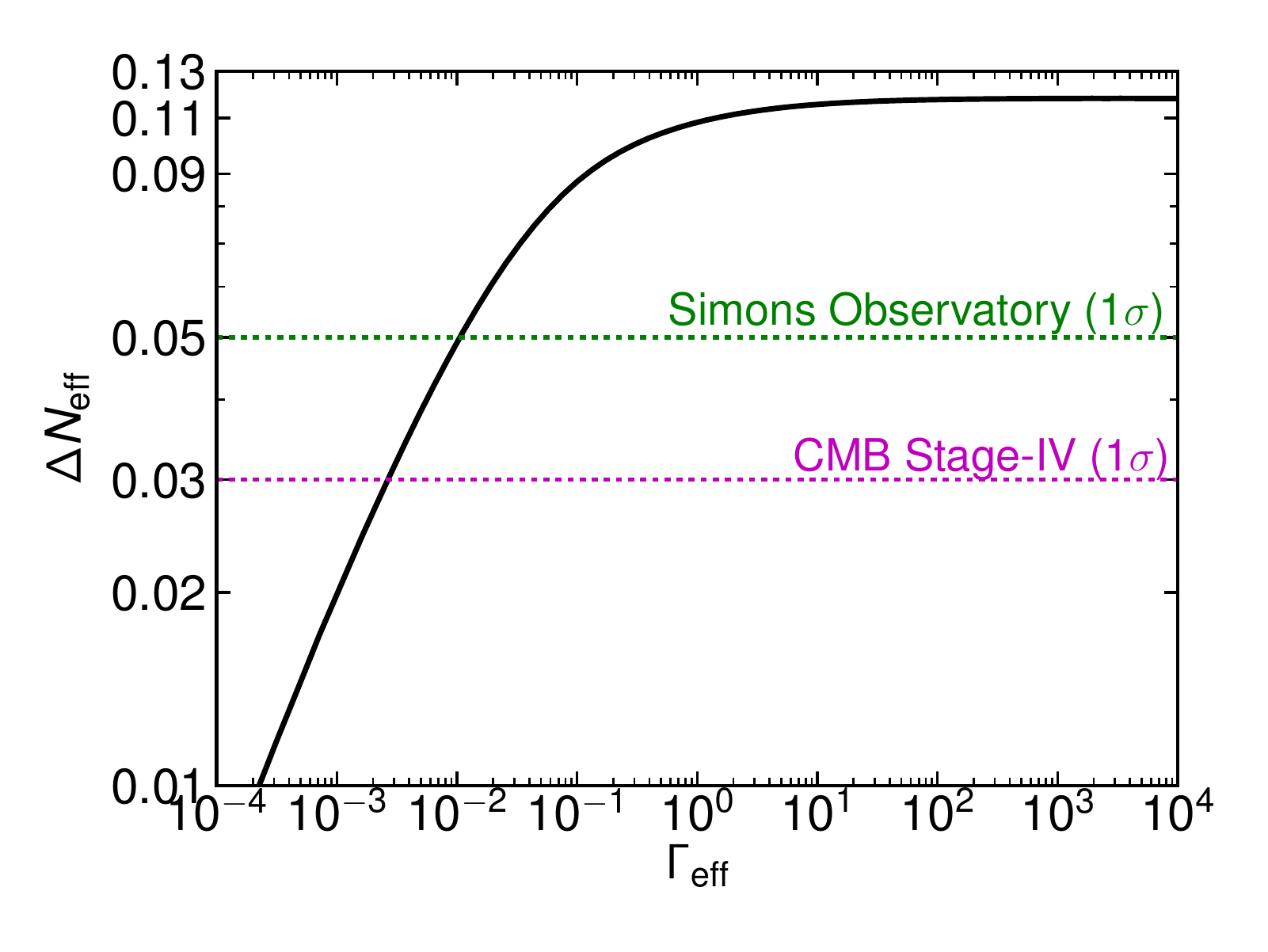}
\vspace{-0.4cm}
\caption{$\Delta N_{\rm eff}$ as a function of $\Gamma_{\rm eff}$ as relevant for CMB observations and as obtained by solving Equations~\eqref{eq:full_system_maj}. The $1\sigma$ sensitivities for the Simons Observatory~\cite{Ade:2018sbj} and CMB-S4~\cite{Abazajian:2019eic} are shown for illustration. We appreciate that when $\Gamma_{\rm eff} \gg 1$ $\Delta N_{\rm eff}$ matches the perfect thermal equilibrium prediction in~\eqref{eq:DNeff_majoron_thermal}.  }\label{fig:Majoron_Thermo_Neff}
\end{figure}

In Figure~\ref{fig:Majoron_Thermo_Neff} we show the resulting value of $\Delta N_{\rm eff}$ as a function of $\Gamma_{\rm eff}$. We can clearly appreciate that when $\Gamma_{\rm eff}\gg 1$, $\Delta N_{\rm eff}$ asymptotically reaches the value obtained in~\eqref{eq:DNeff_majoron_thermal}. We see that wide regions of parameter space in this scenario are within the reach of next generation of CMB experiments~\cite{Abazajian:2019eic,Ade:2018sbj}. We note that for all the cases presented in this study, we have explicitly checked that the continuity equation~\eqref{eq:rho_continuity} is fulfilled at each integration step with a relative accuracy of better than $10^{-5}$. Therefore the resulting values of $\Delta N_{\rm eff}$ are numerically accurate to a level of $\sim0.0001$.

\begin{figure}[t]
\centering
\begin{tabular}{cc}
\hspace{-0.8cm} \includegraphics[width=0.525\textwidth]{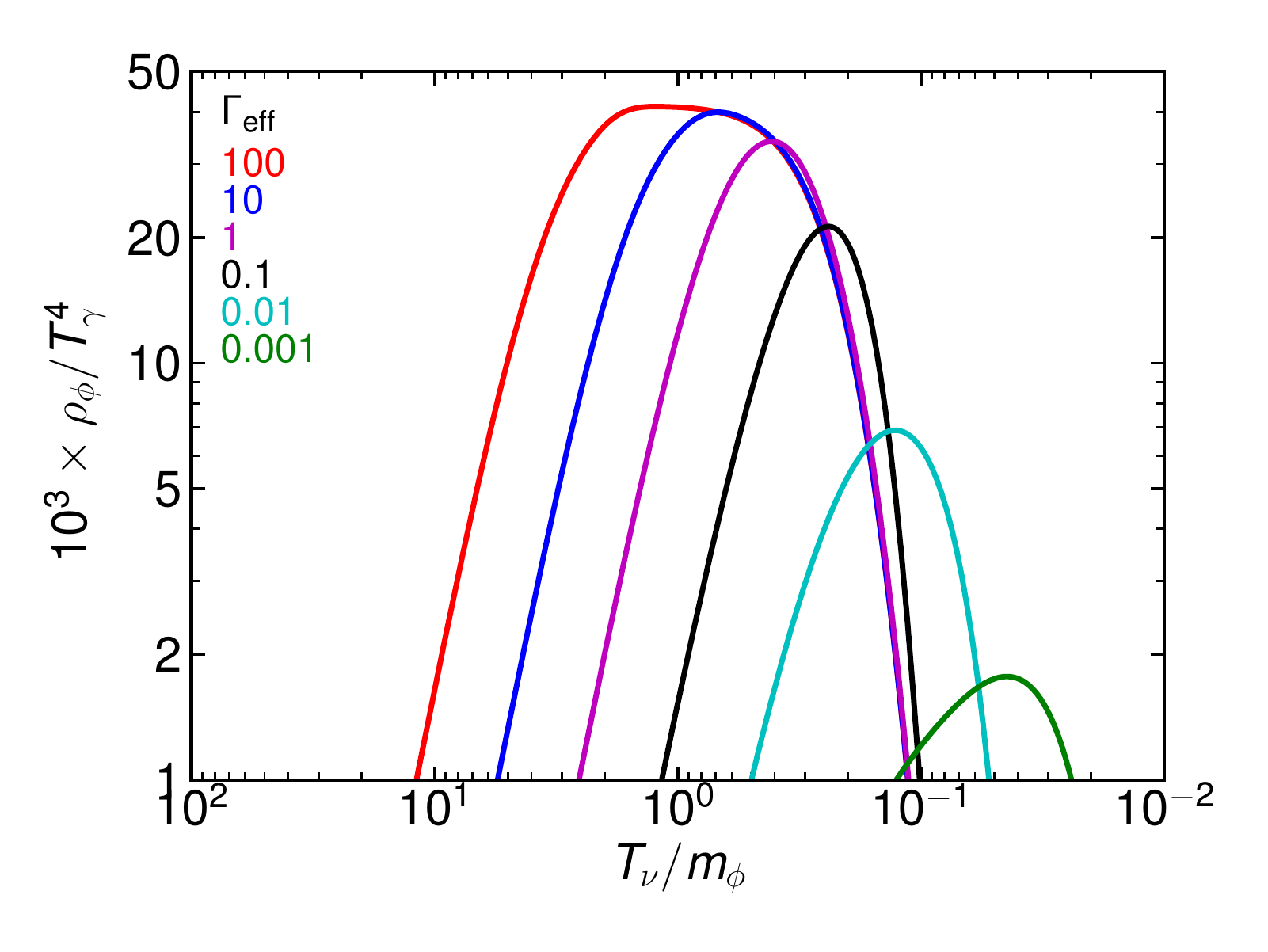} & \hspace{-0.7cm}  \includegraphics[width=0.525\textwidth]{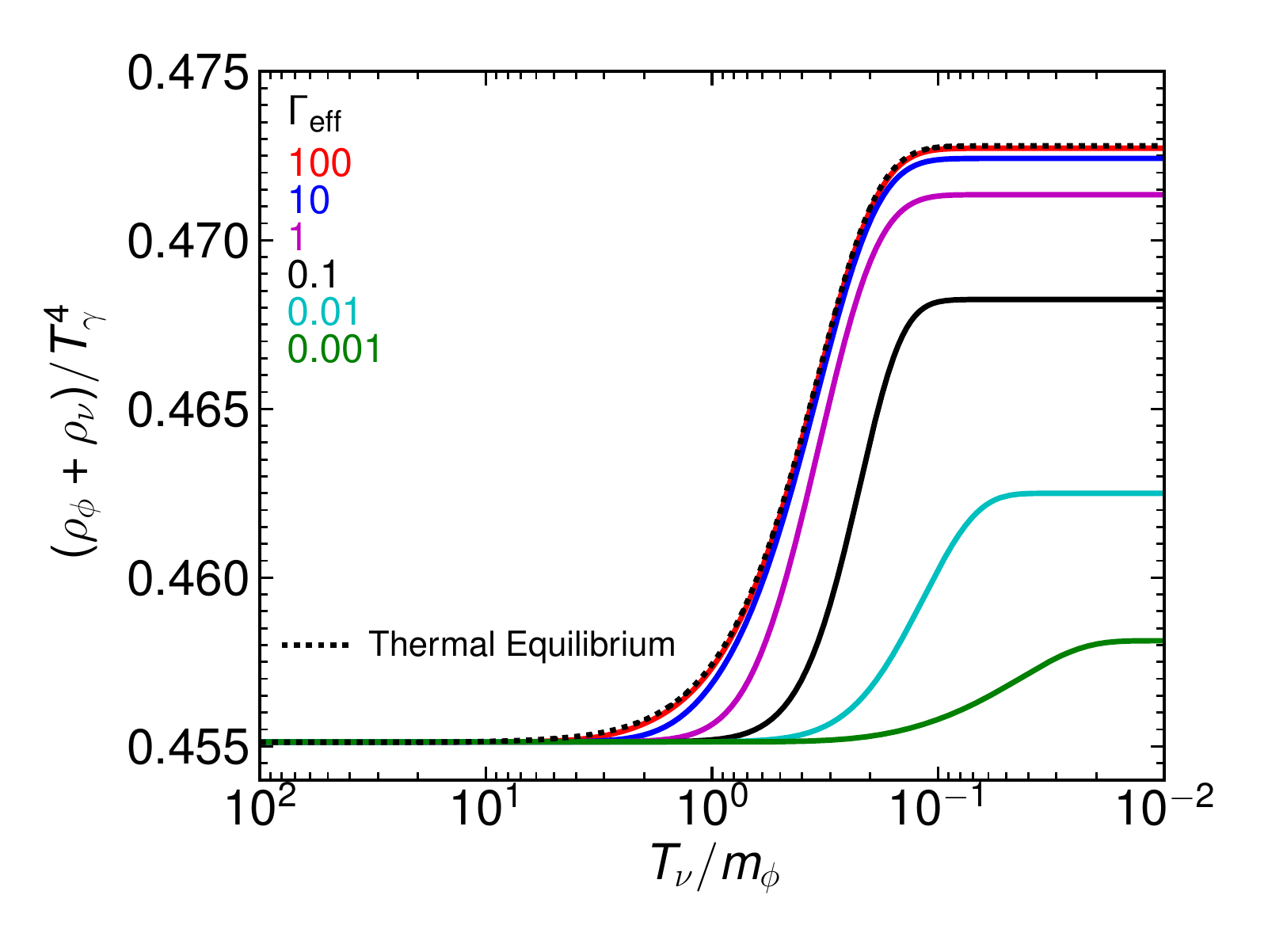}    \\
\end{tabular}
\vspace{-0.4cm}
\caption{\textit{Left panel:} Evolution of $\rho_\phi$ as a function of temperature for various choices of $\Gamma_{\rm eff}$. Note that $\rho_\phi$ never exceeds the thermal equilibrium prediction of~\eqref{eq:thermal_rhophi}. \textit{Right panel:} Evolution of $\rho_\phi +\rho_\nu$ as a function of temperature. The dashed line indicates the evolution as dictated by Equation~\eqref{eq:continuity_nuphi}.  }\label{fig:Majoron_Thermo}
\end{figure}

\vspace{1cm}
In the left panel of Figure~\ref{fig:Majoron_Thermo}, we show the evolution of the energy density of $\phi$ species for representative values of the interaction strength $\Gamma_{\rm eff}$. We can clearly appreciate that for $\Gamma_{\rm eff} \gtrsim 1$ the evolution of the $\phi$ scalar at $T \lesssim m_\phi$ is the same -- as dictated by entropy conservation. Similarly, in the right panel of Figure~\ref{fig:Majoron_Thermo}, we show the evolution of $\rho_\phi + \rho_\nu$ and appreciate that, as expected, for $\Gamma_{\rm eff} \gg 1$ the thermodynamic evolution matches the evolution derived in the perfect thermal equilibrium regime~\eqref{eq:continuity_nuphi}.

\begin{figure}[t]
\centering
\begin{tabular}{cc}
\hspace{-0.8cm}  \includegraphics[width=0.53\textwidth]{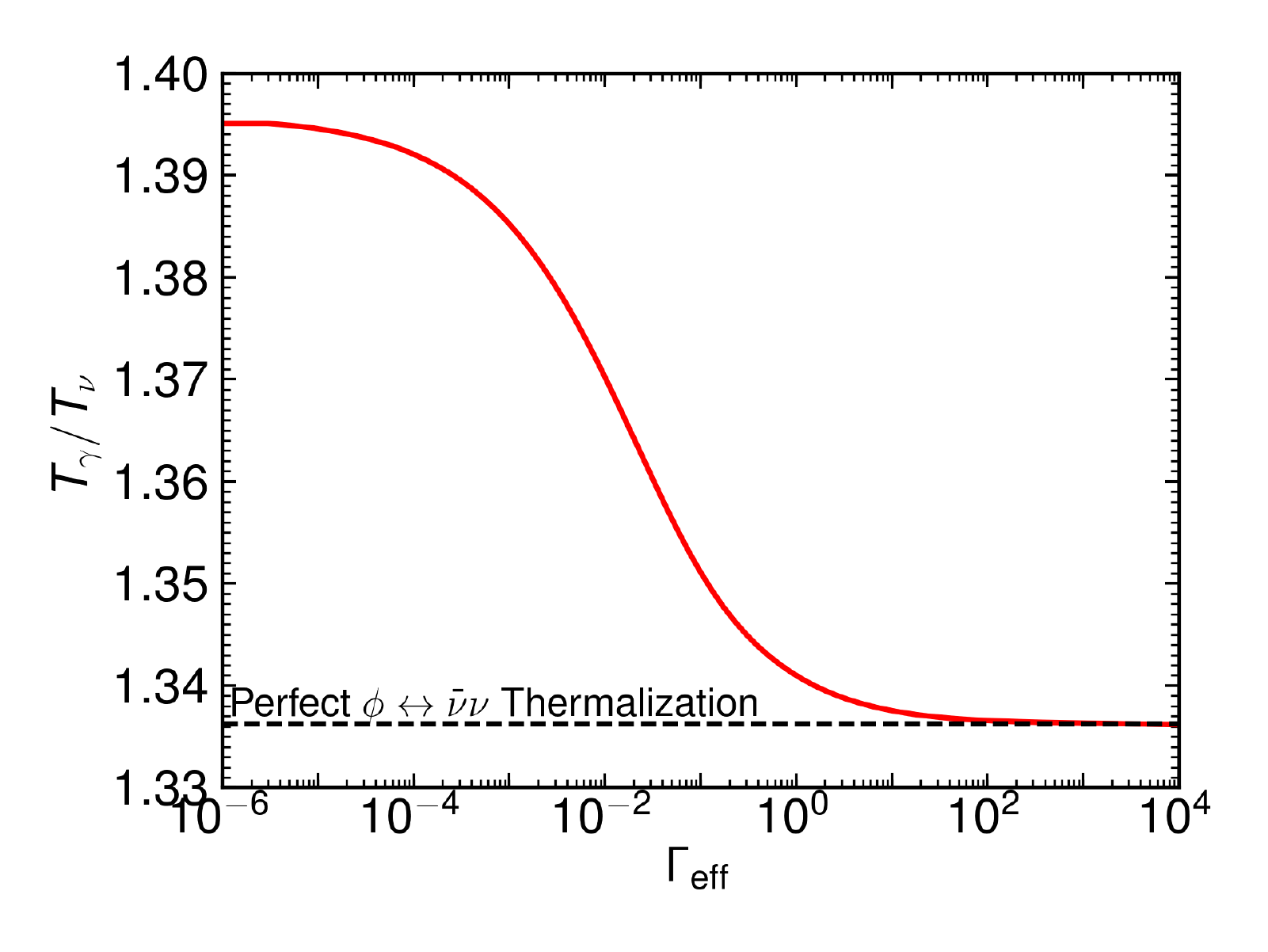} & \hspace{-0.7cm}  \includegraphics[width=0.53\textwidth]{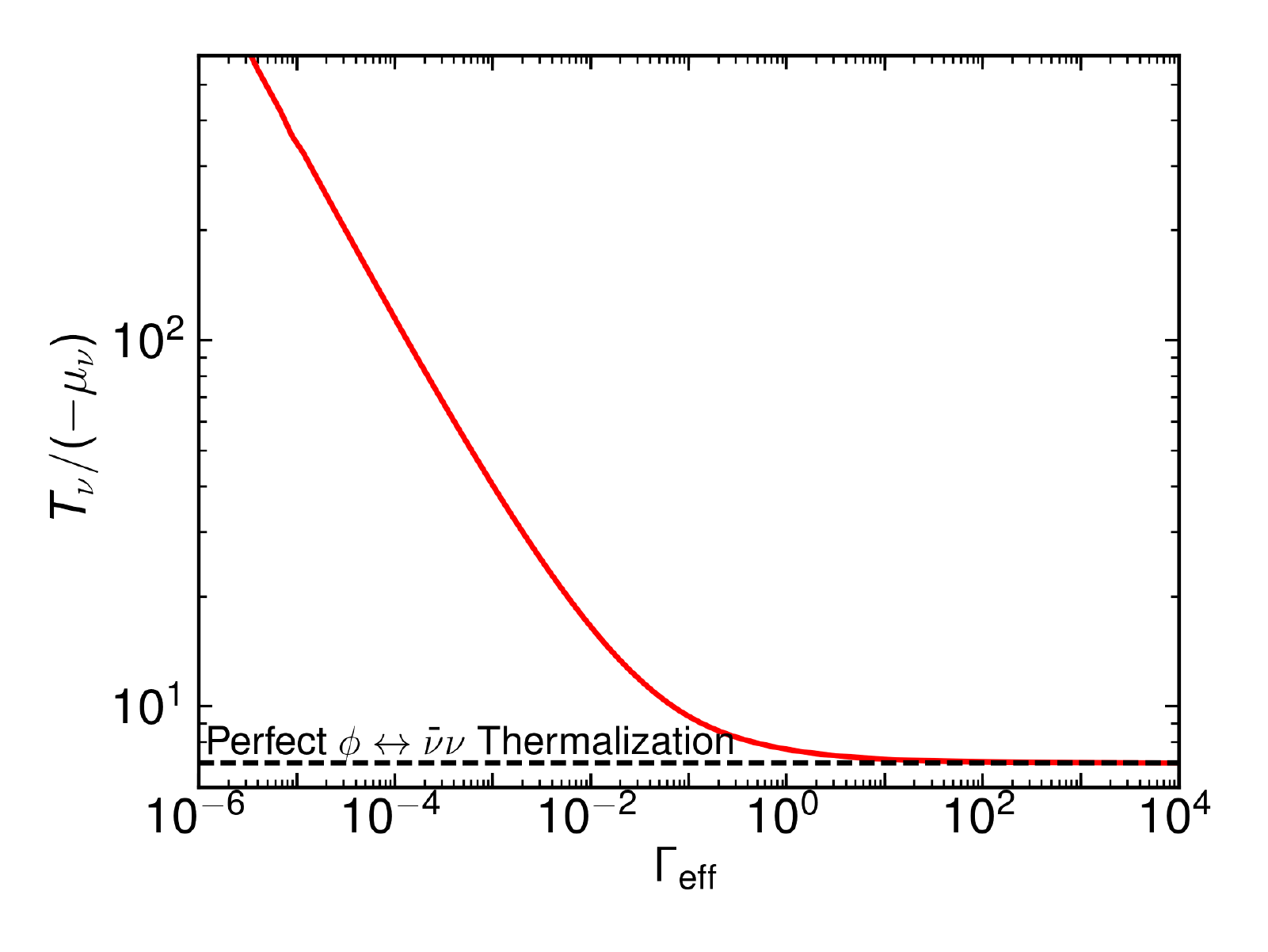} \\
\end{tabular}
\vspace{-0.8cm}
\caption{Values of $T_\nu$ and $\mu_\nu$ after the $\phi$ population has completely decayed away, i.e. $T_\nu \ll m_\phi$. We can appreciate that for $\Gamma_{\rm eff} \gtrsim 10$ both $T_\nu$ and $\mu_\nu$ match those derived assuming perfect thermal equilibrium (dashed lines), as given by Equation~\eqref{eq:full_thermalization}.}\label{fig:Tnumnu_asym}
\end{figure}

 In Figure~\ref{fig:Tnumnu_asym} we plot our results for $T_\gamma/T_\nu$ and $T_\nu/\mu_\nu$ for $T \ll m_\phi$ as a function of $\Gamma_{\rm eff}$. We clearly see that, for $\Gamma_{\rm eff} > 1$, the values of $T_\gamma/T_\nu$ and $T_\nu/m_\nu$ asymptotically reach the values of~\eqref{eq:full_thermalization}. 

\subsection{Comparison with the exact Liouville equation}\label{sec:evol_boson_1to2_exact}
The aim of this section is to solve for the exact background thermodynamics of the scenario outlined in Section~\ref{sec:model_nu}. We do this by means of solving the Liouville equation for the neutrino and $\phi$ distribution functions. As discussed in the introduction, solving the Liouville equation requires one to solve a system of hundreds of \textit{stiff} integro-differential equations that is computationally challenging. We therefore only find solutions for some representative values of $\Gamma_{\rm eff}$. We finally compare the results of solving the exact Liouville equation with the approach presented here and find they are in excellent agreement. 

\subsubsection*{The Liouville equation }\label{sec:evol_boson_1to2_Liouville}
In the region of parameter space in which the only relevant interactions are $ \phi \leftrightarrow \bar{\nu}\nu $, and after the neutrinos have decoupled ($T_\gamma < 2\,\text{MeV}$), the neutrino and $\phi$ distribution functions are characterized by the following system of Boltzmann equations~\cite{Kawasaki:1992kg,Dolgov:1998st}:
\begin{subequations}\label{eq:Boltzmann_maj_exact}
 \begin{align}
 \!\!\! \frac{df_\phi}{dt} - H\,p_\phi\,\frac{\partial f_\phi}{\partial p_\phi} &= \mathcal{C}_\phi^{ \phi \leftrightarrow \bar{\nu}\nu }[f_\phi] = -\frac{m_{\phi} \, \Gamma_{\phi}}{E_{\phi} \, p_{\phi}}  \int_{\frac{E_{\phi} -p_{\phi}}{2}}^{\frac{E_{\phi} + p_{\phi}}{2}} d E_\nu \, F_{\rm dec}\left( E_{\phi}, E_\nu, E_{\phi}-E_\nu \right) \, ,  \label{eq:dfdt_eta} \\
 \!\!\!  \frac{df_\nu}{dt} - H\,p_\nu\,\frac{\partial f_\nu}{\partial p_\nu} &= \mathcal{C}_\nu^{ \phi \leftrightarrow \bar{\nu}\nu }[f_\nu] = \frac{m_{\phi} \, \Gamma_{\phi}}{3\,E_{\nu} \, p_{\nu}} \int_{|(m^2_{\phi}/4p_{\nu})-p_{\nu}|}^{\infty} \frac{d p_{\phi}  p_{\phi}}{E_{\phi}} \, F_{\rm dec}\left( E_{\phi}, E_\nu, E_{\phi}-E_\nu \right) \,, \label{eq:dfdt_nu}
 \end{align}
 \end{subequations}
where $f_\nu$ corresponds to the distribution function of one single neutrino species, and $F_{\rm dec}$ is:
\begin{align}\label{eq:boltzdecay_Fdecay}
F_{\rm dec}\left( E_\phi, E_{\nu}, E_{\nu'} \right) =  f_\phi(E_\phi)  \left[1-f_\nu(E_{\nu})\right]  \left[1-f_\nu(E_{\nu'})\right] -f_\nu(E_{\nu}) f_\nu(E_{\nu'}) \left[ 1+f_\phi(E_\phi)\right]\,.
 \end{align}

\textit{Numerics to solve for $f_\nu$ and $f_\phi$.} We solve the system of integro-differential equations~\eqref{eq:Boltzmann_maj_exact} by binning the neutrino and $\phi$ distribution functions in comoving momentum in the range $y \equiv p \, a/m_\phi = [0.005,20]$ in 200 bins each. This represents a high accuracy setting~\cite{Dolgov:1998sf}. The system thus consists of 400 stiff integro-differential equations that we solve using backward differentiation formulas from \texttt{vode} in \texttt{Python}. We use the default settings for the absolute and relative accuracies of the integrator. \\

\textit{CPU time usage.} The typical time needed to solve the exact Liouville equation~\eqref{eq:full_system_maj} is roughly $t_{\rm CPU}/\text{day} = 0.1+0.4\,\Gamma_{\rm eff}^{1.4} $. This is in sharp contrast with $t_{\rm CPU}\sim 1\,\text{min}$ that takes~\href{https://github.com/MiguelEA/nudec_BSM}{NUDEC\_BSM} to solve the simplified system of equations~\eqref{eq:full_system_maj}. Of course, the system of equations~\eqref{eq:Boltzmann_maj_exact} can be parallelized thereby reducing the actual amount of time significantly. Perhaps a \texttt{C} or \texttt{Fortran} implementation of the solver could potentially yield an order of magnitude improvement in the speed but still would be orders of magnitude far from $1\,\text{min}$.

\textit{Initial Conditions and Integration Range.} We use as initial conditions for $f_\nu$ and $f_\phi$ Fermi-Dirac and Bose-Einstein formulas respectively with temperature and chemical potentials as in~\eqref{eq_initial_con}. We run the integrator until the largest time between $T_\nu = m_\phi/15$ and $t  = 5/\Gamma_\phi$ ($T_\nu \sim m_\phi/16 \sqrt{\Gamma_{\rm eff}/0.02}) $. This ensures that the $\phi$ population has completely decayed away by then. Namely, $\rho_\phi/\rho_\nu < 10^{-5}$. 

\vspace{-0.2cm}
\subsubsection*{Comparison}\label{sec:compa_boson_1to2}
\vspace{-0.1cm}

Here we compare the results of solving the Liuoville equation~\eqref{eq:Boltzmann_maj_exact} and the system of equations~\eqref{eq:full_system_maj} that describe the same system: a light neutrinophilic scalar where $\phi \leftrightarrow \bar{\nu}\nu$ interactions are cosmologically relevant. We shall denote the solution to the exact Liouville equation as ``Full'' while we denote the solution to the system of differential equations~\eqref{eq:full_system_maj} as ``Fast". We explicitly compare the solution of the two approaches in terms of the time evolution of $\rho_\phi + \rho_\nu$ and $\Delta N_{\rm eff}$. The reader is referred to Appendix~\ref{app:1_2_cases} for a comparison in terms of $\rho_\phi$, $\rho_\nu$, $n_\phi$, $n_\nu$, $f_\nu$ and $f_\phi$ where very good agreement is also found. 

We focus the comparison for interaction strengths in the range $ 10^{-3}< \Gamma_{\rm eff} <20 $. For $  \Gamma_{\rm eff} >20 $ solving the Liouville equation is too time consuming ($t_{\rm CPU} \sim 25\, (\Gamma_{\rm eff}/20)^{1.4} \,\text{days} $), while for $\Gamma_{\rm eff} < 10^{-3}$ the cosmological consequences of the very light neutrinophilic scalar we consider are negligible (as can be seen from Figure~\ref{fig:Majoron_Thermo_Neff}). 

\begin{figure}[t]
\centering
\begin{tabular}{cc}
\hspace{-0.8cm} \includegraphics[width=0.50\textwidth]{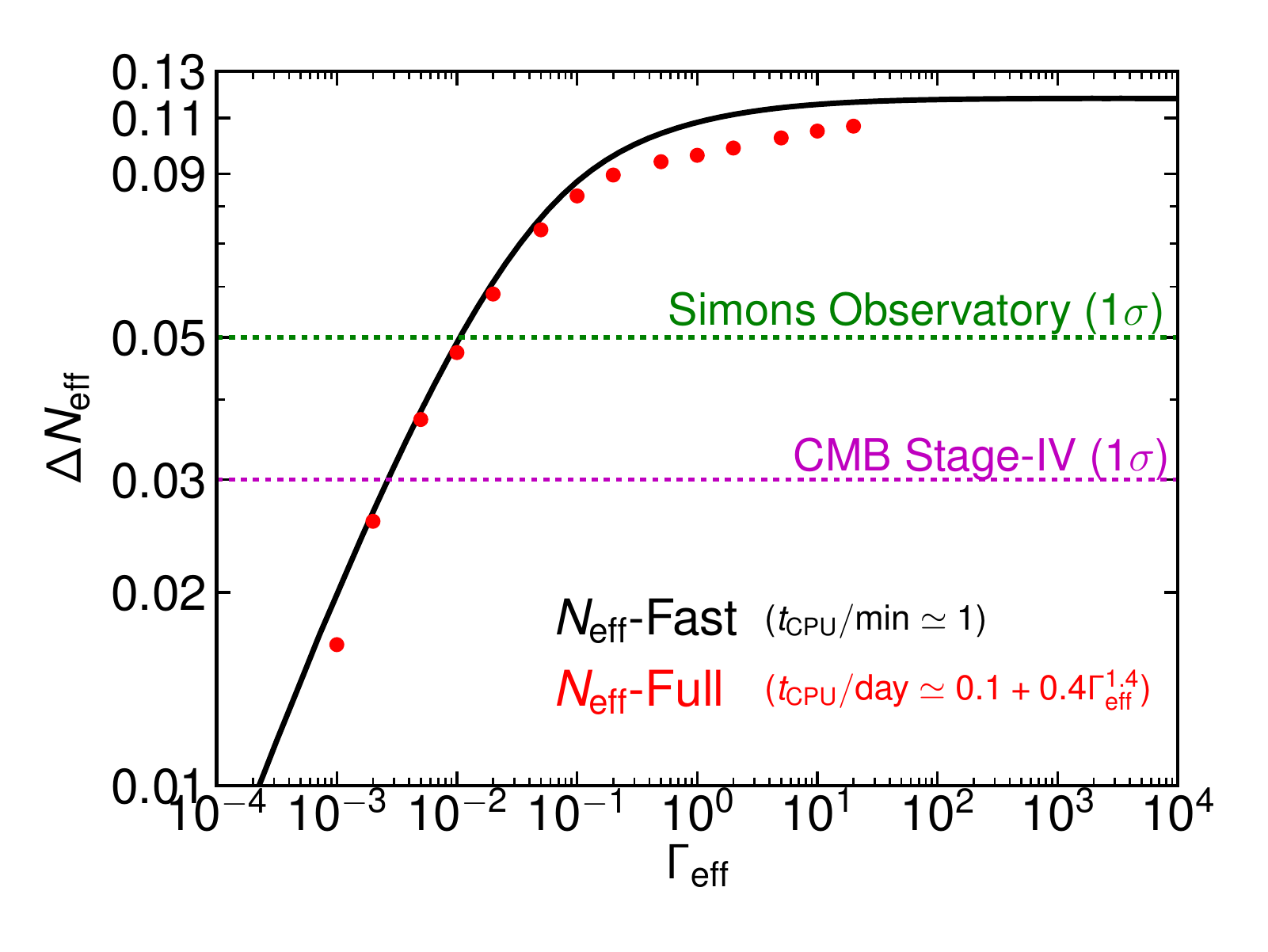} & \hspace{-0.7cm}   \includegraphics[width=0.50\textwidth]{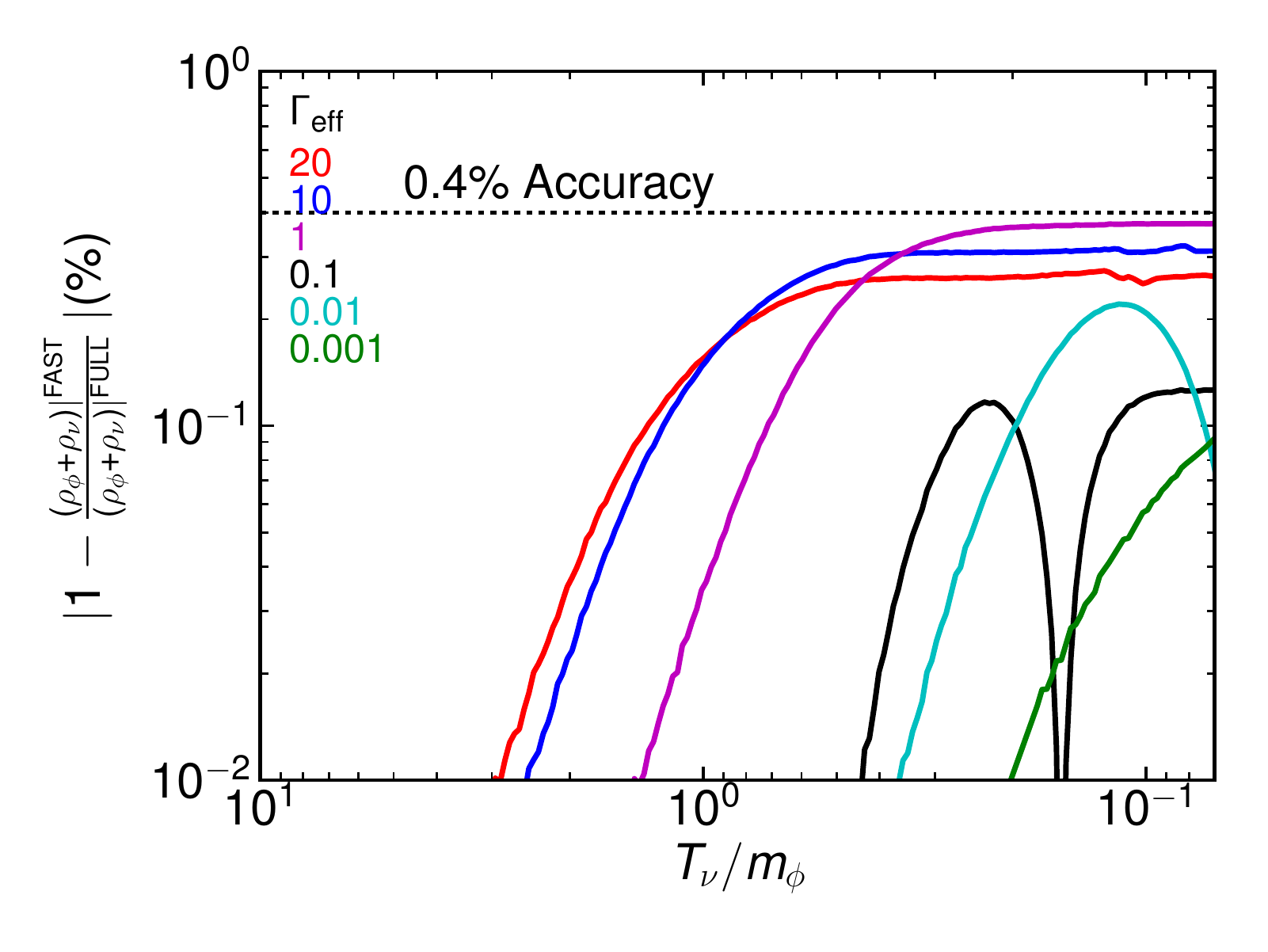}   \\
\end{tabular}
\vspace{-0.65cm}
\caption{\textit{Left panel:} $\Delta N_{\rm eff}$ as a function of $\Gamma_{\rm eff}$. In black (Fast) the results from solving~\eqref{eq:full_system_maj} and in red (Full) those by solving~\eqref{eq:Boltzmann_maj_exact}. The agreement is very good and the small differences are well below any expected CMB experiment sensitivity. \textit{Right panel:} Relative difference between the Fast and Full solution for $\rho_\nu+\rho_\phi$. The agreement between the two is better than 0.4\% in all cases considered in this study.  }\label{fig:Majoron_precision}
\end{figure}

In the right panel of Figure~\ref{fig:Majoron_precision} we show the total energy density of the system, $\rho_\phi + \rho_\nu$. We appreciate an excellent agreement between the two approaches. For $10^{-3} < \Gamma_{\rm eff} < 20$ the relative difference between the two approaches in terms of $\rho_\phi + \rho_\nu$ at any given temperature is always better than $0.4\%$.

We show the comparison in terms of $\Delta N_{\rm eff}  = N_{\rm eff}-3.045$ in the left panel of Figure~\ref{fig:Majoron_precision}. Very similar results are found between the two approaches and only when $\Gamma_{\rm eff} \gtrsim 1$ we can see a small disagreement (at the level of 0.01) for $\Delta N_{\rm eff}$. This difference is well below any expected future sensitivity from CMB observations. In fact, we believe that in the regime $\Gamma_{\rm eff} \gtrsim 1$ the actual value of $\Delta N_{\rm eff}$ is that given by the Fast approach and not the Full one. The reason is twofold: \textit{i)} the Fast approach does converge to the value of $\Delta N_{\rm eff}$ predicted by assuming pure thermal equilibrium, that must hold for $\Gamma_{\rm eff} \gg 1$~\eqref{eq:DNeff_majoron_thermal}. \textit{ii)} the larger $\Gamma_{\rm eff}$ is, the more numerically unstable the Liouville equation becomes. On the one hand, as $\Gamma_{\rm eff}$ gets larger, $f_\nu$ and $f_\phi$ are more accurately described by perfect thermal equilibrium distribution functions. On the other hand, if $f_\nu$ and $f_\phi$ resemble equilibrium distribution functions, Equation~\eqref{eq:boltzdecay_Fdecay} tends to 0. Thus, we believe that for $\Gamma_{\rm eff} \gtrsim 1$, small numerical instabilities prevent perfect thermalization of $f_\nu$ and $f_\phi$ leading to a slightly underestimated value of $\Delta N_{\rm eff}$ by the Full solution.

Thus, we have demonstrated that the set of differential equations~\eqref{eq:dTdmu_generic_simple} describe very well the thermodynamic evolution of the scalar-neutrino system considered in this work. Importantly, the total energy density of the system ($\rho_\phi+\rho_\nu$) agrees between the Full and Fast solutions to better than 0.4\% at any given temperature. In addition, the asymptotic values of $\Delta N_{\rm eff}$ agree within 0.01 or better. Therefore, the Fast approach developed in this study describes the thermodynamics of the system very accurately and can be used to study the reach of ultrasensitive CMB experiments~\cite{Abazajian:2019eic,Hanany:2019lle,DiValentino:2016foa,Sehgal:2019ewc} to this particular scenario.

\section{Discussion}\label{sec:discussion}
\vspace{-0.25cm}
In this work we have shown that thermal equilibrium distributions can accurately track the thermal history of the Universe in various scenarios. We have explicitly demonstrated this statement in two specific cases: \textit{i)} neutrino decoupling in the Standard Model (see Section~\ref{sec:SM_neutrinodec}) and \textit{ii)} a SM extension featuring a very light and weakly coupled neutrinophilic scalar (see Section~\ref{sec:boson_1to2}). This poses the question of why this occurs given that out-of-equilibrium processes are relevant in the two cases. We believe that the reason is twofold:

\begin{enumerate}[leftmargin=1cm,itemsep=0.4pt]\vspace{-0.2cm}
\item The evolution equations that dictate the temperature and chemical potential evolution~\eqref{eq:dTdmu_generic_simple} describing each species arise from conservation of energy and number density -- Equations~\eqref{eq:n_general} and~\eqref{eq:rho_general} respectively. Given that they result from conservation equations for $\rho$ and $n$, the main thermodynamic quantities should be well described.
\item Departures from thermal equilibrium are not too severe in either of the two scenarios. 

\begin{enumerate}[leftmargin=1cm,itemsep=0.5pt]\vspace{-0.2cm}
\item \textit{Neutrino Decoupling in the Standard Model}. The out-of-equilibrium injection of neutrinos from $e^+e^-$ annihilations mainly occurs at $T \sim m_e$. The interaction rate for this process is only mildly sub-Hubble: $\Gamma/H \sim (m_e/T_{\nu}^{\rm dec})^3 \sim 0.02$, and the neutrinos produced from $e^+e^-$ annihilations have an energy $E_\nu \sim 3\,T_\gamma +m_e$ which is similar to the mean neutrino energy $\left< E_\nu\right> \sim 3\, T_\nu$. Hence, a Fermi-Dirac distribution function for the neutrinos describes well the thermodynamics -- provided that the temperature evolution accounts for the relevant interactions. 
\item \textit{Light and Weakly Coupled Neutrinophilic Scalar}. The interaction rate is a free parameter in this scenario and we have found an excellent description of the thermodynamics for all situations in which decays and inverse decays fulfilled $\Gamma_{\rm eff} \sim \Gamma/H > 10^{-3}$. In this case, $\Gamma_{\rm eff}$ also controls the typical energy of the neutrinos injected from $\phi \to \bar{\nu}\nu$ decays. For $\Gamma_{\rm eff }\gtrsim 1$ the neutrinos that are produced and decayed away from $\phi \leftrightarrow \bar{\nu}\nu$ interactions have a typical energy $E_\nu \sim 3\, T_\nu$ and hence a thermal equilibrium distribution function characterizes very well the system. For $\Gamma_{\rm eff }\ll 1$, $E_\nu^{\phi \to \bar{\nu}\nu} \sim 3\,T_\nu \sqrt{0.03/\Gamma_{\rm eff}}$, and therefore for $\Gamma_{\rm eff} \gtrsim 10^{-3}$ the neutrinos injected have an energy sufficiently similar to $\left< E_\nu\right> \sim 3\, T_\nu$ and perfect thermal distribution functions can characterize the system accurately. 
\end{enumerate}\vspace{-0.2cm}
\end{enumerate}\vspace{-0.2cm}
Thus, given these examples, we believe that the thermodynamics of any system where departures from thermal equilibrium are moderate\footnote{This is, if interactions are present, when the different species interact efficiently enough ($\Gamma/H  \gtrsim  10^{-3}$) and when particles are produced with energies that are similar to the mean energy of the fluid ($\left<E\right> \sim 3 \,T$). } can be accurately described by the method presented in Section~\ref{sec:BSM_neutrinodec}. Furthermore, the larger the interaction rates are, the better the system will be described by this approach. For very small interaction rates ($\Gamma/H \lesssim 10^{-3}$) a given scenario should still be approximately described by our system of equations~\eqref{eq:dTdmu_generic_simple} but the accuracy of the approach could be substantially reduced. Of course, in this regime, solving the exact Liouville equation is not as challenging given the smallness of the interaction rates. This occurs in relevant cases such as sterile neutrino dark matter~\cite{Dodelson:1993je,Shi:1998km}, see e.g.~\cite{Venumadhav:2015pla}, or freeze-in dark matter~\cite{Hall:2009bx}, see e.g.~\cite{Dvorkin:2019zdi}. We finish by mentioning that, in fact, very out-of-equilibrum processes are particularly simple to solve and can be worked out at the level of the energy and number density of particles, see for example Chapter 5.3 of~\cite{Kolb:1990vq}.
 
\section{A Recipe to Model Early Universe BSM Thermodynamics }\label{sec:BSM_recipe}

Here we provide a four steps recipe to model the early Universe thermodynamics in generic BSM scenarios. The steps are the following:
\begin{enumerate}[leftmargin=1cm,itemsep=0.4pt]
\item \textit{Identify what are the relevant species and group them in sectors.} \\
If interactions within a sector are strong, then one can write down evolution equations for the joint system rather than for each species. For example, this is the case in the Standard Model neutrino decoupling for electrons, positrons and photons. 
\item \textit{Determine whether chemical potentials can be neglected or not.} \\
Chemical potentials can be neglected if processes that do not conserve particle number are active and there is no primordial asymmetry between particles and antiparticles. This leads to a considerable simplification of the evolution equations. 
\item \textit{Include all relevant interactions among sectors.} \\
Calculate the relevant energy and number density transfer rates from decay, annihilation and scattering processes. The expressions for decay and annihilation processes in the Maxwell-Boltzmann approximation can be found in Section~\ref{sec:interaction_rates}. 
\item \textit{Write down evolution equations for $T$ and $\mu$ for each sector within the scenario.}  \\
Use Equations~\eqref{eq:dTdmu_generic_simple} if chemical potentials cannot be neglected and Equation~\eqref{eq:dTdt_nochem} otherwise. In order to implement numerically the recipe, one can use as a starting point~\href{https://github.com/MiguelEA/nudec_BSM}{NUDEC\_BSM} where a suite of BSM scenarios have already been coded up. 
\end{enumerate}
\subsection*{An example: Massless Neutrinophilic Scalar Thermalization during BBN}

We illustrate the use of this recipe by considering the following scenario: a massless neutrinophilic scalar that thermalizes during BBN. For concreteness, the scenario is described by the same interaction Lagrangian as in Equation~\eqref{eq:L_majoron}: $\mathcal{L}_{\rm int} = \lambda/2\, \phi\, \sum_i \,\bar{\nu}_i \gamma_5 \nu_i$. Given this interaction Lagrangian and since $m_\phi = 0$, kinematically, the only possible processes in which $\phi$ can participate are $\bar{\nu}\nu \leftrightarrow \phi \phi$, $\nu \phi \leftrightarrow \nu \phi$ and $\bar{\nu} \phi \leftrightarrow \bar{\nu} \phi$.

\begin{enumerate}[leftmargin=1cm,itemsep=0.4pt]
\item \textit{Identify what are the relevant species and group them in sectors.} \\
There are three different sectors in this example: 1) $e^+,\,e^-,\,\gamma$, 2) $\nu$ and 3) $\phi$.  
\item \textit{Determine whether chemical potentials can be neglected or not.} \\
Chemical potentials in the electromagnetic sector of the plasma can safely be ignored as a result of the small baryon-to-photon ratio. Since we consider no primordial asymmetry for neutrinos we can also set $\mu_\nu = \mu_{\bar{\nu}} = 0$. We can also safely set $\mu_\phi = 0$ since the rest of the particles have $\mu=0$ and none of the processes $\phi$ participates in can generate a net chemical potential. 
\item \textit{Include all relevant interactions among sectors.} \\
We should calculate the relevant annihilation and scattering energy density transfer rates. The rates for $\nu$-$e$ interactions in the Standard Model can be found in~\eqref{eq:energyrates_nu_SM}. The $ \phi \phi \leftrightarrow \bar{\nu}\nu $ rate for this scenario can be obtained from Equation~\eqref{eq:deltan_dt_ann_mless} given $\sigma(s) = \frac{\lambda^4}{32\pi s} \log \left(s/m_\nu^2\right)$. The energy transfer rate for $\nu \phi \leftrightarrow \nu \phi$ processes cannot be easily calculated. However, it can be neglected on the grounds that scattering interactions transfer substantially less energy than annihilation interactions, see Section~\ref{sec:BSM_recipe_scatt}.
\item \textit{Write down evolution equations for $T$ and $\mu$ for each sector within the scenario.}  \\
Since chemical potentials are neglected for all the species, the temperature evolution of each sector is given by Equation~\eqref{eq:dTdt_nochem} sourced with the relevant energy transfer rates. The actual system of differential equations for $T_\gamma$, $T_\nu$ and $T_\phi$ can be found in~\cite{Escudero:2019gfk}.
\end{enumerate}

\section{Summary and Conclusions}\label{sec:summary}

Precision measurements of key cosmological observables such as $N_{\rm eff}$ and the primordial element abundances represent a confirmation of the thermal history of the Standard Model and a stringent constraint on many of its extensions. In this work -- motivated by the complexity of accurately modelling early Universe thermodynamics and by building upon~\cite{Escudero:2018mvt} -- we have developed a simple, accurate and fast approach to calculate early Universe thermodynamics in the Standard Model and beyond. 

In Section~\ref{sec:BSM_neutrinodec}, we have detailed the approximations upon which the method is based on and developed the relevant equations that track the early Universe thermodynamics of any given system. In Section~\ref{sec:SM_neutrinodec}, the method was applied to study neutrino decoupling in the Standard Model. In Section~\ref{sec:boson_1to2}, we have used this approach to model the early Universe thermodynamics of a BSM scenario featuring a light neutrinophilic boson. We have found excellent agreement between our approach and previous literature on the SM, and between our BSM results and those obtained by solving the exact Liouville equation. In Section~\ref{sec:discussion}, we have discussed theoretical arguments supporting why the approach presented in this study is precise and also have discussed its limitations. 

\vspace{0.05cm}
\noindent The main results obtained in this study are:
\begin{itemize}
\item Neutrino decoupling in the Standard Model can be accurately described with perfect Fermi-Dirac distribution functions for the neutrinos that evolve according to~\eqref{eq:dTdt_nu_SM}. By accounting for finite temperature corrections, and for spin-statistics and the electron mass in the $\nu$-$\nu$ and $\nu$-$e$ interaction rates we find $N_{\rm eff}^{\rm SM} = 3.045$. A result that is in excellent agreement with previous precision determinations~\cite{deSalas:2016ztq,Mangano:2005cc}. Very good agreement is also found for other relevant cosmological observables as summarized in Table~\ref{tab:SM_summary}.
\item By solving the exact Liouville equation we have explicitly demonstrated that the thermodynamic evolution of a very light ($1\,\text{eV}<m_\phi < 1\,\text{MeV}$) and weakly coupled ($\lambda \lesssim 10^{-9}$) neutrinophilic scalar can be accurately described by Equations~\eqref{eq:dTdmu_generic_simple}. 
\end{itemize}

\vspace{0.5cm}

\noindent These results allow us to draw the main conclusion of this work:
\vspace{-0.05cm}

\begin{itemize}
\item The early Universe thermodynamics of any given system in which departures from thermal equilibrium are not severe can be accurately modelled by a few simple differential equations for the temperature and chemical potentials describing each of the relevant species. These equations are~\eqref{eq:dTdmu_generic_simple}. One can (and should) use Equation~\eqref{eq:dTdt_nochem} if chemical potentials can be neglected. These equations are simple and fast to solve, and including interactions among the various particles is straightforward as described in Section~\ref{sec:BSM_neutrinodec}. A recipe to implement the method in generic BSM scenarios is outlined in Section~\ref{sec:BSM_recipe}.
\end{itemize}
\vspace{-0.05cm}

\noindent To summarize, the method developed in this study and in~\cite{Escudero:2018mvt} can be used to accurately track the early Universe evolution in the Standard Model and beyond. Thereby enabling one to find precise predictions for $N_{\rm eff}$ and the primordial element abundances in BSM theories. We note that the method has already been applied to study: \textit{i)} the BBN effect of MeV-scale thermal dark sectors~\cite{Escudero:2018mvt,Sabti:2019mhn}, \textit{ii)} the impact on BBN of invisible neutrino decays~\cite{Escudero:2019gfk}, \textit{iii)} the cosmological consequences of light and weakly coupled flavorful gauge bosons~\cite{Escudero:2019gzq}, \textit{iv)} $N_{\rm eff}$ constraints on dark photons~\cite{Ibe:2019gpv}, \textit{v)} variations of Newton's constant at the time of BBN~\cite{Alvey:2019ctk}, and \textit{vi)} the CMB implications of sub-MeV neutrinophilic scalars~\cite{Escudero:2019gvw}. 

We expect the proposed method to prove useful to study many other extensions of the Standard Model and in other contexts not necessarily restricted to neutrino decoupling or BBN. For example, the method could be used to study: late dark matter decoupling~\cite{Diacoumis:2018nbq}, generic dark sectors~\cite{Buen-Abad:2015ova,Chacko:2015noa}, dark sectors equilibrating during BBN~\cite{Berlin:2019pbq}, or the BBN era in low-scale Baryogenesis set-ups~\cite{Aitken:2017wie,Elor:2018twp,Nelson:2019fln,Kane:2019nes}.

We conclude by highlighting that we publicly release a \texttt{Mathematica} and \texttt{Python} code: \href{https://github.com/MiguelEA/nudec_BSM}{NUDEC\_BSM} containing the codes developed in this study. The code is fast ($t_{\rm CPU} = \mathcal{O}(10)\,\text{s}$) and we believe that it could serve as a useful tool for particle phenomenologists and cosmologists interested in calculating the early Universe thermodynamics of BSM scenarios.

\vspace{-0.2cm}
\section*{Acknowledgments}
\vspace{-0.2cm}
I am grateful to Sam Witte, Chris McCabe, Sergio Pastor, Stefano Gariazzo and Pablo F. de Salas for their very helpful comments and suggestions over a draft version of this paper, and to Toni Pich for useful correspondence. This work is supported by the European Research Council under the European Union's Horizon 2020 program (ERC Grant Agreement No 648680 DARKHORIZONS).

\vspace{-0.25cm}
\section*{v2 Bonus: $N_{\rm eff}^{\rm SM}$ with NLO $e^+e^- \leftrightarrow \bar{\nu}\nu$ rates}
\vspace{-0.25cm}

In our $N_{\rm eff}^{\rm SM}$ calculation in Section~\ref{sec:SM_neutrinodec} we have accounted for the $e^3$ correction to the energy and pressure density of the electromagnetic plasma. The reader may be wondering what is the impact of radiative corrections to the $\nu\!-\!e$ interaction rates. The weak corrections to the rates are tiny since $G_F E^2 = G_F (10\,\text{MeV})^2 < 10^{-9}$. However, at finite temperature and for temperatures relevant for neutrino decoupling, QED corrections have been shown to reduce the $e^+e^- \leftrightarrow \bar{\nu}\nu$ rates by $\sim -4\%$~\cite{Esposito:2003wv}. We estimate the impact on $N_{\rm eff}$ of such corrections by using \texttt{NUDEC\_BSM} and by approximating the NLO results of~\cite{Esposito:2003wv} with a simple formula (and extrapolating for $T\gtrsim 1\,\text{MeV}$). By doing so, we find that $N_{\rm eff}^{\rm SM}$ including such corrections is expected to be shifted by $\simeq -0.0007$. Namely,
\begin{align*}
\left. \frac{\delta \rho}{\delta t}\right|_{e^+ e^-\leftrightarrow \bar{\nu}\nu}^{\rm NLO} \! =\! \left. \frac{\delta \rho}{\delta t}\right|_{e^+ e^- \leftrightarrow \bar{\nu}\nu}^{\rm LO}\! \left[1 - 0.038 \log_{10}\left( \frac{T_\gamma}{0.11\,\text{MeV}}\right)\!\right] \longrightarrow \left. N_{\rm eff}^{\rm SM} \right|_{\rm NLO} -\left. N_{\rm eff}^{\rm SM} \right|_{\rm LO} \simeq -0.001 \,,
\end{align*}
and hence, it seems that using LO $\nu\!-\!e$ rates limits the accuracy of $N_{\rm eff}^{\rm SM}$ to be $\simeq 0.001$.

\newpage
\bibliographystyle{JHEP}
\bibliography{biblio}

\newpage
\appendix   
\section{Appendices}\label{sec:app_TOTAL}

In these appendices we provide: details about Standard Model interaction rates including spin-statistics and the electron mass~\ref{app:rates_SM}. A detailed comparison between our SM results and previous SM calculations~\ref{app:SM_comparison_indetail}. The relevant equations and results from solving neutrino decoupling in the SM including neutrino chemical potentials~\ref{app:SM_chemical}.  A comparison between the solution to the Liouville equation and the proposed method for a very light and weakly coupled neutrinophilic scalar~\ref{app:1_2_cases}. The temperature and chemical potential evolution equations in the Maxwell-Boltzmann approximation~\ref{app:MB_limit}. Thermodynamic formulae~\ref{app:Thermo_formulae}. An explicit calculation of the number and energy density transfer rates in the Maxwell-Boltzmann approximation for decay, annihilation and scattering processes together with spin-statistics corrections to them~\ref{app:rates_MB}.

\subsection{SM $\nu$-$e$ and $\nu$-$\nu$ interaction rates}\label{app:rates_SM}
In this appendix we outline the results of the neutrino-electron and neutrino-neutrino energy and number transfer rates including spin-statistics and a non-negligible electron mass. For that purpose, we have calculated the exact collision terms following the phase space integration method of~\cite{Hannestad:1995rs} (see also the appendices of~\cite{Fradette:2018hhl} and~\cite{Kreisch:2019yzn}).

\subsubsection*{Interaction rates in the Maxwell-Boltzmann approximation and $m_e=0$}
We begin by outlining the results obtained in~\cite{Escudero:2018mvt}, the reader is referred to Appendix A.2 of that reference for more details. In the Maxwell-Boltzmann approximation and by neglecting the electron mass, the neutrino energy density transfer rates read:
\begin{subequations}\label{eq:energyrates_nu_SM_MB}
\begin{align}
\left. \frac{\delta \rho_{\nu_e}}{\delta t} \right|_{\rm SM}^{\rm MB} &= \frac{G_F^2}{\pi^5}\left[4\left(g_{eL}^2+g_{eR}^2\right) \, F_{\rm MB}(T_\gamma,T_{\nu_e}) + 2 \, F_{\rm MB}(T_{\nu_\mu},T_{\nu_e}) \right] \, ,\\
\left. \frac{\delta \rho_{\nu_\mu}}{\delta t} \right|_{\rm SM}^{\rm MB} &= \frac{G_F^2}{\pi^5}\left[4\left(g_{\mu L}^2+g_{\mu R}^2\right)  \, F_{\rm MB}(T_\gamma,T_{\nu_\mu}) -   F_{\rm MB}(T_{\nu_\mu},T_{\nu_e}) \right] \, ,
\end{align}
\end{subequations}
where $g_{eL},\,g_{eR},\,g_{\mu L},\,g_{\mu R}$ are defined in Equation~\eqref{eq:LowEnergyCouplings} and where we have defined:
\begin{align}
F_{\rm MB}(T_1,T_2) = 32 \, (T_1^9-T_2^9) + 56 \, T_1^4\,T_2^4 \, (T_1-T_2)\, .
\end{align}
Similarly, the number density transfer rates read:
\begin{subequations}
\begin{align}
\left. \frac{\delta n_{\nu_e}}{\delta t} \right|_{\rm SM}^{\rm MB} &= 8\frac{G_F^2}{\pi^5}\left[4\left(g_{eL}^2+g_{eR}^2\right) \, (T_\gamma^8 - T_{\nu_e}^8) + 2 \, (T_{\nu_\mu}^8 - T_{\nu_e}^8) \right] \, ,\\
\left. \frac{\delta n_{\nu_\mu}}{\delta t} \right|_{\rm SM}^{\rm MB} &= 8\frac{G_F^2}{\pi^5}\left[4\left(g_{\mu L}^2+g_{\mu R}^2\right) \, (T_\gamma^8 - T_{\nu_\mu}^8) -  (T_{\nu_\mu}^8 - T_{\nu_e}^8) \right] \, ,
\end{align}
\end{subequations}
where $\delta n_i/\delta t \equiv \sum_j \left<\sigma v\right>_{ij} (n_i^2-n_j^2)$.

\subsubsection*{Electron-neutrino rates with quantum statistics}
By numerically evaluating the energy transfer rates between neutrinos and electrons we have found that the rates including quantum statistics are:
\begin{subequations}
\begin{align}
\left. \frac{\delta \rho_{\nu_e}}{\delta t} \right|_{\rm SM}^{\rm FD} &= \frac{G_F^2}{\pi^5}\left[4\left(g_{eL}^2+g_{eR}^2\right) \, F(T_\gamma,T_{\nu_e}) + 2 \, F(T_{\nu_\mu},T_{\nu_e}) \right] \, ,\\
\left. \frac{\delta \rho_{\nu_\mu}}{\delta t} \right|_{\rm SM}^{\rm FD} &= \frac{G_F^2}{\pi^5}\left[4\left(g_{\mu L}^2+g_{\mu R}^2\right)  \, F(T_\gamma,T_{\nu_\mu}) -  F(T_{\nu_\mu},T_{\nu_e}) \right] \, ,
\end{align}
\end{subequations}
where we have defined:
\begin{align}
F(T_1,T_2) &=  32 \, f_a^{\rm FD} \,  (T_1^9-T_2^9) + 56 \, f_s^{\rm FD}  \,T_1^4\,T_2^4 \, (T_1-T_2) \, ,
\end{align} 
where $f_a^{\rm FD} = 0.884$, $f_s^{\rm FD} = 0.829$. Therefore, the factors of $f_a^{\rm FD}$ and $f_s^{\rm FD}$ account for the Pauli blocking effects in the rates for annihilations and scatterings processes, respectively. 

For the number density transfer rates we find:
\begin{align}
\left. \frac{\delta n_{\nu}}{\delta t} \right|_{\rm SM}^{\rm FD} &=  0.852 \times \left. \frac{\delta n_{\nu}}{\delta t} \right|_{\rm SM}^{\rm MB} \, .
\end{align}
The annihilation cross section taking into account Fermi-Dirac statistics was first calculated in~\cite{Enqvist:1991gx}. Since $\left< \sigma v \right>_\nu = \frac{1}{3} (\left< \sigma v \right>_{\nu_e} + 2 \left< \sigma v \right>_{\nu_\mu}) \equiv \left. \frac{1}{n_\nu^2(T_\gamma) } \frac{\delta n}{\delta t}\right|_{T_\nu =0} $, we find the following neutrino annihilation cross sections:
\begin{align}
\left< \sigma v \right>_{\bar{\nu} \nu \leftrightarrow e^+e^-}^{\rm MB} &=  3.45\, G_F^2  \, T^2 \,, \\
\left< \sigma v \right>_{\bar{\nu} \nu \leftrightarrow e^+e^-}^{\rm FD} &=  2.94\, G_F^2 \, T^2 \, .
\end{align}
In the MB approximation we find an annihilation cross section that is 6\% higher than the one reported in~\cite{Enqvist:1991gx}. The annihilation cross section taking into account Pauli blocking precisely matches the one reported in~\cite{Enqvist:1991gx}.

The Fermi-Dirac suppression factors for the energy transfer rates in~\eqref{eq:G_FermiDirac} of 0.884 and 0.829 have been calculated assuming that $(T_\gamma-T_\nu)/T_\gamma \equiv \Delta T/T = 0.01$. However, we have explicitly checked that even if $\Delta T/T = \pm 0.3$ (which is a very unlikely scenario when the processes are efficient, namely for $T\gtrsim 3\,\text{MeV}$), these numbers do not vary by more than 1\%. Thus, they accurately take into account the Pauli blocking suppression in the $\nu$-$e$ and $\nu$-$\nu$ rates in any realistic BSM temperature evolution scenario. 

\begin{figure}[t]
\centering
\begin{tabular}{cc}
\hspace{-0.8cm}  \includegraphics[width=0.53\textwidth]{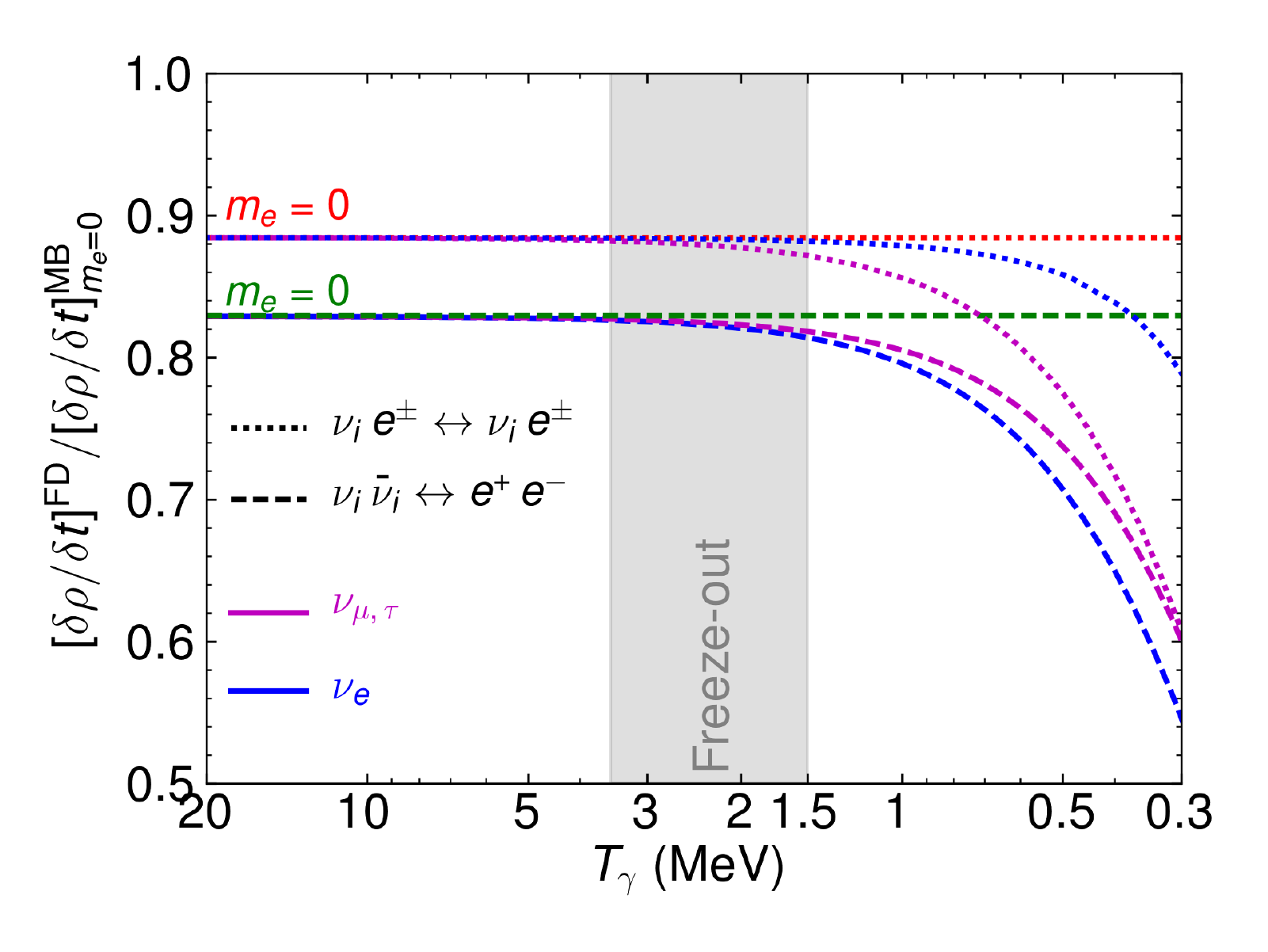} & \hspace{-0.7cm}  \includegraphics[width=0.53\textwidth]{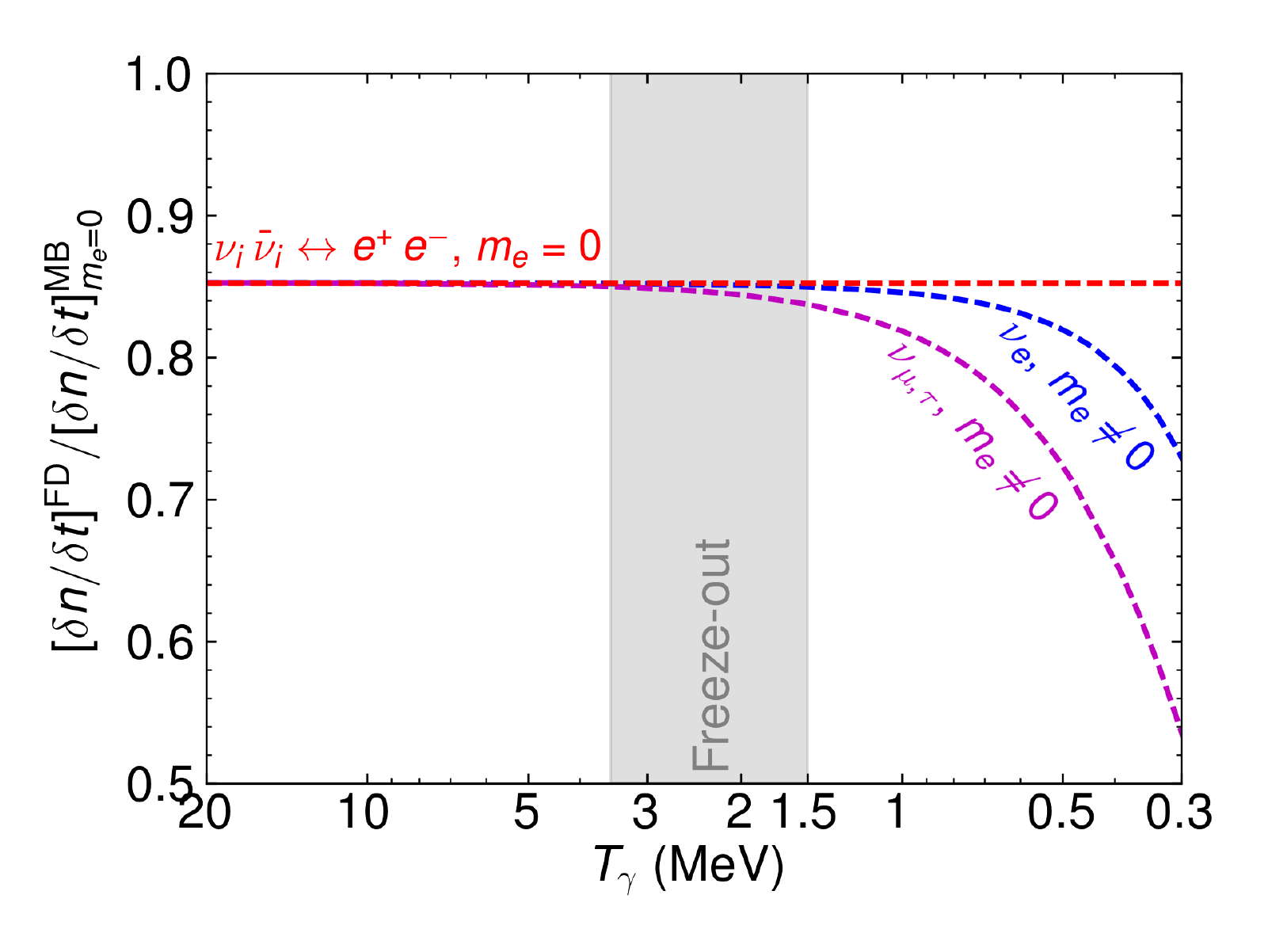} \\
\end{tabular}\vspace{-0.4cm}
\caption{Neutrino-electron energy (\textit{left panel}) and number (\textit{right panel}) density transfer rates taking into account spin statistics in the collision term and the electron mass as compared to the ones obtained in the Maxwell-Boltzmann approximation with $m_e = 0$~\eqref{eq:energyrates_nu_SM_MB}. The dotted lines correspond to scatterings and the dashed lines correspond to annihilations. Clearly, the impact of the electron mass is small for the relevant range of temperatures. Yet, we take it into account in~\href{https://github.com/MiguelEA/nudec_BSM}{NUDEC\_BSM}.   }\label{fig:Rates_comparison}
\end{figure}

\subsubsection*{Electron-neutrino rates with quantum statistics and the electron mass}
We have numerically calculated the effect of the electron mass in the energy and number density transfer rates. The effect of the electron mass is small when the $e$-$\nu$ interactions are efficient in the early Universe, namely for $T > 1\,\text{MeV}$. This can be appreciated from Figure~\ref{fig:Rates_comparison}. In terms of $N_{\rm eff}^{\rm SM}$, a non-negligible electron mass leads to a decrease of $ \simeq -0.003$, see Table~\ref{tab:Neff_SM}.  We incorporate the effect of the electron mass in the numerical code by interpolating over the precomputed rates. In order to compute the effect of the electron mass we assumed $(T_\gamma-T_\nu)/T_\gamma \equiv \Delta T/T = 0.01$. We have explicitly checked that the resulting rates considering the actual $T_\nu(T_\gamma)$ evolution in the early Universe does not lead to relevant changes in $N_{\rm eff}$.

\subsection{Comparison with previous literature on the SM: supplementary material}\label{app:SM_comparison_indetail}
In this appendix we provide detailed comparison between our calculation of neutrino decoupling with previous accurate calculations. We compare at the level of the energy carried by each neutrino species, neutrino number density, entropy density of the Universe, yields of the primordial elements, the resulting frozen neutrino number and energy contributions, and the time evolution of the entropy released from out of equilibrium $e$-$\nu$ interactions.

In order to make contact with previous literature, it is useful not to work in terms of $T_\gamma$ and $T_\nu$ but to do so in terms of comoving dimensionless ratios. We define
\begin{align}\label{eq:Tcomoving}
z_\gamma \equiv  a\,T_\gamma  /m_0 \, ,\qquad z_\nu \equiv a\, T_\nu  /m_0\,,
\end{align}
where $a$ is the \textit{dimensionless} scale factor, and $m_0$ is a normalization factor with dimensions of energy. For concreteness, we shall choose $m_0 = m_e$. Given these definitions we can easily find their scale factor evolution:
\begin{align}\label{eq:dzda_dxida}
\frac{dz}{da} &= \frac{z}{a} \left[1+ \frac{1}{H \, T} \frac{dT}{dt} \right]\,.
\end{align}

By plugging Equations~\eqref{eq:T_gamma_QED} and~\eqref{eq:dTdt_nu_SM} into~\eqref{eq:dzda_dxida} we can explicitly write down the evolution equation for $z_\gamma$ and $z_\nu$ in the Standard Model:
\begin{subequations}\label{eq:comoving_T}
\begin{align}
\frac{dz_{\nu_\alpha}}{da} &= \frac{z_{\nu_\alpha}}{a} \frac{\frac{\delta \rho_{\nu_\alpha}}{\delta t} }{4 H \rho_{\nu_\alpha} }  \, ,\\
\frac{dz_\gamma}{da} &= \frac{z_\gamma}{a}\frac{\left[T_\gamma \frac{d\rho_e}{dT_\gamma} - 3 (p_e+\rho_e)\right]  + T_\gamma\left[T_\gamma \frac{d^2P_{\rm int}}{dT_\gamma^2} - 3   \frac{d P_{\rm int}}{d T_\gamma}\right]  -\frac{3}{H}\frac{\delta \rho_{\nu}}{\delta t}}{ T_\gamma \left(T_\gamma \frac{d^2 P_{\rm int}}{d T_\gamma^2}+\frac{d\rho_\gamma}{dT_\gamma}+\frac{d\rho_e}{dT_\gamma}\right)} \, . 
\end{align}
\end{subequations}
We solve these equations in the SM by taking into account all relevant interactions and NLO finite temperature corrections. We use the same initial conditions as in Section~\ref{sec:SM_results} and find that once neutrinos have completely decoupled and electrons and positrons annihilated away:
\begin{align}
T_{\nu_e} \neq T_{\nu_{\mu,\tau}} \rightarrow z_\gamma = 1.39791, \, z_{\nu_e} =1.00237,\, z_{\nu_{\mu,\tau}} = 1.00098\,,\rightarrow N_{\rm eff} =3.0443\,, \\
T_{\nu_e} = T_{\nu_{\mu,\tau}} \rightarrow z_\gamma = 1.39786, \, z_{\nu_e} =1.00149,\, z_{\nu_{\mu,\tau}} = 1.00149\,, \rightarrow N_{\rm eff} =3.0453\,. 
\end{align}

\subsubsection*{$N_{\rm eff}$}\label{sec:comparison_Neff}

\begin{table}[t]
\begin{center}
\begin{tabular}{l|c|cccc}
\hline\hline
\multicolumn{6}{c}{Neutrino Decoupling in the Standard Model}  \\ \hline \hline
No oscillations, no QED      	   & $N_\text{eff}$   &$z_\gamma $	& $\delta\rho_{\nu_e}/\rho_{\nu_e}\,(\%)$ &$\delta\rho_{\nu_{\mu}}/\rho_{\nu_\mu}\,(\%)$ &$\delta\rho_{\nu_{\tau}}/\rho_{\nu_\tau}\,(\%)$	 \\ \hline 
This work   &3.0350 &1.39903  &	0.971       &  0.407  &  0.407  	 \\
Dolgov {et. al.}~\cite{Dolgov:1998sf} &3.0340 &	1.39910               &	0.946       &     0.398          & 0.398	 \\ 
Grohs {et. al.}~\cite{Grohs:2015tfy} &3.0340 &	1.39904               &	0.928       &     0.377          & 0.377	 \\ \hline\hline
No oscillations, LO-QED      	   & $N_\text{eff}$   &$z_\gamma $	& $\delta\rho_{\nu_e}/\rho_{\nu_e}\,(\%)$ &$\delta\rho_{\nu_{\mu}}/\rho_{\nu_\mu}\,(\%)$ &$\delta\rho_{\nu_{\tau}}/\rho_{\nu_\tau}\,(\%)$	 \\ \hline
This work   &3.0453 &1.39782  &	 0.959       &  0.401  &  0.401  	 \\
de Salas \& Pastor~\cite{deSalas:2016ztq} &3.0446 &	1.39784               &	0.920       &     0.392          & 0.392	 \\ \hline\hline
 With oscillations, LO-QED      & $N_\text{eff}$   	       &$z_\gamma $	& $\delta\rho_{\nu_1}/\rho_{\nu_1}\,(\%)$ &$\delta\rho_{\nu_{2}}/\rho_{\nu_2}\,(\%)$ &$\delta\rho_{\nu_{3}}/\rho_{\nu_3}\,(\%)$	 \\\hline
 This work   &3.0462 &1.39777  &	0.603        &  0.603  &  0.603    	 \\ 
NH, de Salas \& Pastor~\cite{deSalas:2016ztq}  &	3.0453 &     1.39779         &	  0.636     &   0.574            & 0.519	 \\
IH, $\,\!$ de Salas \& Pastor~\cite{deSalas:2016ztq}  &3.0453	 &     1.39779         &	  0.635     &   0.574           & 0.520	 \\
  \hline \hline
\end{tabular}
\end{center}\vspace{-0.5cm}
\caption{$N_{\rm eff}^{\rm SM}$ as obtained in this work and compared with the calculations of de Salas \& Pastor~\cite{deSalas:2016ztq} that includes neutrino oscillations and LO-finite temperature corrections, and that of Dolgov {et. al.}~\cite{Dolgov:1998sf} and of Grohs {et. al.}~\cite{Grohs:2015tfy} that do not. When neutrino oscillations are neglected, the difference in $N_{\rm eff}^{\rm SM}$ is $0.0007$ as compared to~\cite{deSalas:2016ztq} and $0.001$ as compared to~\cite{Dolgov:1998sf,Grohs:2015tfy}. In our case, $\delta\rho_{\nu_\alpha}/\rho_{\nu_\alpha} \equiv z_{\nu_\alpha}^4-1$ (see Equation~\eqref{eq:Tcomoving}). Note also that the relative non-instantaneous contribution to the energy density of each neutrino species is accurate to the per mille level. When neutrino oscillations are considered (and here we simply assume that $T_{\nu_e} = T_{\nu_\mu} = T_{\nu_\tau}$) the difference in terms of $N_{\rm eff}^{\rm SM}$ is $0.0009$.  }\label{tab:SM_compa}
\end{table}

In Table~\ref{tab:SM_compa} we compare the results from our solution to neutrino decoupling to those of the latest and most accurate determination in the SM by de Salas \& Pastor~\cite{deSalas:2016ztq}, and also to those of Dolgov et. al.~\cite{Dolgov:1998sf} and Grohs {et. al.}~\cite{Grohs:2015tfy} that do not account for neutrino oscillations or finite temperature corrections. We can appreciate that the resulting values of $N_{\rm eff}$ only differ by at most $0.001$, thus showing the high accuracy of the calculation performed in this work. Note also that in our calculation in which neutrino oscillations are neglected, the amount of energy carried out by each neutrino species agrees with that obtained in Ref.~\cite{deSalas:2016ztq} at the remarkable level of $0.04\%$, and at the level of $0.03\%$ as compared to the results of Ref.~\cite{Dolgov:1998sf}. This appears to be a consequence of the starting point of the derivation of our temperature time evolution equations: energy conservation~\eqref{eq:rho_general}. We therefore estimate the accuracy of our reported value of $N_{\rm eff}^{\rm SM}$ to be $0.001$. 

\subsubsection*{$\Omega_{\nu}h^2$}\label{sec:comparison_omega_nu}
A parameter that is of relevance to cosmology is the energy density encoded in non-relativistic neutrino species: $\Omega_\nu h^2$. Since we describe neutrinos with a perfect Fermi-Dirac distribution function, we find that for degenerate and non-relativistic neutrinos:
\begin{align}
\Omega_\nu h^2 \equiv \frac{\sum_\nu m_\nu  \,n_{\nu} }{\rho_c/h^2} = \frac{\sum_\nu m_\nu}{\rho_c/h^2} \frac{3}{2} \frac{\zeta(3)}{\pi^2} \, T_{\nu}^3  = \frac{\sum_\nu m_\nu}{93.05 \,\text{eV}}\,,
\end{align}
where we have used $T_\gamma/T_\nu = 1.39578$ and $T_\gamma^0 = 2.7255\,\text{K}$~\cite{Fixsen:2009ug}. We notice that our determination accounting for LO-QED finite temperature corrections has an accuracy of 0.09\% as compared to the calculations that account for non-thermal distribution functions and neutrino oscillations~\cite{Mangano:2005cc}, $\Omega_\nu h^2 = \sum_\nu m_\nu/(93.12\,\text{eV})$. An instantaneous neutrino decoupling would correspond to $\Omega_{\nu}h^2 = \sum_\nu m_\nu/(94.11\,\text{eV})$~\cite{Dodelson:2003ft}.

When neutrino chemical potentials are allowed to vary (see Appendix~\ref{app:SM_chemical}), we find that the resulting number density encoded in non-relativistic degenerate neutrinos is:
\begin{align}
\Omega_\nu h^2 = \frac{\sum_\nu m_\nu}{93.14 \,\text{eV}}\,,
\end{align}
and hence in excellent agreement with the results of~\cite{Mangano:2005cc}. 

\subsubsection*{Primordial nuclei abundances }\label{sec:comparison_BBN}
\begin{table}[t]
\begin{center}
\begin{tabular}{l|cccc}
\hline\hline
\multicolumn{5}{c}{Primordial nuclei abundances in the Standard Model}  \\ \hline
Relative uncertainty      	   & $\Delta_{\rm rel} Y_p $   &$\Delta_{\rm rel} (\text{D/H}) $	& $\Delta_{\rm rel} (^{3}{\rm He}/\text{H}) $ &$ \Delta_{\rm rel} (^{7}{\rm Li}/\text{H})$  \\ \hline
Neutrino evolution, $T_{\nu_e} = T_{\nu_\mu}$   &  $8\times10^{-5}$    & $8\times 10^{-4}$  &	 $2\times 10^{-4}$      & $-8\times 10^{-4}$   	 \\
Neutrino evolution, $T_{\nu_e} \neq  T_{\nu_\mu}$   & $4\times10^{-5}$     & $1\times 10^{-3}$ &$4\times 10^{-4}$	       & $-1\times 10^{-3}$   	 \\
  \hline 
   Measurements, PDG~\cite{pdg}& $1.2\times10^{-2}$     & $10^{-2}$ &-	       & $1.9\times 10^{-1}$   	 \\
 Nuclear rates,~\texttt{PRIMAT}~\cite{Pitrou:2018cgg} & $6.8\times10^{-4}$     & $1.49\times 10^{-2}$ &$2.43\times 10^{-2}$	       & $4.4\times 10^{-2}$   	 \\
\hline\hline
\end{tabular}
\end{center}\vspace{-0.4cm}
\caption{$\Delta_{\rm rel} X\equiv\sigma_X/X$. The first two rows correspond to the relative change in the primordial nuclei abundances using the thermodynamic evolution obtained in this work to that compared to the default neutrino evolution used in~\texttt{PArthENoPE} which is based on the SM calculation of~\cite{Mangano:2005cc}. We appreciate that the relative difference between the evolution used in~\texttt{PArthENoPE} and that obtained here yields differences in the primordial element abundances that are at least one order of magnitude smaller than the error associated with current measurements or with nuclear uncertainties.    }\label{tab:Ynuclei_SM_compa}
\end{table}

 Neutrino decoupling occurs at $T\sim 2\,\text{MeV}$ and hence soon before proton-to-neutron interactions freeze-out: $T\sim 0.7\,\text{MeV}$, see e.g.~\cite{Pospelov:2010hj,Iocco:2008va,Sarkar:1995dd} and Figure~\ref{fig:SM_temperature}. The effect of neutrinos in primordial nucleosynthesis is threefold~\cite{Fields:1992zb,Serpico:2004gx}: \textit{i)} neutrinos participate in proton-to-neutron reactions and hence the neutrino number density enters the proton-to-neutron conversion rates, \textit{ii)} neutrinos affect the expansion rate of the Universe, and \textit{iii)} residual electron-neutrino interactions generate a net amount of entropy in the Universe. 
 
 State-of-the-art BBN codes, such as~\texttt{PArthENoPE}~\cite{Pisanti:2007hk,Consiglio:2017pot} and now also~\texttt{PRIMAT}~\cite{Pitrou:2018cgg}, account for each of these effects by taking the relevant neutrino evolution in the SM as obtained in~\cite{Mangano:2005cc}. 
 
 Here, we compare the relative changes to the nuclei abundances of some relevant nuclei given our neutrino evolution by implementing it in~\texttt{PArthENoPE}. In Table~\ref{tab:Ynuclei_SM_compa} we compare the relative changes in the nuclei abundances from our neutrino evolution to those obtained in the default mode of~\texttt{PArthENoPE}. We can clearly appreciate that the relative changes to the primordial nuclei abundances from our evolution in the SM as compared to the default mode of~\texttt{PArthENoPE} are smaller -- at least by an order of magnitude -- to current observational~\cite{pdg} and nuclear reaction~\cite{Pitrou:2018cgg} uncertainties. Finally, we note that we provide a file with the relevant Standard Model evolution that could be used in BBN codes. 

\subsubsection*{Entropy }\label{sec:comparison_entropy}
As a result of the out-of-equilibrium energy exchange between electrons and neutrinos at the time of neutrino decoupling, a small amount of entropy is generated in the Universe, see e.g.~\cite{Grohs:2015tfy}. We can parametrize the entropy of the Universe in terms of the usual number of degrees of freedom contributing to entropy density: $g_{\star s} \equiv s \,45/(2\pi^2T_\gamma^3)$, where $s$ is the entropy density. We obtain:
\begin{align}\label{eq:entropy_density_SM}
g_{\star s} = 2 + \frac{21}{4}\left( \frac{T_\nu}{T_\gamma}\right)^3 = 3.931\,.
\end{align}
If neutrinos decouple instantaneously~\cite{Dodelson:2003ft}: $g_{\star s} = 3.909$. From the results of~\cite{deSalas:2016ztq} one can infer $g_{\star s} = 3.933$. We appreciate that our resulting value of $g_{\star s}$ is accurate at the per mille level as compared to that obtained from~\cite{deSalas:2016ztq}.

\subsubsection*{$d\rho_\nu/dp_\nu$ and $dn_\nu/dp_\nu$}
\begin{figure}[t]
\centering
\begin{tabular}{cc}
\hspace{-0.8cm} \includegraphics[width=0.5\textwidth]{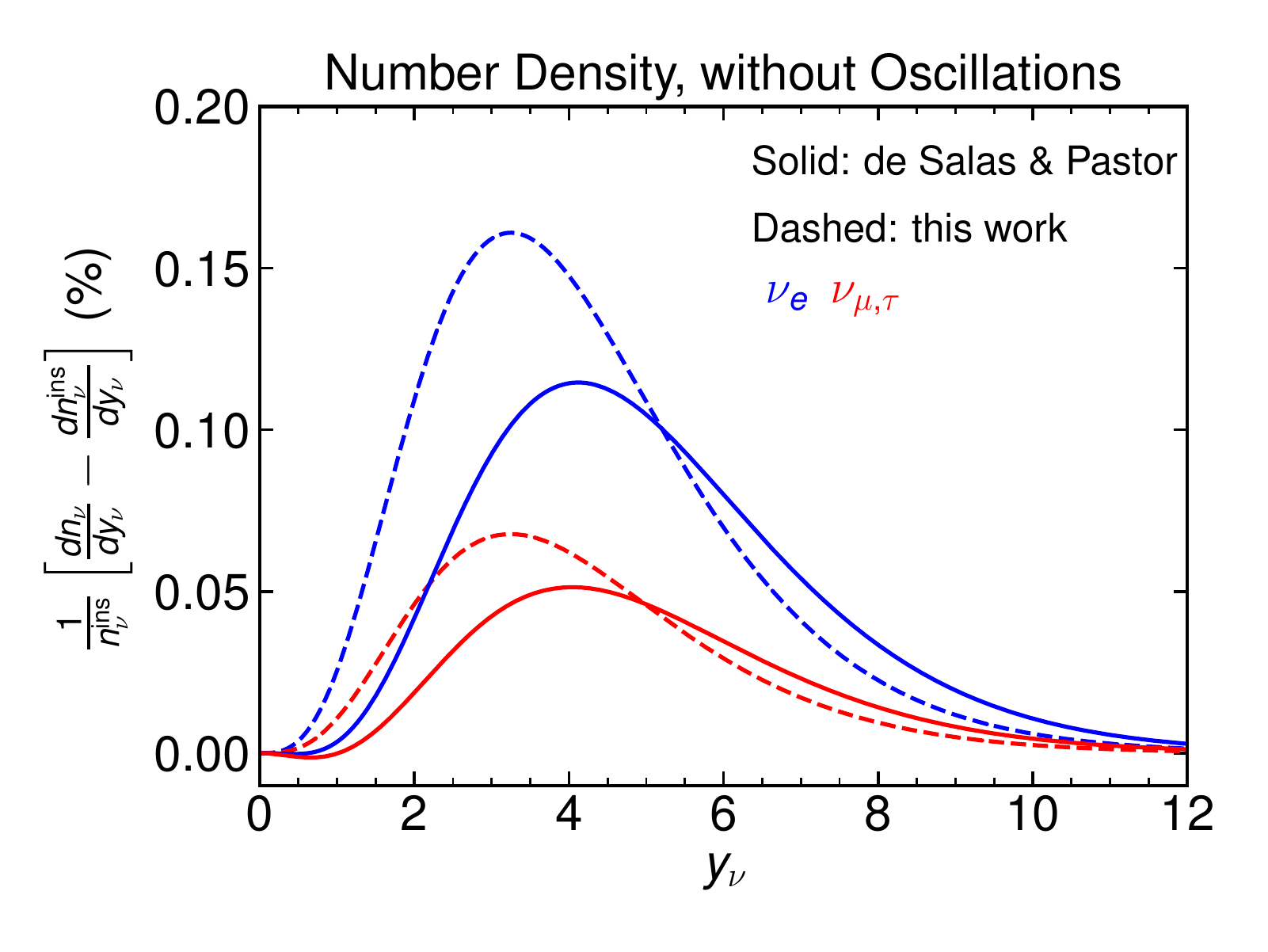} & \hspace{-0.7cm} \includegraphics[width=0.5\textwidth]{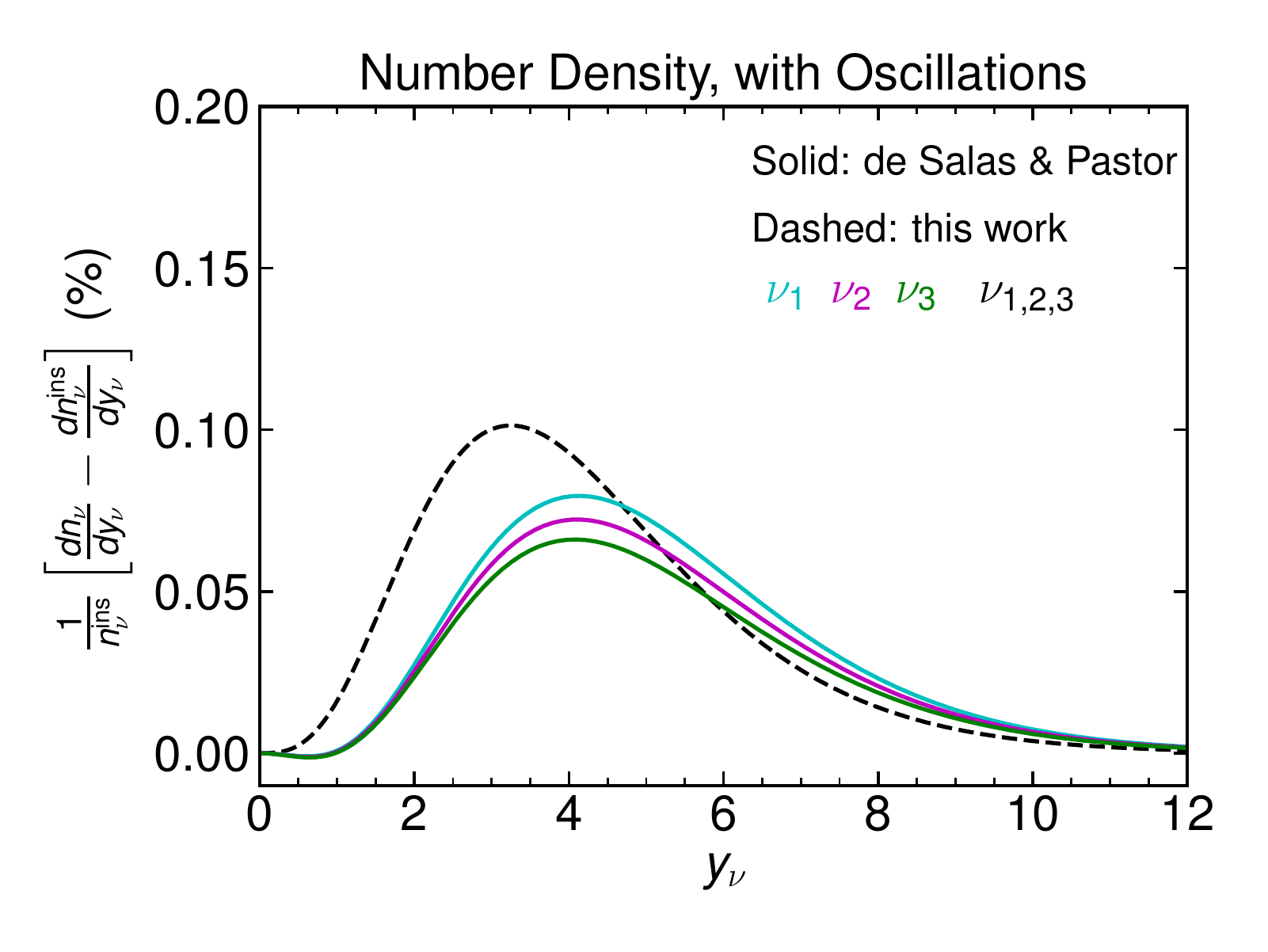} \\
\hspace{-0.8cm} \includegraphics[width=0.5\textwidth]{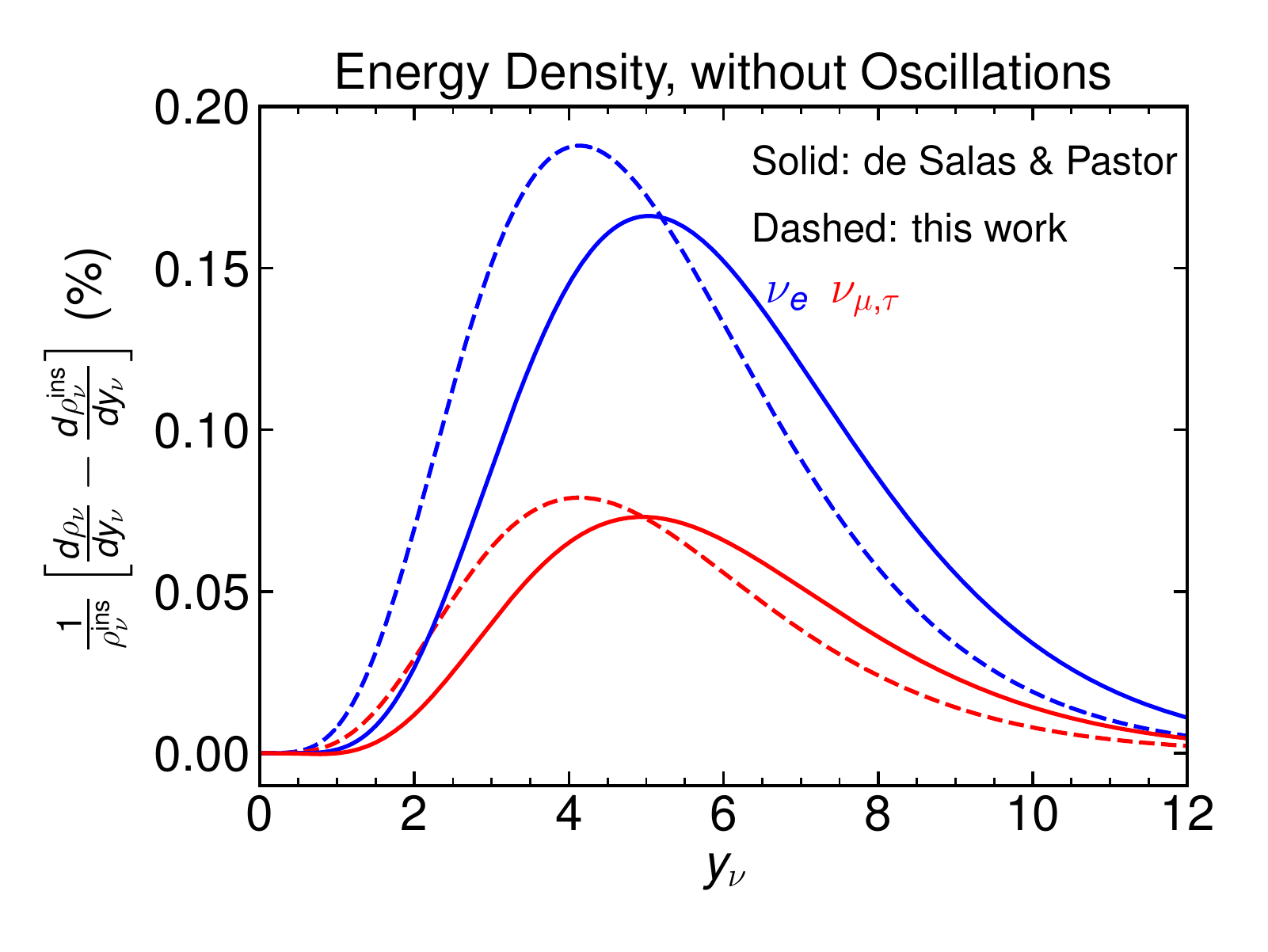} & \hspace{-0.7cm} \includegraphics[width=0.5\textwidth]{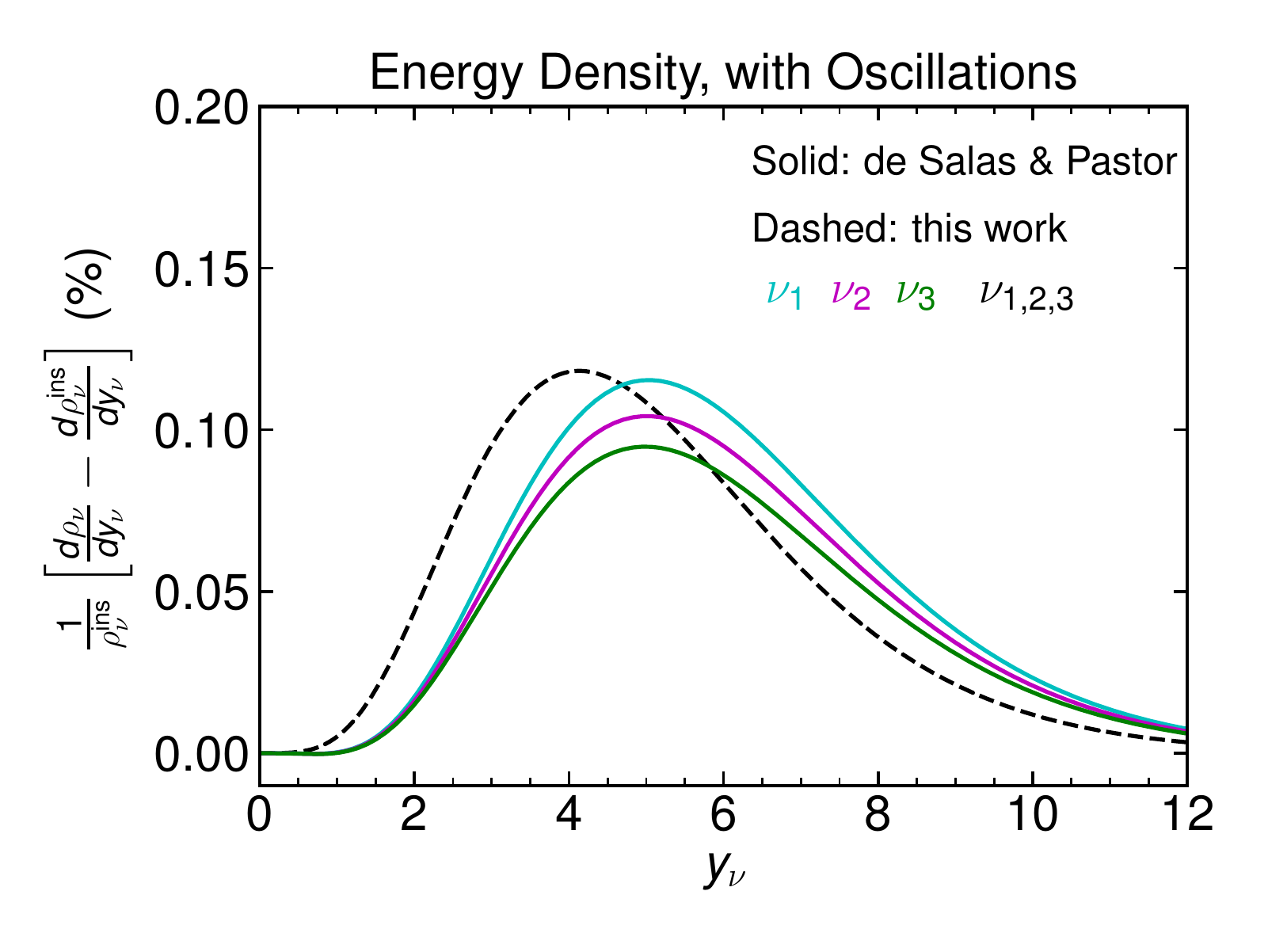}
\end{tabular}\vspace{-0.4cm}
\caption{Comparison between the results of de Salas \& Pastor~\cite{deSalas:2016ztq} (solid) and those obtained here (dashed). $y_\nu$ is the comoving momentum that is related to the photon temperature and the physical neutrino momentum as $y_\nu \equiv z_\gamma \times p_\nu/T_\gamma$, where $z_\gamma = 1.39777$. $f_\nu^{\rm ins}(y) \equiv (1+e^y)^{-1}$ is the neutrino distribution function of an instantaneously decoupled neutrino. $dn_\nu/dy \propto y^2 f_\nu$ and $d\rho_\nu/dy \propto y^3 f_\nu$. Our results in the left panels are obtained by evolving separately $T_{\nu_e}$ and $T_{\nu_{\mu,\,\tau}}$, while our results for the right panels are obtained by solving for a common neutrino temperature, $T_\nu$. The upper panels correspond to the contribution to the number density in neutrino species from residual electron-positron annihilations. The lower frames correspond to the resulting change in the energy density.  }\label{fig:dfdEcompa}
\end{figure}
Figure~\ref{fig:dfdEcompa} shows the neutrino distribution functions at $T<m_e/20$ in the case obtained here and compared to that obtained by solving the full Liouville equation as found in Ref.~\cite{deSalas:2016ztq}. We appreciate that the resulting non-instantaneous neutrino decoupling corrections to the number density in~\cite{deSalas:2016ztq} peak at $y_\nu \equiv p_\nu/T_\nu \sim 4$, while ours do at $y_\nu \sim 3$. In the case of $d\rho_\nu/dy_\nu$, those from~\cite{deSalas:2016ztq} peak at $y_\nu \sim 5$ while ours do at $y_\nu \sim 4$. Our results for $d\rho_\nu/dy_\nu$ and $dn_\nu/dy_\nu$ do not precisely match those of Ref.~\cite{deSalas:2016ztq} (as is to be expected, because the distribution functions that we consider are pure Fermi-Dirac distributions), however, they are accurate to the 0.05\% level for any momenta $y_\nu$. In addition, the agreement between the two calculations to the total neutrino energy density in each neutrino species is remarkable (see the first five rows of Table~\ref{tab:SM_compa}).

\subsubsection*{Evolution of entropy release: implications for BBN}

\begin{figure}[t]
\centering
\begin{tabular}{cc}
\hspace{-0.8cm} \includegraphics[width=0.53\textwidth]{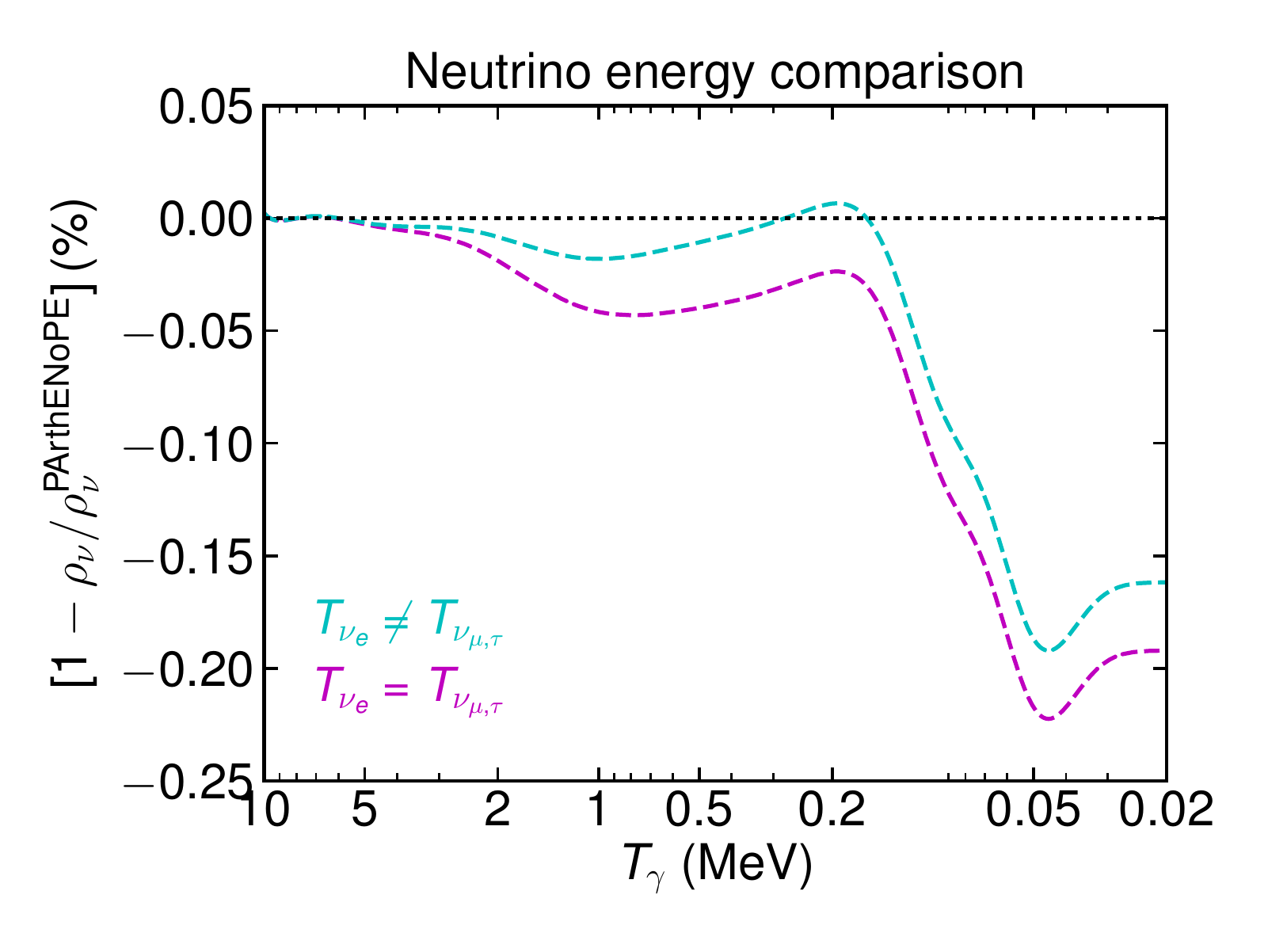} & \hspace{-0.7cm} \includegraphics[width=0.53\textwidth]{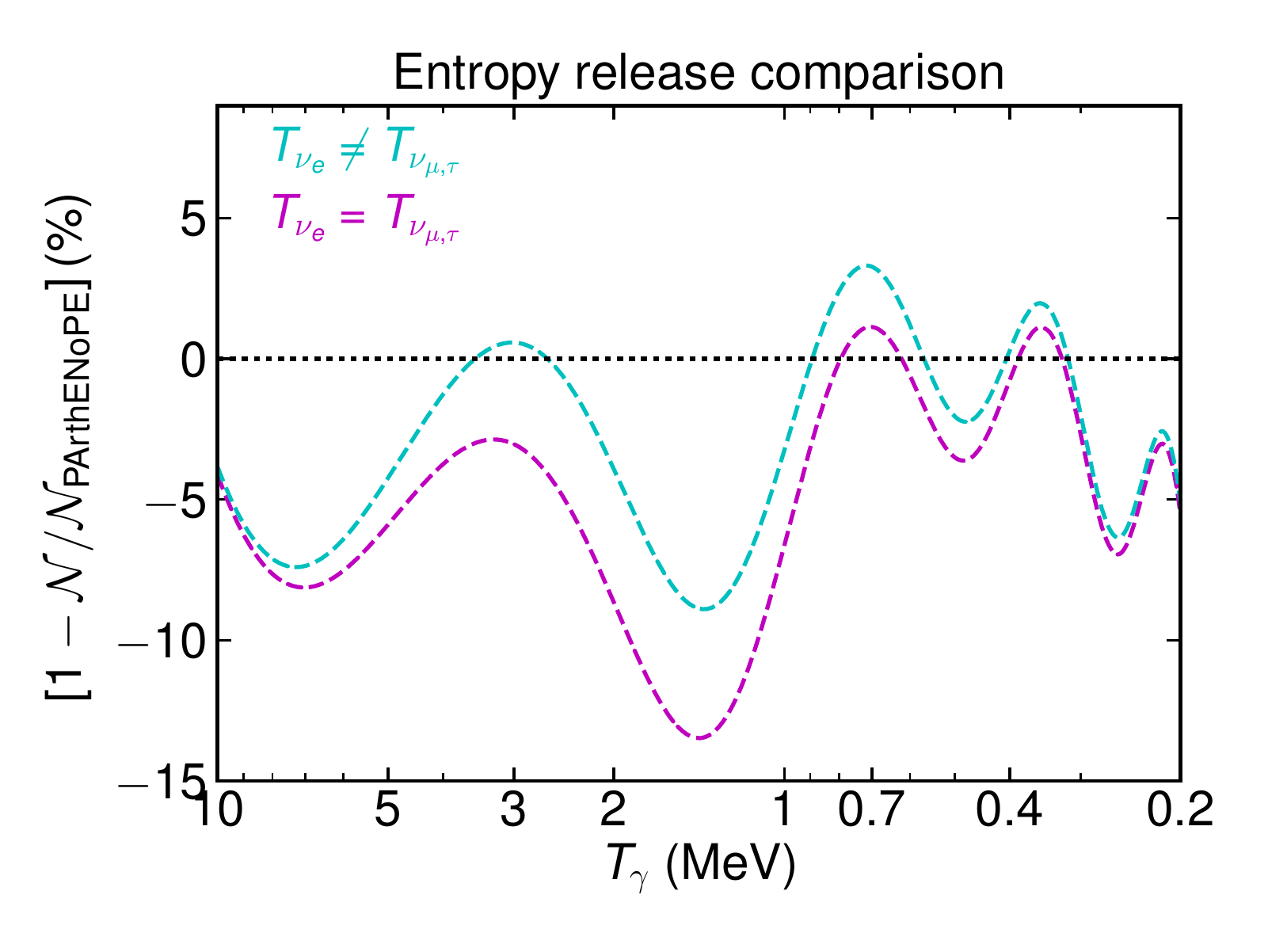}
\end{tabular}\vspace{-0.4cm}
\caption{\textit{Left panel:} neutrino energy evolution in the Standard Model as obtained in this work and compared with that used in the BBN code~\texttt{PArthENoPE}~\cite{Pisanti:2007hk,Consiglio:2017pot}. The difference between the two is at most 0.25\%. \textit{Right panel:} entropy release as a result of residual electron-positron annihilations into neutrinos in the Standard Model. This is encoded in terms of the function $\mathcal{N}$ that controls the time dependence of the energy exchange between electrons and neutrinos in the early Universe. The difference between the two is at the level of $\simeq 10\%$. We note, however, that our results for $\mathcal{N}$ in the case $T_{\nu_e} \neq T_{\nu_{\mu,\,\tau}}$ agree to the $\sim 3\%$ level with $\mathcal{N}$ as recently calculated in~\cite{Froustey:2019owm} by solving for the full Liouville equation but neglecting neutrino oscillations. }\label{fig:N_compa}
\end{figure}

We make contact with previous SM neutrino evolution results by considering the variable $\mathcal{N}$, introduced in Ref.~\cite{Pisanti:2007hk} (\texttt{PArthENoPE}), that accounts for the entropy release of electron-positron annihilation to the neutrino sector. 

Given our definitions (see Equation~\ref{eq:Tcomoving}), $\mathcal{N}$ can be written as:
\begin{align}
\mathcal{N}(a) \equiv \frac{1}{z_\gamma^4} \, \left(a \frac{d}{da} \left[ \left(\frac{a}{m_0}\right)^4 \rho_\nu\right] \right)\,,
\end{align}
where $\rho_\nu = \rho_{\nu_e} + 2 \rho_{\nu_\mu}$. We can work out this expression to write it as:
\begin{align}
\mathcal{N}(a) &=  \frac{4}{z_\gamma^4} \, \frac{7}{8}\frac{\pi^2}{15}\,a\, \sum_\alpha  z_{\nu_\alpha}^3 \, \frac{d{z_{\nu_\alpha}}}{da}  \,,
\end{align}
and hence, it is clear that if $\delta \rho_\nu/\delta t = 0$ then $\mathcal{N} =0$ (see Equation~\eqref{eq:comoving_T}). Ref.~\cite{Pisanti:2007hk} provides a fitting function for $\mathcal{N}$ as a function of $m_e/T_\gamma$. In our case, $m_e/T_\gamma = \frac{m_e}{m_0}\frac{a}{z_\gamma} = \frac{a}{z_\gamma}$. 

The right panels of Figure~\ref{fig:N_compa} show the comparison between $\mathcal{N}$ as obtained from our approach and that used by default in \texttt{PArthENoPE}. We can appreciate that the difference is at the level of $10\%$ or less. The left panels of Figure~\ref{fig:N_compa} show the comparison in terms of the neutrino energy density. We see that difference in terms of  $\rho_\nu$ is at most 0.25\%. The collective effect of these two variables in terms of the prediction for the primordial abundances is shown in Table~\ref{tab:Ynuclei_SM_compa} where one can see that the change in the relevant nuclei abundances is smaller -- at least by an order of magnitude -- to current observational and nuclear rates uncertainties.

\newpage

\subsection{$N_{\rm eff}$ in the Standard Model including neutrino chemical potentials}\label{app:SM_chemical}
In this appendix we consider the case of neutrino decoupling in the Standard Model by allowing chemical potentials to vary. We do so by solving equations~\eqref{eq:dT_dt_simple} and~\eqref{eq:dmu_dt_simple} for the neutrino temperature and chemical potential respectively. For the electromagnetic component of the plasma we still neglect chemical potentials since they are highly suppressed by the small baryon-to-photon ratio $\mu_e/T \sim 10^{-9}$. Therefore, the photon temperature still evolves according to~\eqref{eq:T_gamma_QED}. In addition, we consider that there is no primordial lepton asymmetry, so that $\mu_\nu = \mu_{\bar{\nu}}$.

\subsubsection*{Neutrino-neutrino and electron-neutrino interactions}\label{app:SM_chemical_inter}

We include the effect of the neutrino chemical potentials in the neutrino-neutrino and neutrino-electron energy and number density transfer rates, that allowing for neutrino chemical potentials explicitly read:
\begin{align}
\left. \frac{\delta \rho_{\nu_e}}{\delta t} \right|_{\rm SM}^{\rm FD} &= \frac{G_F^2}{\pi^5}\left[ 4\left(g_{eL}^2+g_{eR}^2\right) \, G(T_\gamma,0,T_{\nu_e},\mu_{\nu_e}) + 2 \, G(T_{\nu_\mu},\mu_{\nu_\mu},T_{\nu_e},\mu_{\nu_e}) \right] \, ,\\
\left. \frac{\delta \rho_{\nu_\mu}}{\delta t} \right|_{\rm SM}^{\rm FD} &= \frac{G_F^2}{\pi^5}\left[ 4\left(g_{\mu L}^2+g_{\mu R}^2\right) \, G(T_\gamma,0,T_{\nu_\mu},\mu_{\nu_\mu}) -  G(T_{\nu_\mu},\mu_{\nu_\mu},T_{\nu_e},\mu_{\nu_e}) \right] \, , \\
\left. \frac{\delta n_{\nu_e}}{\delta t} \right|_{\rm SM}^{\rm FD} &= 8f_n^{\rm FD}\frac{G_F^2}{\pi^5}\left[ 4\left(g_{eL}^2+g_{eR}^2\right) (T_\gamma^8 - T_{\nu_e}^8e^{2\frac{\mu_{\nu_e}}{T_{\nu_e}}})\! + \!2  (T_{\nu_\mu}^8  e^{2\frac{\mu_{\nu_\mu}}{T_{\nu_\mu}}} - T_{\nu_e}^8 e^{2\frac{\mu_{\nu_e}}{T_{\nu_e}}}) \right] ,\\
\left. \frac{\delta n_{\nu_\mu}}{\delta t} \right|_{\rm SM}^{\rm FD} &= 8f_n^{\rm FD}\frac{G_F^2}{\pi^5}\left[4\left(g_{\mu L}^2+g_{\mu R}^2\right)  (T_\gamma^8 - T_{\nu_\mu}^8e^{2\frac{\mu_{\nu_\mu}}{T_{\nu_\mu}}}) -  (T_{\nu_\mu}^8e^{2\frac{\mu_{\nu_\mu}}{T_{\nu_\mu}}} - T_{\nu_e}^8e^{2\frac{\mu_{\nu_e}}{T_{\nu_e}}}) \right] \, ,
\end{align}
where $f_n^{\rm FD} = 0.852$ and 
\begin{align}
G(T_1,\mu_1,T_2,\mu_2) &=  32 \, f_a^{\rm FD} \,  (T_1^9\,e^{2\frac{\mu_1}{T_1}}-T_2^9\,e^{2\frac{\mu_2}{T_2}}) + 56 \, f_s^{\rm FD} \, e^{\frac{\mu_1}{T_1}}\,e^{\frac{\mu_2}{T_2}} \,T_1^4\,T_2^4 \, (T_1-T_2) \, .
\end{align} 
The effect of a non-negligible electron mass is also included as in~\eqref{sec:SM_interactions}.

\subsubsection*{Results}\label{app:SM_chemical_results}

We solve Equations~\eqref{eq:dTdmu_generic_simple} for the neutrinos and Equation~\eqref{eq:T_gamma_QED} for the electromagnetic plasma using as initial conditions $T_\nu = T_\gamma = 10\,\text{MeV}$ and $\mu_\nu/T_\nu = -10^{-4}$. Such choice of initial value for the neutrino chemical potentials does not affect any result, since the energy density in neutrinos scales as $\rho_\nu \sim e^{\mu_\nu/T_\nu}$.

Our results are outlined in Table~\ref{tab:Neff_SM_chem}. We can appreciate that when neutrino chemical potentials are allowed to vary and their value is dictated by the dynamics, they are still negligible: $T_\nu/\mu_\nu \sim -240$. These values of the neutrino-chemical potentials in the Standard Model are in excellent agreement with the results reported in~\cite{Birrell:2014uka}. Since $\rho_\nu \sim e^{\mu_\nu/T_\nu}$ and  $1-e^{-1/240} \sim 0.004$, the effect of neutrino chemical potentials on $N_{\rm eff}$ is small. This therefore justifies solving for neutrino decoupling assuming vanishing neutrino chemical potentials.  

When comparing the results in terms of $N_{\rm eff}$, $g_{\star S}$, $\Omega_\nu h^2$ with previous state-of-the-art calculations in the Standard Model~\cite{deSalas:2016ztq,Mangano:2005cc} we find an excellent agreement. Specially, we notice that allowing chemical potentials to vary yields an even better value of the neutrino number density and therefore of $\Omega_\nu h^2$. This was to be expected, since neutrino chemical potentials arise as a result of the freeze-out of number changing processes. 

\begin{table}[t]
\begin{center}
\begin{tabular}{l|c|c|c|c|c}
\hline\hline
\multicolumn{6}{c}{$\,\, $Neutrino Decoupling in the SM including neutrino chemical potentials $\,\, $}  \\ \hline\hline
 $\,$Case	     		         &$T_\gamma/ T_{\nu}$ &$T_{\nu}/\mu_{\nu}$ &$N_{\rm eff}$	& $g_{\star s}$	& $\sum m_\nu/\Omega_\nu h^2$	 \\ \hline 
      Instantaneous 							& 1.40102 & $\infty$  &3.000	& 3.909	& 94.11 eV	 \\  \hline
            $\,\,\mu_\nu = 0$, LO-QED 	     		         &1.39567	& $\infty$ &{3.046}	& {3.931}	& {93.03} eV	 \\ \hline 
      $\mu_\nu \neq 0$, LO-QED 	     		        &1.39434  & -236.5   &3.045	& 3.931	& 93.12 eV	 \\ \hline 
       \,Refs~\cite{deSalas:2016ztq,Mangano:2005cc} (LO-QED)	& - 		& - 		&3.045	& 3.933	& 93.12 eV	 \\  \hline\hline
          $\,\,\mu_\nu = 0$, NLO-QED 	     		         &1.39578	& $\infty$ &{3.045}	& {3.931}	& {93.05} eV	 \\ \hline 
      $\mu_\nu \neq 0$, NLO-QED 	     		         &1.39445 & -236.5   &\textbf{3.045}	& \textbf{3.930}	& \textbf{93.14 eV}	 \\ \hline
      \hline
\end{tabular}
\end{center}\vspace{-0.3cm}
\caption{The resulting neutrino temperature and chemical potentials as obtained in this work with and without allowing for non-negligible neutrino chemical potentials. We also show the resulting values of $N_{\rm eff}$, $g_{\star s}$ and $\sum m_\nu/\Omega_\nu h^2$ for degenerate neutrinos.   }\label{tab:Neff_SM_chem}
\end{table}

\subsection{Detailed neutrinophilic scalar thermodynamics}\label{app:1_2_cases}
In this appendix we present a detailed comparison between the Full solution to the Liouville equation and the Fast approach presented in this work. 

\begin{figure}[b]
\centering
\begin{tabular}{cc}
\hspace{-0.8cm} \includegraphics[width=0.53\textwidth]{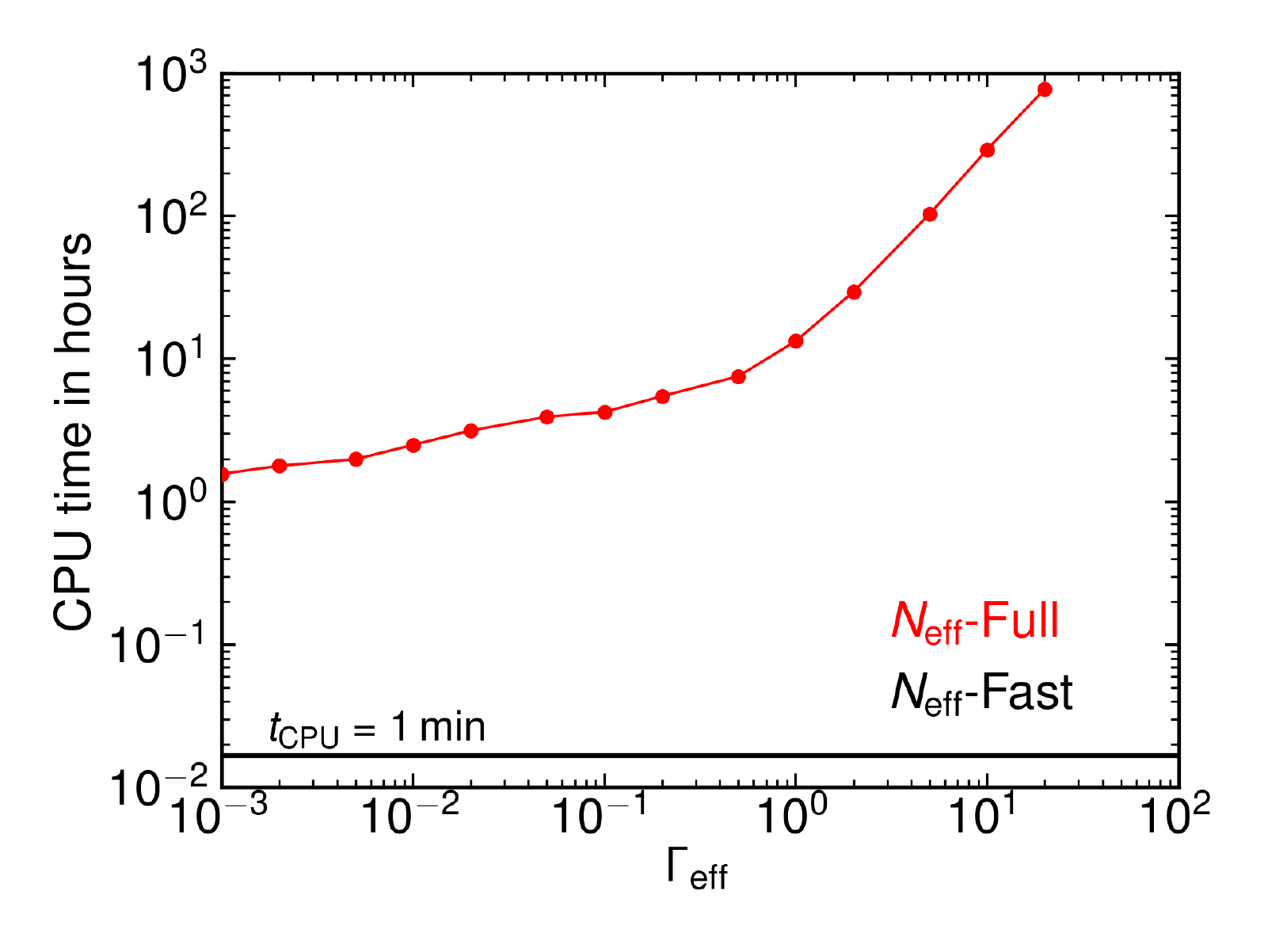} & \hspace{-0.7cm} \includegraphics[width=0.53\textwidth]{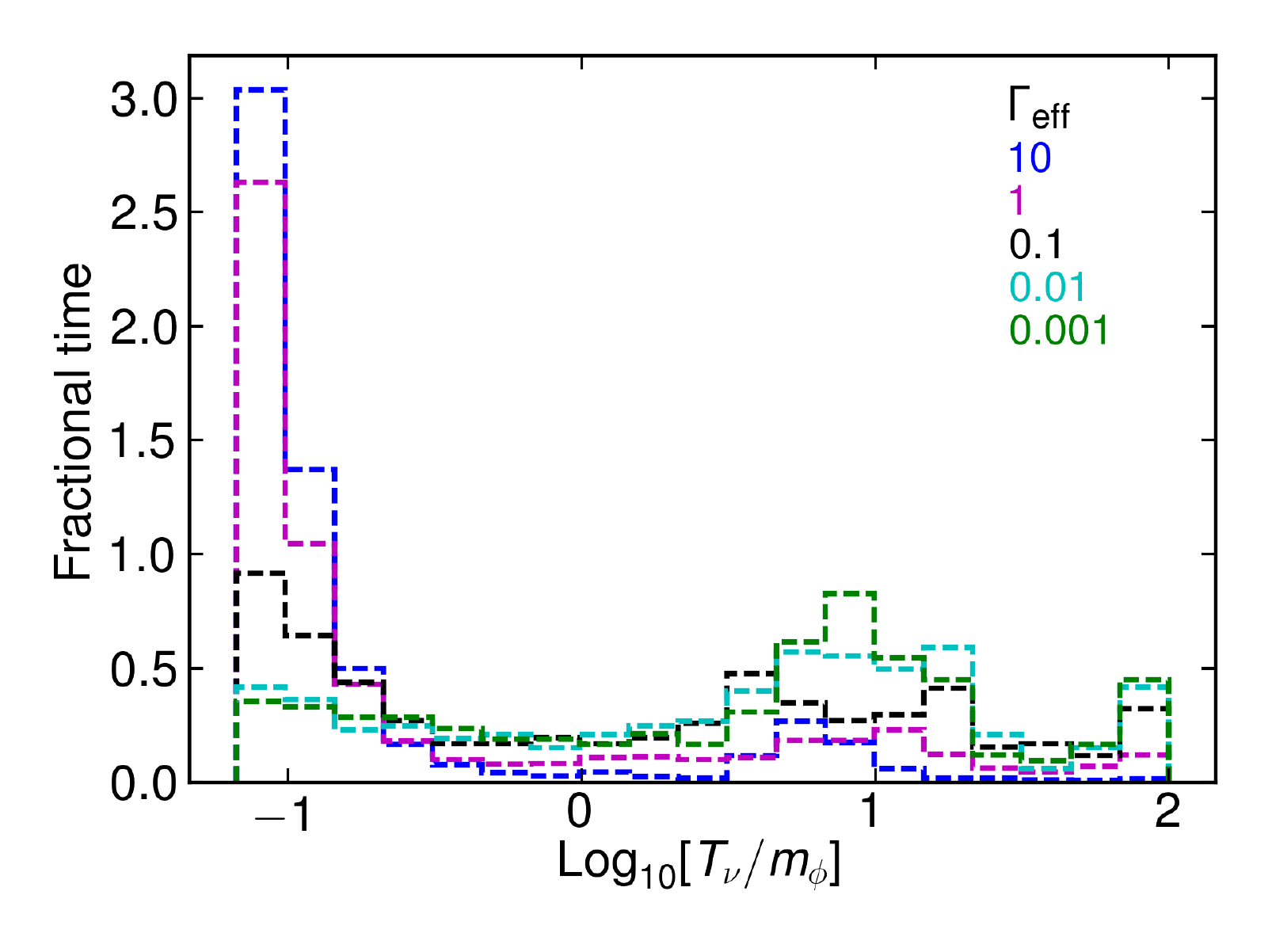} 
\end{tabular}
\vspace{-0.8cm}
\caption{\textit{Left:} CPU time spent by the integrator in solving the Liouville equation for the neutrino-$\phi$ system~\eqref{eq:Boltzmann_maj_exact} (red) and by solving the system of differential equations~\eqref{eq:full_system_maj} (black). \textit{Right:} fractional time spent by the integrator when solving the Liouville equation for $f_\nu$ and $f_\phi$ as a function of $T_\nu$.}\label{fig:CPUtime}
\end{figure}

\begin{figure}[t]
\centering
\begin{tabular}{cc}
\hspace{-0.8cm} \includegraphics[width=0.53\textwidth]{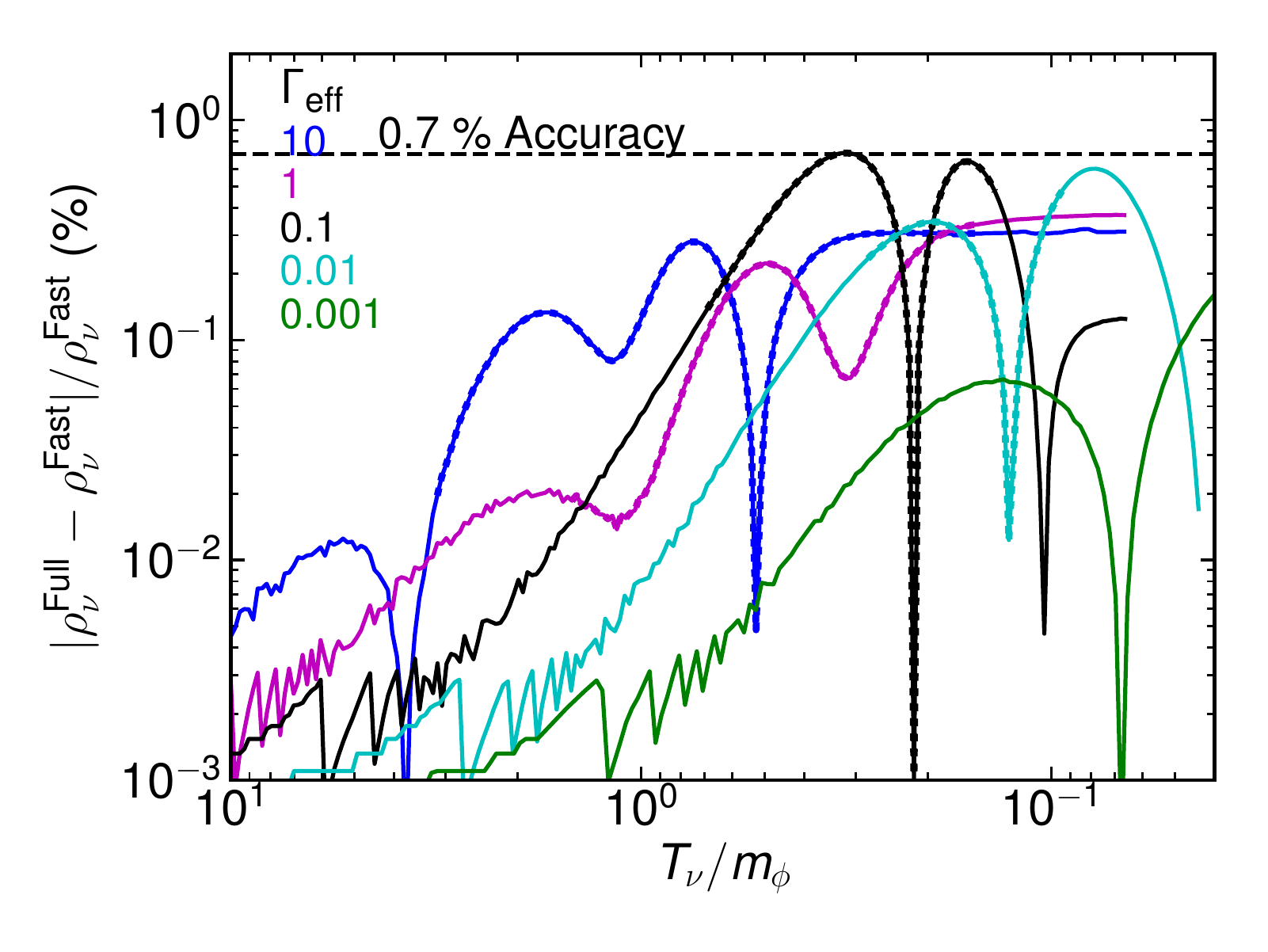} & \hspace{-0.7cm} \includegraphics[width=0.53\textwidth]{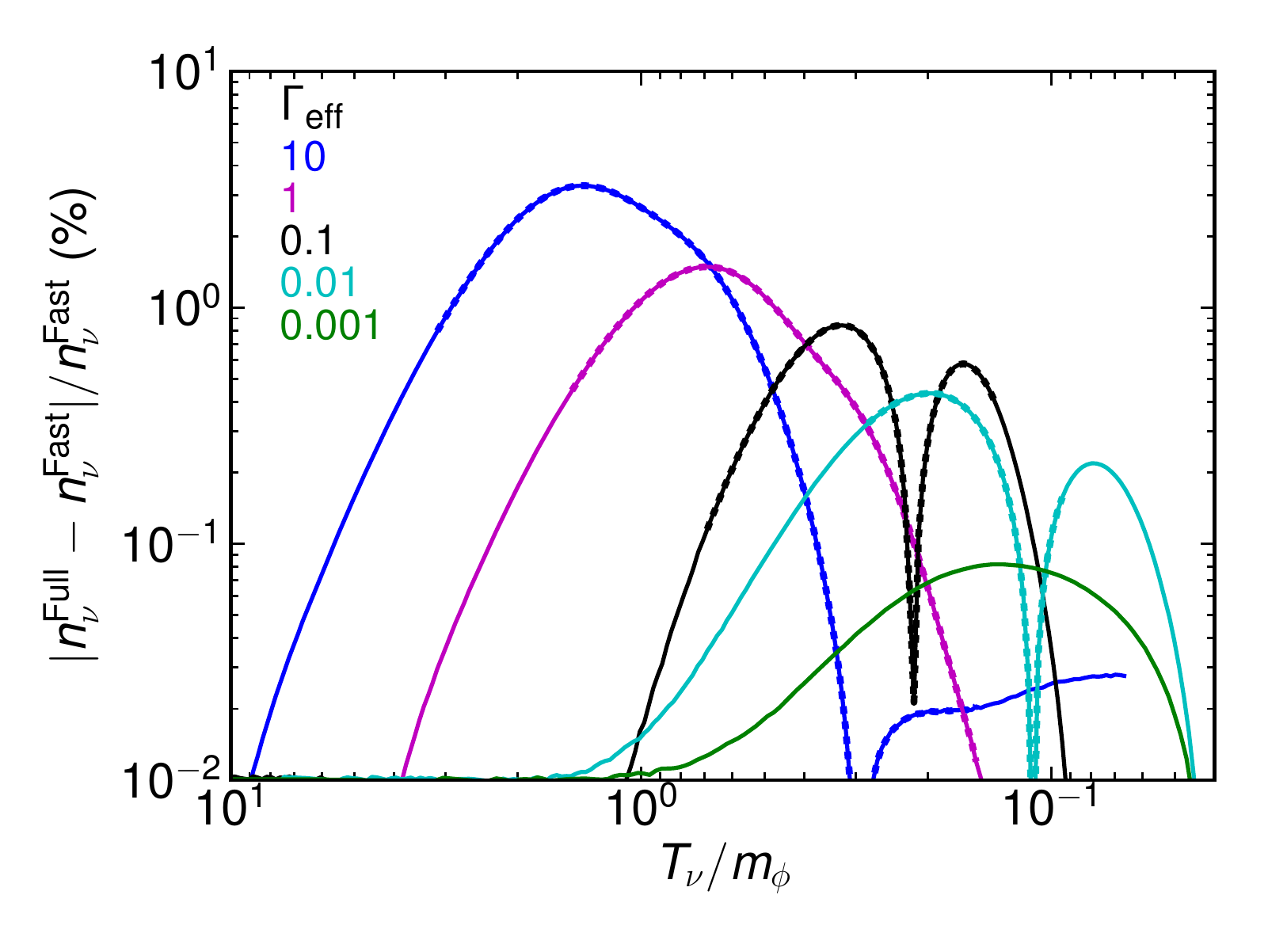} \\
\hspace{-0.8cm} \includegraphics[width=0.53\textwidth]{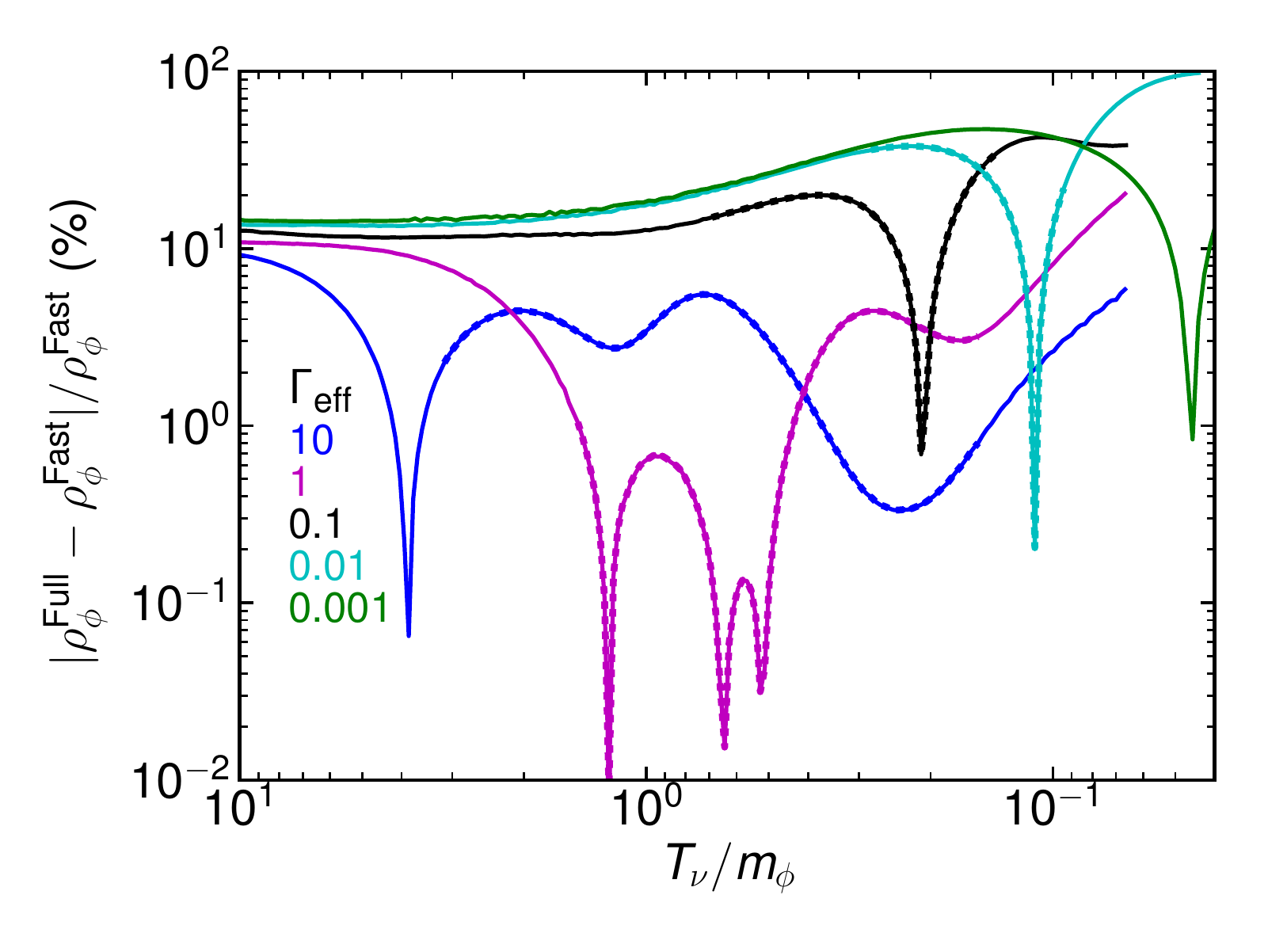} & \hspace{-0.7cm} \includegraphics[width=0.53\textwidth]{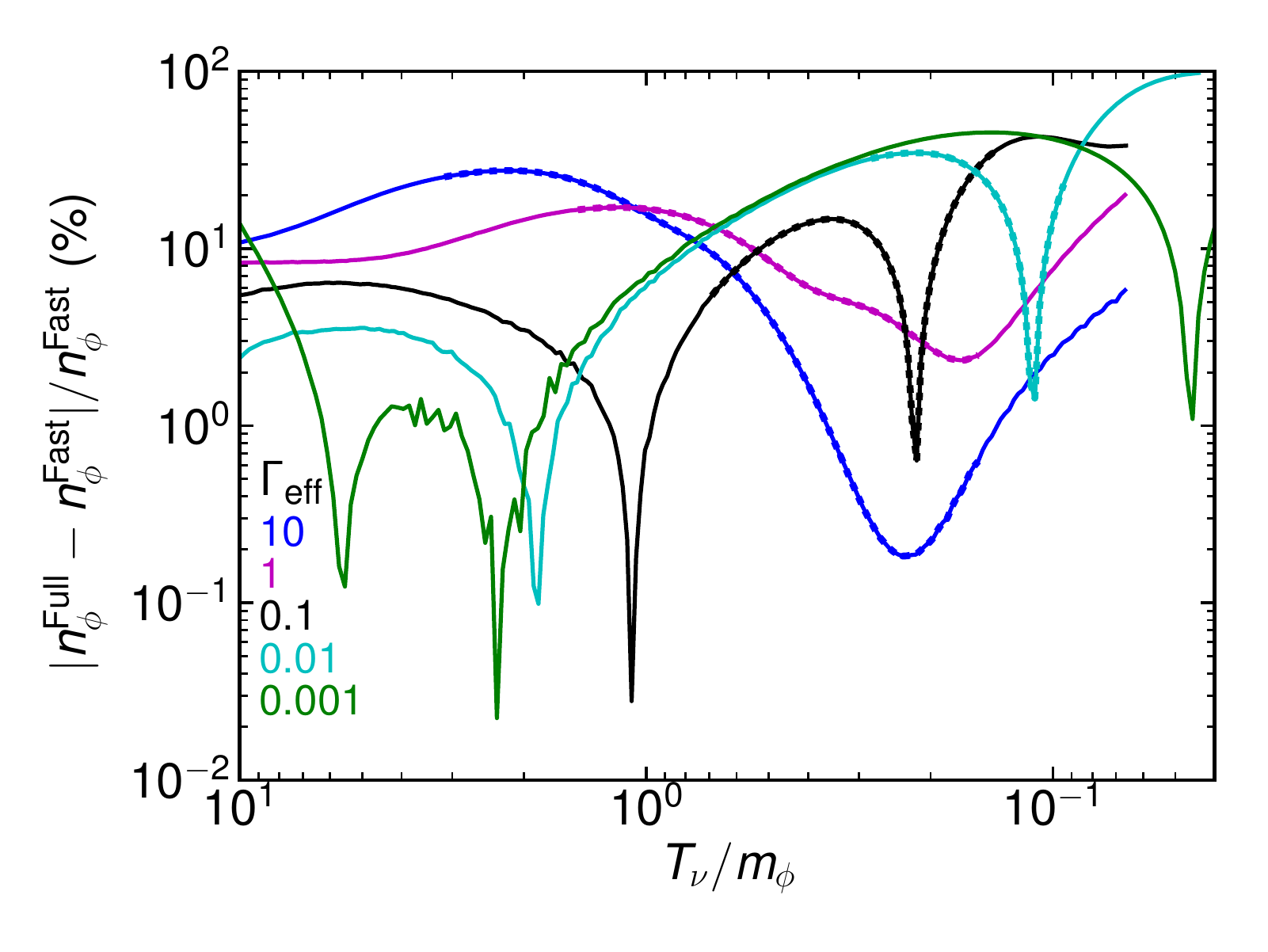} \\
\end{tabular}
\vspace{-0.5cm}
\caption{Relative difference of $\rho_\nu$, $n_\nu$, $\rho_\phi$ and $n_\phi$ as a function of $T_\nu$ between the full solution to the Liouville equation~\eqref{eq:Boltzmann_maj_exact} and the simplified and fast approach to it developed in this study~\eqref{eq:dTdmu_generic_simple}. The regions of the lines that are dashed represent the cosmologically relevant ones, in which $\rho_\phi/\rho_\nu > 10^{-3}$. }\label{fig:rhoandnu_compa}
\end{figure}

 First, in Figure~\ref{fig:CPUtime} we show the CPU time used for each computation as a function of $\Gamma_{\rm eff}$.
The comparison in terms of $\rho_\phi$, $\rho_\nu$, $n_\phi$ and $n_\nu$ a function of $T_\nu$ is displayed in Figure~\ref{fig:rhoandnu_compa}. We can appreciate and excellent overall agreement.  

\begin{figure}[t]
\centering
\begin{tabular}{cc}
\hspace{-0.8cm} \includegraphics[width=0.53\textwidth]{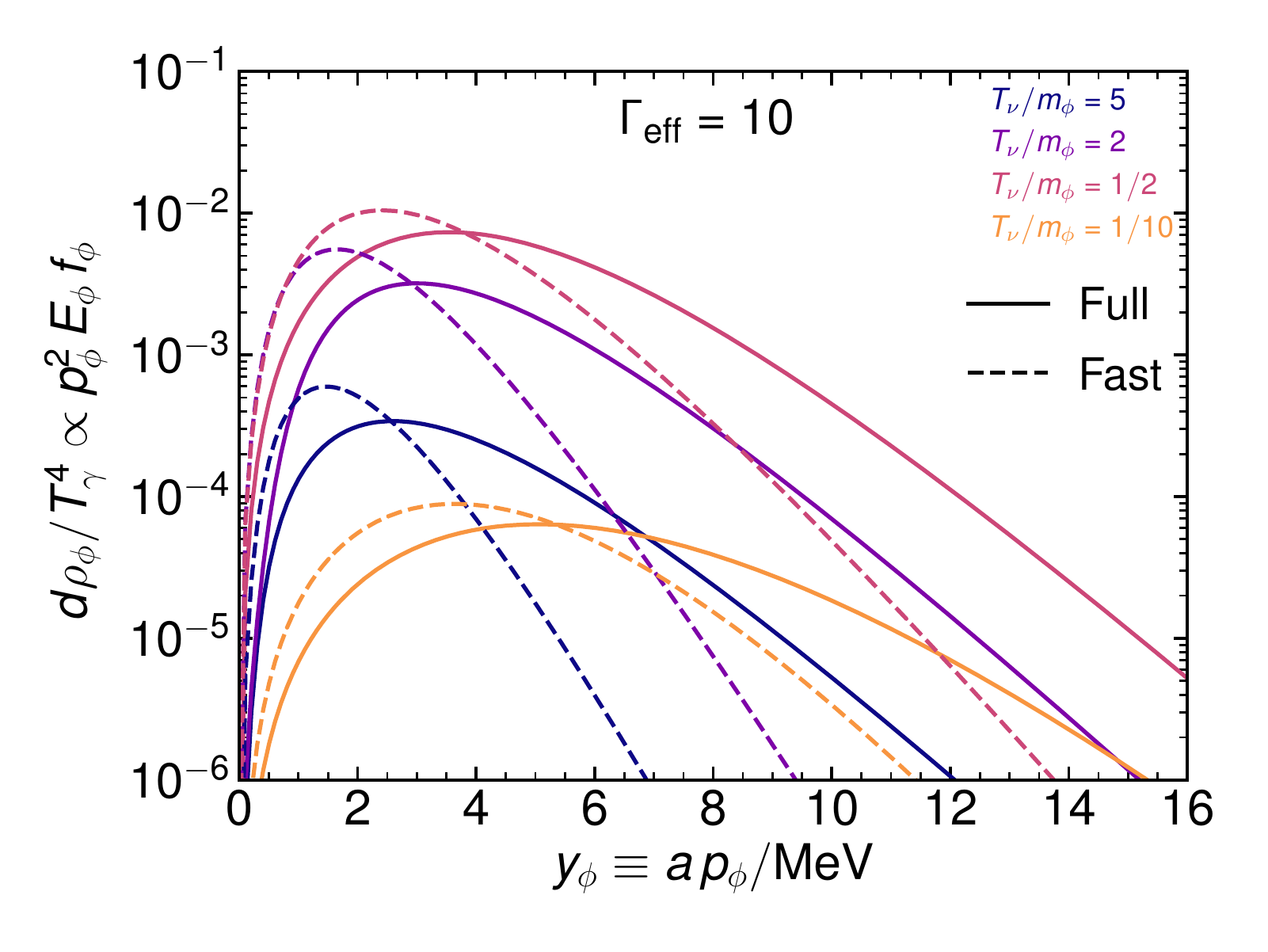} & \hspace{-0.7cm} \includegraphics[width=0.53\textwidth]{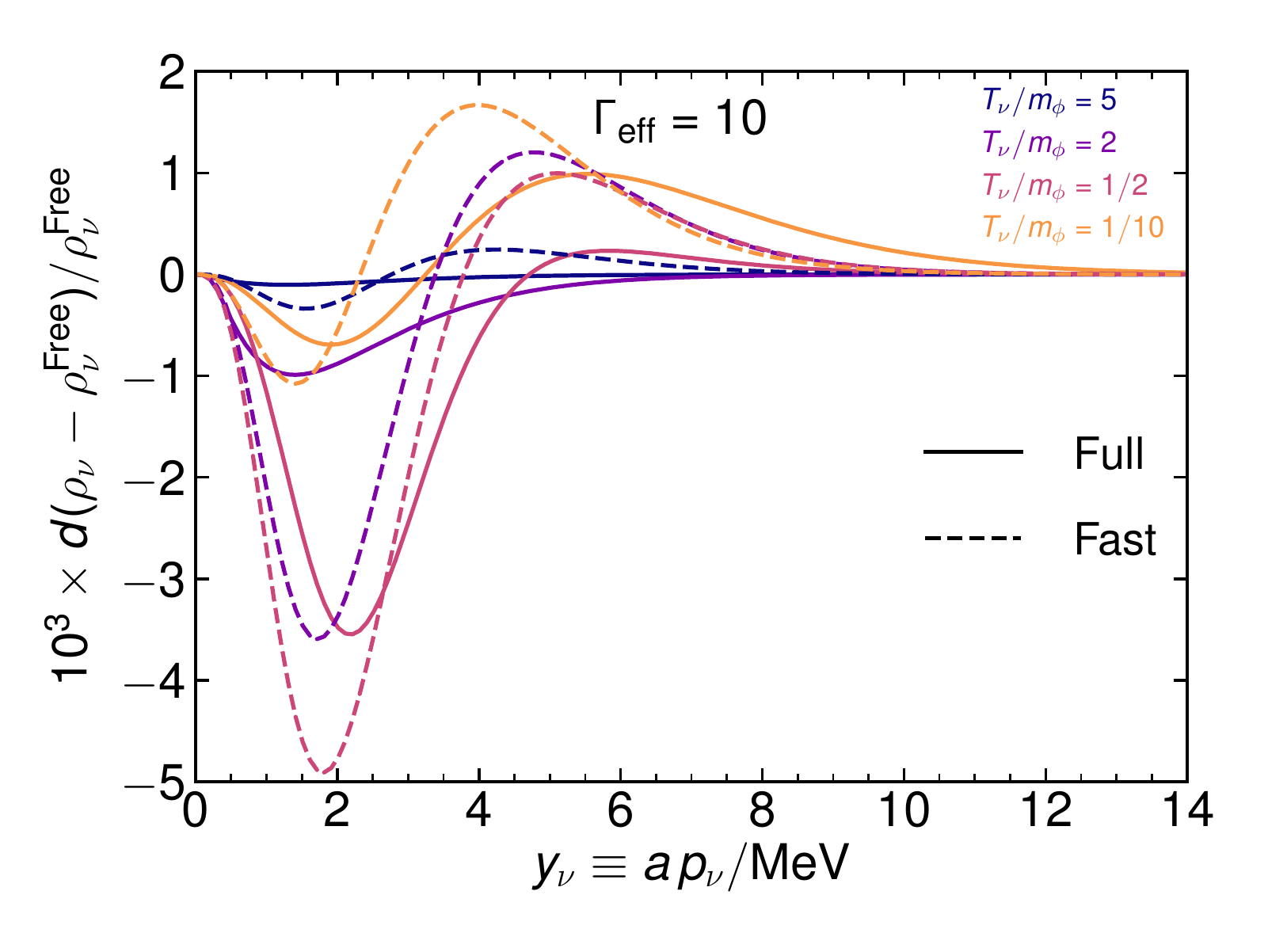} \\
\hspace{-0.8cm} \includegraphics[width=0.53\textwidth]{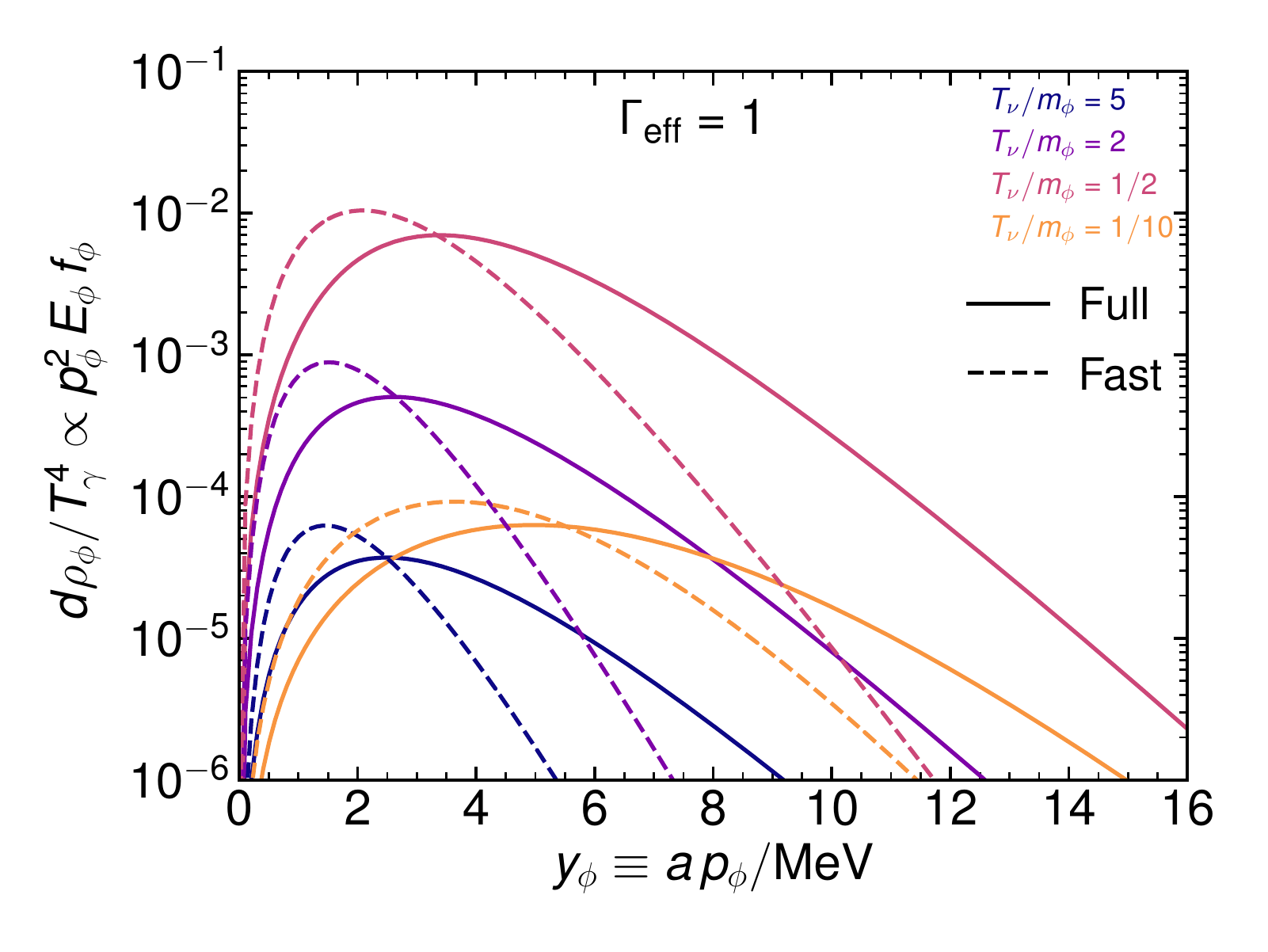} & \hspace{-0.7cm} \includegraphics[width=0.53\textwidth]{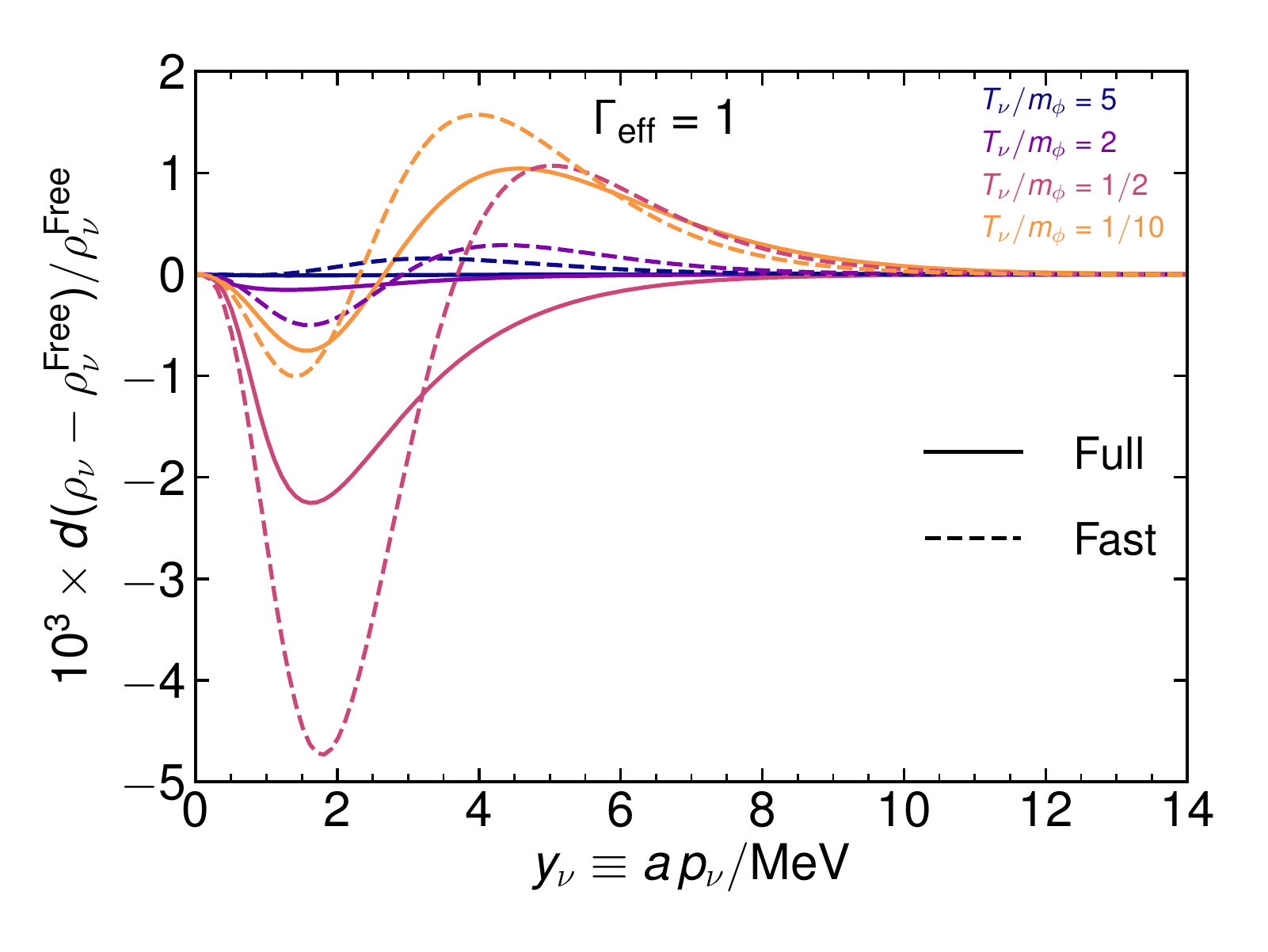} \\
\hspace{-0.8cm} \includegraphics[width=0.53\textwidth]{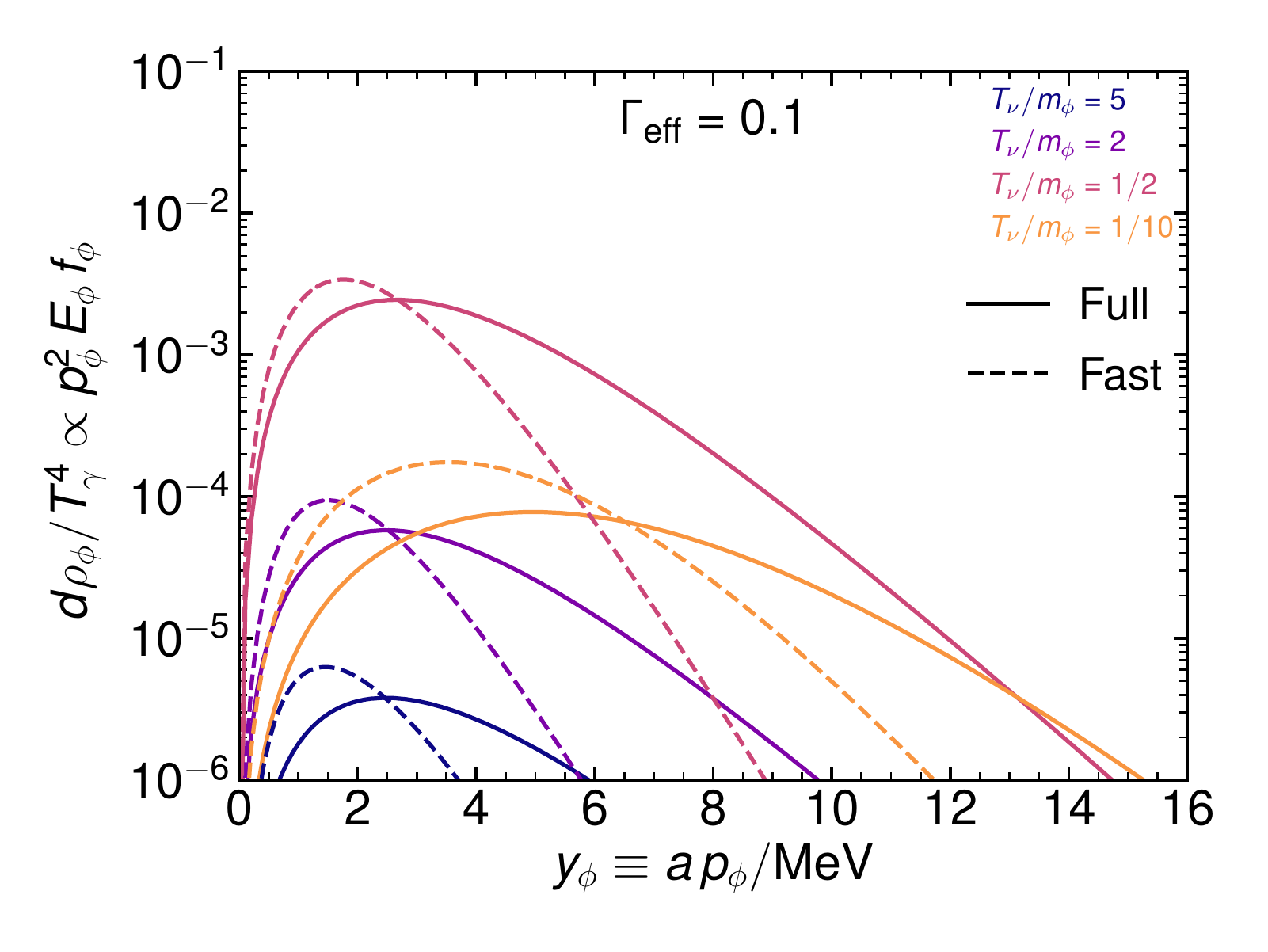} & \hspace{-0.7cm} \includegraphics[width=0.53\textwidth]{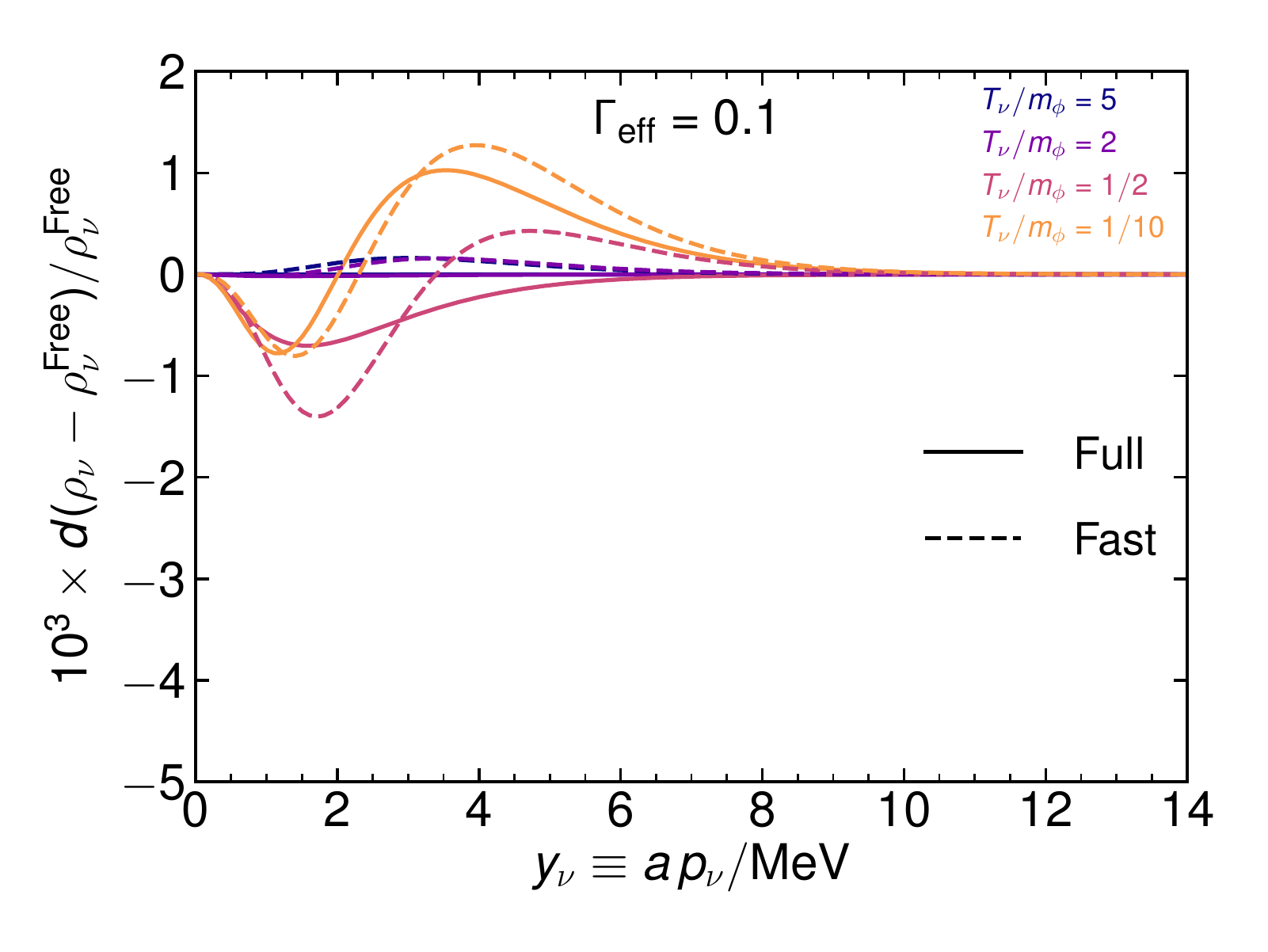} \\
\end{tabular}
\vspace{-0.4cm}
\caption{Snapshots of $f_\nu$ and $f_\phi$ for $\Gamma_{\rm eff} = 0.1,\,1,\,10$ as a function of $T_\nu/m_\phi$. Solid lines correspond to the solution of the exact Liouville equation~\eqref{eq:Boltzmann_maj_exact} while dashed lines correspond to the solution of the simplified and fast approach to it developed in this study~\eqref{eq:dTdmu_generic_simple}. \textit{Left panel:} $d\rho_\phi(y_\phi)/T^4$ as a function of comoving momenta $y$. This quantity is proportional to $E_\phi p_\phi^2 f_\phi$. \textit{Right panel:} relative difference between the distribution function of neutrinos as compared with purely freely streaming ones with $T_\gamma/T_\nu = 1.39573$.}\label{fig:dfdEcompa_nu}
\end{figure}

Finally, in Figure~\ref{fig:dfdEcompa_nu} we show the comparison between the two approaches at the level of the neutrino and $\phi$ distribution functions. We clearly see that there is good agreement between the two although there are some differences. One appreciates that the Fast solutions overestimate the energy/number density of $\phi$ particles at low momenta $ y \lesssim 3-4$ while they underestimate it for $y \gtrsim 4$. However, this mismatch is compensated when integrating over momenta since the total energy and number densities agree very well in the two cases. With respect to the neutrino distribution function, we notice that the relative difference at the level of $d \rho_\nu/dy_\nu$ is below 0.3\% independently of momenta and of temperature. In particular, we notice that for $T\ll m_\phi$, the neutrino distributions agree with a remarkable $0.1\%$ accuracy. Note that these very small differences have no impact for CMB observations~\cite{Cuoco:2005qr}.

\newpage
\subsection{Evolution equations in the Maxwell-Boltzmann approximation}\label{app:MB_limit}
In this appendix we particularize the generic system of Equations~\eqref{eq:dTdmu_generic_simple} for the case in which the species follows Maxwell-Boltzmann statistics. We explicitly write down the results in the limits in which $T \gg m$ and $T\ll m$ and compare the temperature evolution of a non-interacting particle with negligible chemical potential with the Bose-Einstein and Fermi-Dirac cases. Finally, as an illustrative example, we re-derive the well-known baryon temperature evolution as relevant for recombination. 

\subsubsection*{Simplified Evolution Equations}
We can substantially simplify Equations~\eqref{eq:dTdmu_generic_simple} for a species that follows MB statistics, since in that scenario:
\begin{align}
\frac{d n(T,\mu)}{d T} &= \frac{\rho -\mu n }{T^2} \, ,\qquad  \frac{d n(T,\mu)}{d \mu} = \frac{n}{T}  \, ,\\
\frac{d \rho(T,\mu)}{d T} &= \frac{\sigma -\mu \rho}{T^2}  \, ,\qquad  \frac{d \rho(T,\mu)}{d \mu} = \frac{\rho}{T} \, ,
\end{align} 
where we have defined
\begin{align}
\sigma  \equiv  \frac{g}{2\pi^2} \int_{m}^{\infty}  dE \, {E^3 \sqrt{E^2-m^2}}\,e^{-(E-\mu)/T} \, .
\end{align} 
In this case, we can easily work out the temperature and chemical potential evolution~\eqref{eq:dTdmu_generic_simple} to find:
\begin{align}
\frac{dT}{dt} &=  {T^2 \left[-3 H n p-\frac{\delta n}{\delta t} \rho +\frac{\delta \rho}{\delta t} n\right]}\bigg/\left[n \sigma -\rho ^2\right] \,,\label{eq:generalized_evol_MB_T}\\
\frac{d\mu}{dt} &= T \left[3 H (n (\mu  p+\sigma )-\rho  (p+\rho )) + \frac{\delta n}{\delta t} \left(\mu  \rho -\sigma\right) +\frac{\delta \rho}{\delta t}\left( \rho-\mu n \right)\right]\bigg/\left[\rho ^2-n \sigma \right]\,.\label{eq:generalized_evol_MB_mu}
\end{align}
In the Maxwell-Boltzmann approximation, the quantities $n$, $\rho$, $p$ and $\sigma$ can be written in terms of modified Bessel functions as outlined in~\eqref{eq:Thermo_MB}. Since we are working in the MB approximation, Equations~\eqref{eq:generalized_evol_MB_T} and \eqref{eq:generalized_evol_MB_mu} can be written in a very compact manner for the case of ultrarelativistic species ($T\gg m$) or non-relativistic ones ($T \ll m$). 

\subsubsection*{The $T\gg m$ and $T\ll m$ cases}

\noindent For ultrarelativistic species ($T\gg m$), the evolution equations read:
\begin{subequations}\label{eq:MB_Tlarge}
\begin{align}
\frac{dT}{dt} &=-H \, T+ \left[\frac{1}{\rho}  \frac{\delta \rho}{\delta t}-\frac{1}{n} \frac{\delta n}{\delta t}  \right] \,{T}\,,\label{eq:dTdt_MB_Tlarge} \\
\frac{d\mu}{dt} &=-H \, \mu + \left[\frac{1}{\rho} \frac{\delta \rho}{\delta t} - \frac{1}{n}\frac{\delta n}{\delta t}\right] \, \mu + \left[ \frac{4}{n}\frac{\delta n}{\delta t}- \frac{3}{\rho} \frac{\delta \rho}{\delta t} \right] \, T\,, \label{eq:dmudt_MB_Tlarge}
\end{align}
\end{subequations}
where $\rho = g \, \frac{3}{\pi^2} \, T^4 \, e^{\mu/T}$. 

\noindent While, for non-relativistic particles, ($T \ll m$):
\begin{subequations}\label{eq:MB_Tsmall}
\begin{align}
\frac{dT}{dt} &=-2\, H\, T +  \frac{2}{3 \, n}\left[ \frac{\delta \rho}{\delta t} - m \frac{\delta n}{\delta t} \right] \,,  \label{eq:dTdt_MB_Tsmall} \\
\frac{d\mu}{dt} &=- 2 \, H \, (\mu -m)+ \frac{2}{3 \, n} \left[ \frac{\delta \rho}{\delta t} -m\frac{\delta n}{\delta t} \right]  \frac{\mu -m}{T}\,, \label{eq:dmudt_MB_Tsmall}
\end{align}
\end{subequations}
since $n\sigma -\rho^2 \simeq \frac{3}{2} T n p$, $\sigma \simeq m \rho$, $p \simeq T n$ and $\rho \simeq m n$, and $n \simeq g \left( \frac{T m}{2\pi} \right)^{3/2} e^{(\mu-m)/T}$ in the $T \ll m$ limit. 
Clearly the set of equations~\eqref{eq:MB_Tlarge} and~\eqref{eq:MB_Tsmall} reproduce the correct scale factor behaviour in the absence of interactions since $d/dt = a H d/da$. See for instance Equations 3.80, 3.81 and 3.82 of~\cite{Kolb:1990vq}.

\begin{figure}[t]
\centering
\begin{tabular}{cc}
\hspace{-0.8cm} \includegraphics[width=0.99\textwidth]{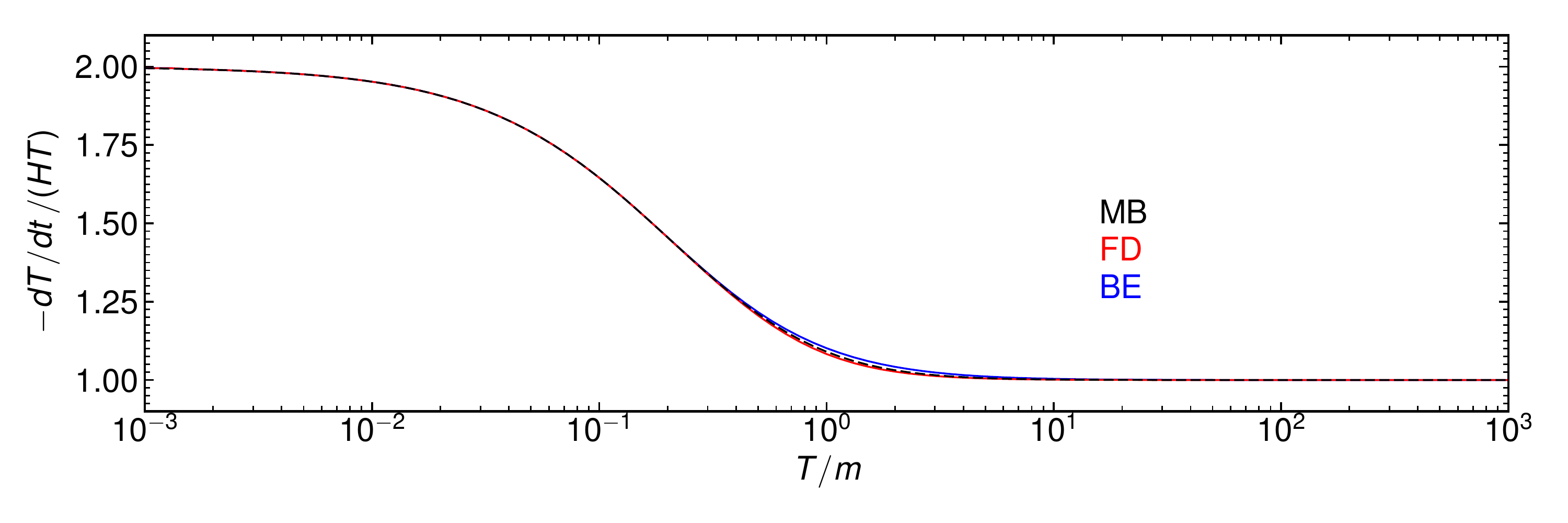} 
\end{tabular}\vspace{-0.0cm}
\caption{$-dT/dt /(H T)$ for a massive decoupled relic as a function of $T/m$. We have chosen $\mu = 0$. The different lines correspond to a particle that follows Fermi-Dirac (red), Bose-Einstein (blue) or Maxwell-Boltzmann (black) statistics.}\label{fig:dT_dt}
\end{figure}

In Figure~\ref{fig:dT_dt} we show $dT/dt\,/(HT)$ for a decoupled species ($\delta \rho/\delta t=0$, and $\delta n/\delta t = 0$) with negligible chemical potential, $\mu=0$. The Fermi-Dirac and Bose-Einstein cases are obtained by numerically evaluating~\eqref{eq:dT_dt_simple}, while the Maxwell-Boltzmann case is a result of~\eqref{eq:generalized_evol_MB_T}. From Figure~\ref{fig:dT_dt} we can clearly appreciate that for $T \gg m$, $dT/dt = -H T$ and for $T \ll m $, $dT/dt = -2H T$. We therefore recover the correct behaviour~\cite{Kolb:1990vq} and also we notice that the Fermi-Dirac and Bose-Einstein cases are relatively similar. 

\newpage
\subsubsection*{An application: Baryon temperature evolution}

The evolution equations~\eqref{eq:MB_Tsmall} allow us to calculate, for example, how the background baryon temperature evolves with time. This is clearly not a new result~\cite{1968ApJ...153....1P}, but serves as an application of the formulae derived above. Since baryons are tightly coupled to electrons, and electrons are tightly coupled to photons via highly efficient elastic scatterings: $\delta n/\delta t = 0$ and hence, we can safely neglect chemical potentials. From Equation~\eqref{eq:dTdt_MB_Tsmall}, we obtain that the temperature evolution for the electron-baryon fluid is:
\begin{align}\label{eq:Tem_baryon}
\frac{dT_b}{dt} &=  - 2 \,H \, T_b + \frac{2}{3} \frac{1}{n_b} \frac{\delta \rho_{b\gamma}}{\delta t} \, .
\end{align}
Then, we are simply left with calculating $\delta \rho_{b\gamma}/\delta t$. Since baryons are strongly coupled to electrons via Rutherford scattering, we can assume that baryons and electrons share the same temperature, $T_e = T_b$, and therefore $\delta \rho_{b\gamma}/\delta t = \delta \rho_{e\gamma}/\delta t = - \delta \rho_{\gamma e}/\delta t $. We simply calculate $ \delta \rho_{\gamma e}/\delta t$ by considering the collision term for electron-photon Compton scattering~\cite{1965PhFl....8.2112W,Hu:1995em}:
\begin{align}
C_\gamma[f_\gamma] = x_e n_e \sigma_T  \frac{1}{m_e p^2} \frac{\partial}{\partial p} \left[p^4 \left(T_b  \frac{\partial}{\partial p} f_\gamma + f_\gamma(1+f_\gamma) \right) \right] \, , 
\end{align}
where $x_e$ is the free electron fraction and $\sigma_T$ is the Thompson scattering cross section. By considering $f_\gamma = \left[-1+e^{p/T_\gamma}\right]^{-1}$ and integrating over photon momenta, $2/(2\pi^2)p^3 dp$, one finds~\cite{1965PhFl....8.2112W}:
\begin{align}
\frac{\delta \rho_{b\gamma}}{\delta t} = \frac{4 {n_e} \rho_\gamma  \sigma_T  {x_e} }{m_e}(T_\gamma-{T_b})\,.
\end{align}
By plugging this expression into Equation~\eqref{eq:Tem_baryon} and by considering that $n_b = \rho_b/\mu_M$ (where $\mu_M$ is the mean molecular baryonic-electron weight), we recover the well-known baryon temperature evolution equation~\cite{1968ApJ...153....1P,Ma:1995ey}:
\begin{align}
\frac{dT_b}{dt} &=  - 2 \,H \, T_b + \frac{8}{3} \frac{\mu_M}{m_e} \frac{ \rho_{\gamma}}{ \rho_{b} } x_e  n_e \sigma_T (T_\gamma-T_b) \, .
\end{align}

\newpage

\subsection{Thermodynamic formulae}\label{app:Thermo_formulae}
We explicit the relevant thermodynamic formulae for the number, energy and pressure densities and their derivatives for particles, with $g$ internal degrees of freedom, that follow Fermi-Dirac, Bose-Einstein and Maxwell-Boltzmann statistics. We particularize also for the case of $m = 0$. We remind the reader that the distribution functions are given by:
\begin{align}
f_{\rm FD}(E) = \frac{1}{1+e^{(E-\mu)/T }}\,,\qquad f_{\rm BE}(E) = \frac{1}{-1+e^{(E-\mu)/T }} \,,\qquad f_{\rm MB}(E) = e^{-(E-\mu)/T }\,.
\end{align}
Note that the entropy density for a perfect fluid is $s = \frac{\rho + p -\mu n}{T}$~\cite{pdg}. In addition, we note that the formulae for $n$, $\rho$, $p$ and their derivatives can also be written as an infinite sum of Bessel functions. This can be useful to find these quantities without integration, see e.g.~\cite{Arbey:2018zfh}. 

\subsubsection*{Fermi-Dirac}
\begin{subequations}\label{eq:Thermo_FD}
\begin{align}
&n^{\rm FD} = \frac{g}{2\pi^2} \int_{m}^{\infty}  dE \,  \frac{E \sqrt{E^2-m^2}}{e^{(E-\mu)/T} +1 } \,, \\
 &\rho^{\rm FD} = \frac{g}{2\pi^2} \int_{m}^{\infty}  dE \,  \frac{E^2 \sqrt{E^2-m^2}}{e^{(E-\mu)/T} +1 } \, ,\\ 
&p^{\rm FD} = \frac{g}{6\pi^2} \int_{m}^{\infty}  dE \,  \frac{(E^2-m^2)^{3/2}}{e^{(E-\mu)/T} +1 } \, ,\\
 &  \frac{\partial n^{\rm FD}}{\partial T} = \frac{g}{2\pi^2} \int_{m}^{\infty}  dE \, E \sqrt{E^2-m^2}\, \frac{(E-\mu ) }{4 T^2} \cosh^{-2}\left(\frac{E-\mu }{2 T}\right)  \,, \\
  &  \frac{\partial \rho^{\rm FD}}{\partial T} = \frac{g}{2\pi^2} \int_{m}^{\infty}  dE  \, E^2 \sqrt{E^2-m^2} \, \frac{(E-\mu ) }{4 T^2} \cosh^{-2}\left(\frac{E-\mu }{2 T}\right) \,, \\
   &  \frac{\partial n^{\rm FD}}{\partial \mu} = \frac{g}{2\pi^2} \int_{m}^{\infty}  dE \, E \sqrt{E^2-m^2}\, \left[{2 T \cosh \left(\frac{E-\mu }{T}\right)+2 T}\right]^{-1}  \,, \\
  &  \frac{\partial \rho^{\rm FD}}{\partial \mu} = \frac{g}{2\pi^2} \int_{m}^{\infty}  dE  \, E^2 \sqrt{E^2-m^2} \, \left[{2 T \cosh \left(\frac{E-\mu }{T}\right)+2 T}\right]^{-1}  \,. 
\end{align}
\end{subequations}

\subsubsection*{Bose-Einstein}
\begin{subequations}\label{eq:Thermo_BE}
\begin{align}
&n^{\rm BE} = \frac{g}{2\pi^2} \int_{m}^{\infty}  dE \,  \frac{E \sqrt{E^2-m^2}}{e^{(E-\mu)/T} -1 } \,, \\
 &\rho^{\rm BE} = \frac{g}{2\pi^2} \int_{m}^{\infty}  dE \,  \frac{E^2 \sqrt{E^2-m^2}}{e^{(E-\mu)/T} -1 } \, ,\\ 
&p^{\rm BE} = \frac{g}{6\pi^2} \int_{m}^{\infty}  dE \,  \frac{(E^2-m^2)^{3/2}}{e^{(E-\mu)/T} -1 } \, ,\\
 &  \frac{\partial n^{\rm BE}}{\partial T} = \frac{g}{2\pi^2} \int_{m}^{\infty}  dE \, E \sqrt{E^2-m^2}\, \frac{(E-\mu ) }{4 T^2}\sinh^{-2}\left(\frac{E-\mu }{2 T}\right) \,, \\
  &  \frac{\partial \rho^{\rm BE}}{\partial T} = \frac{g}{2\pi^2} \int_{m}^{\infty}  dE  \, E^2 \sqrt{E^2-m^2} \, \frac{(E-\mu ) }{4 T^2}\sinh^{-2}\left(\frac{E-\mu }{2 T}\right) \,, \\
   &  \frac{\partial n^{\rm BE}}{\partial \mu} = \frac{g}{2\pi^2} \int_{m}^{\infty}  dE \, E \sqrt{E^2-m^2}\,\frac{1}{4 T}\sinh^{-2}\left(\frac{E-\mu }{2 T}\right)  \,, \\
  &  \frac{\partial \rho^{\rm BE}}{\partial \mu} = \frac{g}{2\pi^2} \int_{m}^{\infty}  dE  \, E^2 \sqrt{E^2-m^2} \, \frac{1}{4 T} \sinh^{-2}\left(\frac{E-\mu }{2 T}\right)\,.
\end{align}
\end{subequations}
\subsubsection*{Maxwell-Boltzmann}
\begin{subequations}\label{eq:Thermo_MB}
\begin{align}
&n^{\rm MB} =g \frac{m^2 T e^{\mu /T} }{2 \pi ^2} K_2\left(\frac{m}{T}\right)\,, \\
 &\rho^{\rm MB} = g\frac{m^2 T e^{\mu /T}}{2 \pi ^2}  \left[m K_1\left(\frac{m}{T}\right)+3 T K_2\left(\frac{m}{T}\right)\right] \, ,\\ 
&p^{\rm MB} =  g\frac{m^2 T e^{\mu /T} }{2 \pi ^2} \,T \, K_2\left(\frac{m}{T}\right)  \, ,\\
&\sigma^{\rm MB} \equiv  \frac{g}{2\pi^2} \int_{m}^{\infty}  dE \, {E^3 \sqrt{E^2-m^2}}\,e^{-(E-\mu)/T} \,\\
& \qquad =  g \frac{m^2 T e^{\mu /T} }{2 \pi ^2} \left[m^2 K_2\left(\frac{m}{T}\right)+3 T m K_3\left(\frac{m}{T}\right)\right] \,,\\
\frac{d n^{\rm MB}}{d T} &= \frac{\rho^{\rm MB} -\mu n^{\rm MB} }{T^2} \, ,\qquad  \frac{d n^{\rm MB}}{d \mu} = \frac{n^{\rm MB}}{T}  \, ,\\
\frac{d \rho^{\rm MB}}{d T} &= \frac{\sigma^{\rm MB} -\mu \rho^{\rm MB}}{T^2}  \, ,\qquad  \frac{d \rho^{\rm MB}}{d \mu} = \frac{\rho^{\rm MB}}{T} \, .
\end{align}
\end{subequations}
where $K_j$ are modified Bessel functions of the $j$ kind.

\subsubsection*{Massless case}
    \begin{equation}\label{eq:thermo_massless}
\def\arraystretch{1.25}
\begin{array}{cccc}
\toprule
      \multicolumn{4}{c@{}}{ \text{Massless Thermodynamics}, \,\, m = 0 \, ,\,\, x \equiv e^{\mu/T} }   \\
    \toprule
  \,\,\, \,\,\,\,   \text{Quantity} \,\,\,\, \,\,\,&\,\,\,\,\text{Fermi-Dirac} \,\,\,\, & \,\,\,\, \text{Bose-Einstein} \,\,\,\, &\,\,\, \text{Maxwell-Boltzmann}\,\,\, \\
     \toprule
        n   & -g\frac{T^3}{\pi ^2} \, \text{Li}_3\left(-x\right)  & g\frac{T^3}{\pi ^2}  \text{Li}_3\left(x\right)& g\frac{T^3}{\pi ^2} \, x\\
        \rho   &  -g\frac{3T^4}{\pi ^2}  \, \text{Li}_4\left(-x\right) & g\frac{3 T^4 }{\pi ^2} \text{Li}_4\left(x\right)& g \frac{3 T^4 }{\pi ^2} \, x\\
        p   &  \rho/3 & \rho/3  & \rho/3 \\
                \partial n/\partial T   & g\frac{T \left(\mu  \text{Li}_2\left(-x\right)-3 T \text{Li}_3\left(-x\right)\right)}{\pi ^2}  & g\frac{T \left(3 T \text{Li}_3\left(x\right)-\mu  \text{Li}_2\left(x\right)\right)}{\pi ^2}  & g\frac{T  (3 T-\mu )}{\pi ^2} x\\ 
                        \partial \rho/\partial T   & g\frac{3 T^2\left(\mu  \text{Li}_3\left(-x\right)-4 T \text{Li}_4\left(-x\right)\right)}{\pi ^2}   &  g\frac{3 T^2 \left(4 T \text{Li}_4\left(x\right)-\mu  \text{Li}_3\left(x\right)\right)}{\pi ^2}&  g\frac{3 T^2  (4 T-\mu )}{\pi ^2} x\\ 
        \partial n/\partial \mu   & -g\frac{T^2 }{\pi ^2} \text{Li}_2\left(-x\right)& g\frac{T^2 }{\pi ^2} \text{Li}_2\left(x\right) & g\frac{T^2 }{\pi ^2} x \\ 
        \partial \rho/\partial \mu   & -g\frac{3 T^3 }{\pi ^2} \text{Li}_3\left(-x\right)  & g\frac{3 T^3 }{\pi ^2}\text{Li}_3\left(x\right) & g\frac{3 T^3 }{\pi ^2} x \tabularnewline
     \bottomrule 
\end{array}
\end{equation}
where $\text{Li}_j$ are Polylogarithms of order $j$.
\subsubsection*{Massless case with $\mu = 0$}
    \begin{equation}\label{eq:thermo_massless}
\def\arraystretch{1.25}
\begin{array}{cccc}
\toprule
      \multicolumn{4}{c@{}}{ \text{Massless Thermodynamics}, \,\, m = 0 \, ,\,\, \mu = 0  }   \\
    \toprule
  \,\,\, \,\,\,\,   \text{Quantity} \,\,\,\, \,\,\,&\,\,\,\,\text{Fermi-Dirac} \,\,\,\, & \,\,\,\, \text{Bose-Einstein} \,\,\,\, &\,\,\, \text{Maxwell-Boltzmann}\,\,\, \\
     \toprule
        n   & g \frac{3}{4} \frac{\zeta(3)}{\pi ^2} \, T^3   & g\frac{\zeta(3)}{\pi ^2} \, T^3 & g\frac{1}{\pi ^2}\,T^3 \\
        \rho   &  g\frac{7}{8}  \frac{\pi^2}{30} \, T^4 &  g  \frac{\pi^2}{30}  \, T^4   & g \frac{3}{\pi ^2}\, T^4  \\
        p   &  \rho/3 & \rho/3  & \rho/3 \\
       \partial n/\partial T   & 3\,n/T  & 3\,n/T & 3\,n/T \\ 
       \partial \rho/\partial T   & 4\,\rho/T  &  4\,\rho/T &  4\,\rho/T \\
            \bottomrule  
\end{array}
\end{equation}
where $\zeta (3) \simeq 1.20206$.

\newpage
\subsection{Number and energy density transfer rates in the MB approximation}\label{app:rates_MB}
In this appendix we work out the number and energy transfer rates for decays, annihilation and scattering processes in the Maxwell-Boltzmann approximation. By numerically integrating the relevant collision terms we also provide spin-statistics corrections to such rates.   

\subsubsection*{Decays}\label{app:rates_decay}
The collision term for a $1 \leftrightarrow 2$ process can be written analytically if all the relevant distribution functions are thermal~\cite{Kawasaki:1992kg,Starkman:1993ik}. 

In the case of the isotropic decay of $a \to 1 + 2$ (this is the type we consider in Section~\ref{sec:boson_1to2}, see e.g.~\cite{Balantekin:2018azf}) the collision term for the $a$ particle reads~\cite{Starkman:1993ik}:
\begin{align}\label{eq:col_decay}
\mathcal{C}_a (p_a) =  -\Gamma_a\frac{m_a}{m_*}\frac{m_{a}}{E_a p_a}  \int_{E_-}^{E_+} d E_1 \, A(E_a, E_1, E_a-E_1)  \, ,
\end{align}
where $\Gamma_a $ is the rest frame decay width, $m_*^2 = m_a^2-2(m_1^2+m_2^2)+(m_1^2-m_2^2)^2/m_a^2$, 
\begin{align}
E_\pm = \frac{m_*}{2m_H} \left[E_a \sqrt{1+\frac{4m_1^2}{m_*^2}} \pm p_a \right] \, ,
\end{align}
 and 
\begin{align}
A=f_a \left[1 \pm f_1\right] \left[1 \pm f_2\right] - f_1 f_2  \left[1 \pm f_a\right]\,,
\end{align}
where the $+$ sign corresponds to bosons and the $-$ sign corresponds to fermions. 

In the particular case in which the decay products are massless, the kinematical region simplifies to $E_\pm =\frac{1}{2}( E_a \pm p_a)$ with $m_* = m_a$. In addition, if $f_1 = f_2$ and if we neglect the statistical terms in the collision integral and in the distribution functions we simply find
\begin{align}\label{eq:C_MB}
\mathcal{C}_a (p_a) =  -\Gamma_a\frac{m_{a}}{E_a} \left[e^{\frac{\mu_a-E_a}{T_a}} -e^{\frac{2\mu_1-E_a}{T_1}} \right] \,.
\end{align}
If we consider that the $a$ particle is a boson and that the $1$ and $2$ particles are identical massless fermions, we find
\begin{align}
\mathcal{C}_a (p_a) = -\Gamma_a\frac{m_{a}}{E_a} \frac{T}{p_a} \frac{\left(e^{\frac{E_a}{T_a}+\frac{2 \mu }{T}}-e^{\frac{E_a}{T}+\frac{\mu_a}{T_a}}\right)}{\left(e^{\frac{E_a}{T}}-e^{\frac{2 \mu }{T}}\right) \left(e^{\frac{E_a}{T_a}}-e^{\frac{\mu_a}{T_a}}\right)}\log \left(\frac{\left(e^{\frac{E_a-p_a}{2 T}}+e^{\frac{\mu}{T}}\right) \left(e^{\frac{E_a+2 \mu +p_a}{2 T}}+e^{E_a/T}\right)}{\left(e^{\frac{E_a+p_a}{2 T}}+e^{\frac{\mu}{T}}\right) \left(e^{\frac{E_a+2 \mu -p_a}{2 T}}+e^{E_a/T}\right)}\right) 
\end{align}
The integral over momenta, assuming Maxwell-Boltzmann statistics~\eqref{eq:C_MB}, yields:
\begin{align}
\frac{\delta n_a}{\delta t} &= \frac{\Gamma_a  g_a m_a^2 }{2 \pi ^2}\left[T e^{\frac{2 \mu }{T}} K_1\left(\frac{m_a}{T}\right)-T_a e^{\frac{\mu_a}{T_a}} K_1\left(\frac{m_a}{T_a}\right)\right]\,,\\ 
\frac{\delta \rho_a}{\delta t} &= \frac{\Gamma_a  g_a m_a^3 }{2 \pi ^2}\left[T e^{\frac{2 \mu }{T}} K_2\left(\frac{m_a}{T}\right)-T_a e^{\frac{\mu_a}{T_a}} K_2\left(\frac{m_a}{T_a}\right)\right]\,.
\end{align}
From the MB equations above, and taking $T_a = T$ and $\mu_a = \mu = 0$, we recover the well-known cases: 
\begin{align}
\left< \Gamma \right>_{a\to 1+2} \equiv \frac{1}{n_a} \left. \frac{\delta n_a}{\delta t}\right|_{a \to 1+2} &= \Gamma_a \frac{K_1(\frac{m}{T})}{K_2(\frac{m}{T})}\,, \\ 
\left< \Gamma \right>_{1+2 \to a} \equiv \frac{1}{n_1} \left. \frac{\delta n_a}{\delta t}\right|_{1+2 \to a} &= \Gamma_a \frac{g_a}{2g_1}  \left(\frac{m}{T}\right)^2 K_1\left(\frac{m}{T}\right)\,,
\end{align}
which in relevant limits read (see e.g. Equations 6.8 in~\cite{Kolb:1990vq}):
\begin{align}
\left. \left< \Gamma \right>_{a\to 1+2}\right|_{T\gg m} &= \Gamma_a \frac{m}{2T}\,,\qquad \qquad \left. \left< \Gamma \right>_{a\to 1+2}\right|_{T\ll m} = \Gamma_a \,, \label{eq:Decay_Averaged_Simple} \\ 
\left. \left< \Gamma \right>_{1+2\to a}\right|_{T\gg m} &=  \Gamma_a \frac{g_a}{2g_1}  \,\frac{m}{T}\,,\qquad \,\,\, \left. \left< \Gamma \right>_{1+2\to a}\right|_{T\ll m} =  \Gamma_a \frac{g_a}{2g_1} \sqrt{\frac{\pi}{2}} \left(\frac{m}{T}\right)^{3/2} \,e^{-\frac{m}{T}}\,,\label{eq:Inverse_Decay_Averaged_Simple}
\end{align}
where $\left< \right>$ represents thermal averaging. 

\begin{figure}[t]
\centering
\begin{tabular}{cc}
\hspace{-0.8cm} \includegraphics[width=0.52\textwidth]{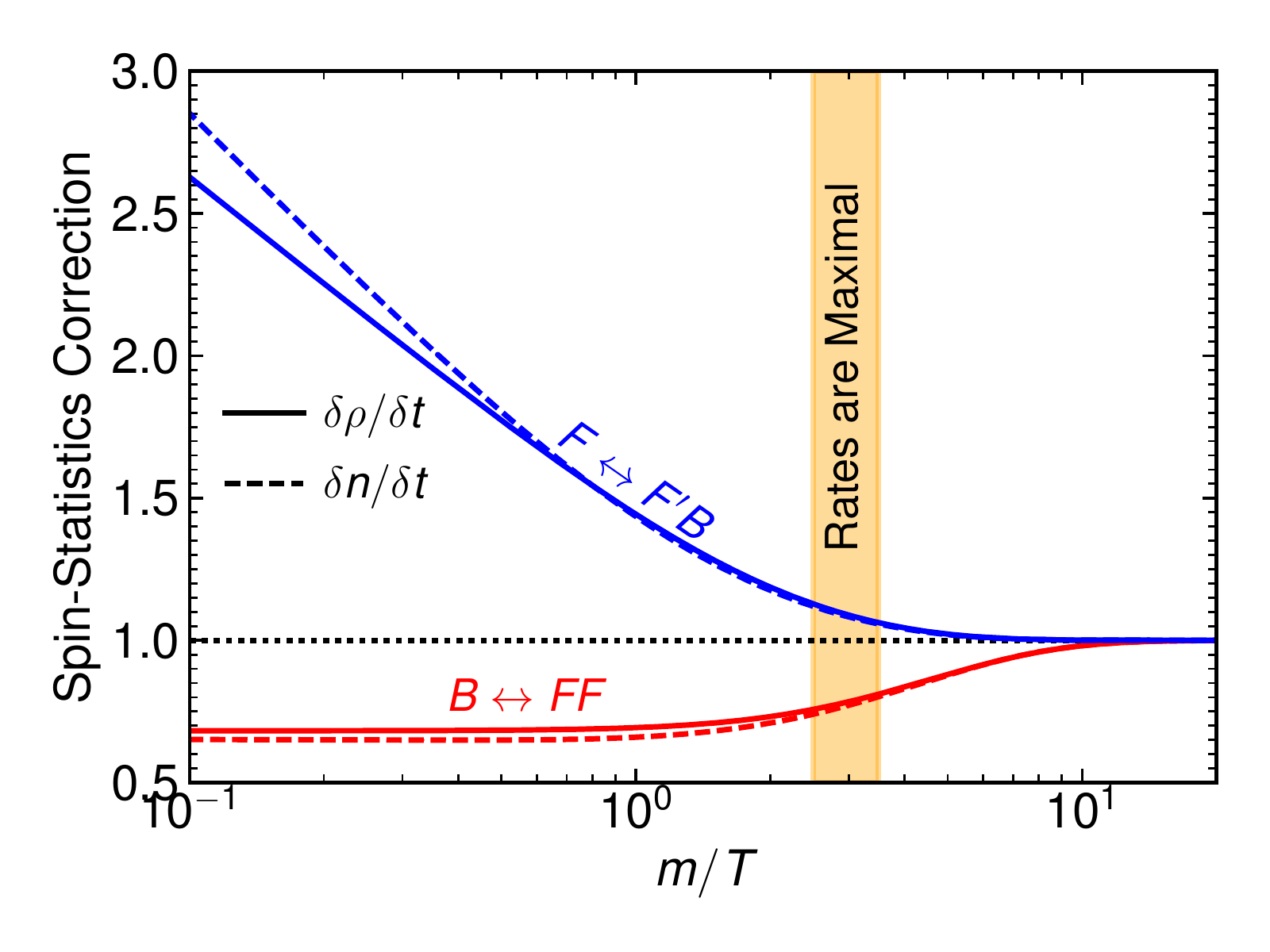} 
\end{tabular}\vspace{-0.4cm}
\caption{Statistical factor correction to the Maxwell-Boltzmann number and energy density rates for decays $B\to FF$ and $F \to F'B$. We have calculated this corrections assuming negligible chemical potentials and $\Delta T/T = 0.01$. We highlight the region in which the rates are maximal, $T\sim m/3$. }\label{fig:Decay_stats}
\end{figure}

The number and energy density exchange rates including quantum statistics cannot be written analytically for decay processes. In Figure~\ref{fig:Decay_stats}, by numerically integrating the relevant collision terms, we show the spin-statistics corrections to the MB transfer rates. We can appreciate that when the rates are maximal (which is the most relevant scenario), the corrections are within 30\%.

\subsubsection*{Annihilations}\label{app:rates_ann}
We calculate the number and energy density transfer rates for $2\leftrightarrow 2$ processes by closely following~\cite{Gondolo:1990dk,Edsjo:1997bg} and neglecting statistical factors in the collision terms. In Table~\ref{tab:suppression} we report the effect of statistical factors correction to the rates in the case of annihilation of massless species interacting via a point-like interaction. 

Neglecting statistical factors and assuming Maxwell-Boltzmann statistics for the distribution functions, the energy density transfer rate explicitly reads~\cite{Gondolo:1990dk}:
\begin{align}\label{eq:deltarho_dt_general}
\frac{\delta \rho}{\delta t} = - \sum_{\rm spins}\int d^3\tilde{p}_1 d^3\tilde{p}_2 d^3\tilde{p}_3 d^3\tilde{p}_4 \,(2\pi)^4\, \delta^4(p_1+p_2-p_3-p_4)\,|\mathcal{M}|^2 \, E_1 \,\left(f_1 f_2 - f_3 f_4\right) \, ,
\end{align}
while the number density transfer reads~\cite{Gondolo:1990dk}:
\begin{align}\label{eq:deltan_dt_general}
\frac{\delta n}{\delta t} = - \sum_{\rm spins}\int d^3\tilde{p}_1 d^3\tilde{p}_2 d^3\tilde{p}_3 d^3\tilde{p}_4 \,(2\pi)^4\, \delta^4(p_1+p_2-p_3-p_4)\,|\mathcal{M}|^2  \,\left(f_1 f_2 - f_3 f_4\right) \, ,
\end{align}
where $d^3\tilde{p_i} = d^3p_i/(2 E_i (2\pi)^3)$ and $\mathcal{M}$ is the amplitude for the $1+ 2 \to 3 +4 $ process (where we have assumed CP conservation). 
\subsubsection*{$\delta \rho/\delta t$ for $1+ 2 \leftrightarrow 3 +4 $ the case: $T_1 = T_2 = T$ and $T_3=T_4 = T'$}\label{app:delta_rho_delta_t_1234}

For the particular process of interest, $T_1 = T_2 = T $ and $T_3 = T_4 = T' $. Hence, within the MB approximation $f_1 f_2 = e^{-(E_1+E_2)/T}$ and $f_3 f_4 = e^{-(E_3+E_4)/T'}= e^{-(E_1+E_2)/T'}$ as a result of energy conservation. Since $f_1 f_2- f_3 f_4$ does not depend upon $p_3$ and $p_4$, the integral over these variables can be performed, leading to~\cite{Gondolo:1990dk,Edsjo:1997bg}:
\begin{align}\label{eq:sigma_red}
\sum_{\rm spins}\int d^3\tilde{p}_3 d^3\tilde{p}_4 \,  (2\pi)^4\, \delta^4(p_1+p_2-p_3-p_4)\,|M|^2 = 4\,g_1\,g_2 \, p_{12}\, \sqrt{s} \, \sigma(s)\, .
\end{align}
Where $g_1$ and $g_2$ are the number of internal degrees of freedom of particles $1$ and $2$ respectively,  $s$ is the centre of mass energy squared, $p_{12} = [s-(m_1+m_2)^2]^{1/2} [s-(m_1-m_2)^2]^{1/2} /(2\sqrt{s})$, and $\sigma(s)$ is the usual cross section of the process (namely, that summed over final spin states and averaged over initial spins). This leads to
\begin{align}
\frac{\delta \rho}{\delta t} = - 4 g_1\,g_2\int d^3\tilde{p}_1 d^3\tilde{p}_2 \, E_1 \,\left[ e^{-\frac{E_1+E_2}{T}} -e^{-\frac{E_1+E_2}{T'}}\right]  \, p_{12}\, \sqrt{s} \, \sigma(s) \, .
\end{align}
By rewriting the integral in terms of the convenient variables $E_+ \equiv E_1+E_2$, $E_- \equiv E_1-E_2$ and $s$, the phase space density is:
\begin{align}
d^3\tilde{p}_1 d^3\tilde{p}_2 =  \frac{1}{(2\pi)^4} \frac{dE_+ dE_- d s}{8} \, ,
\end{align}
with the appropriate limits of integration: 
\begin{align}
s &\geq (m_1+m_2)^2 \, ,\\
E_+& \geq \sqrt{s} \, ,\\
\left|E_- - E_+ \frac{m_2^2-m_1^2}{s}\right|& \leq 2 \, p_{12} \, \sqrt{\frac{E_+^2-s}{s}} \, . 
\end{align}
The energy transfer rate then reads:
\begin{align}
\frac{\delta \rho}{\delta t} = - 4\,g_1\,g_2 \, \int \frac{1}{(2\pi)^4} \frac{dE_+ dE_- d s}{8} \, \frac{E_+ + E_-}{2} \,\left[e^{-\frac{E_+}{T}} -e^{-\frac{E_+}{T'}}\right]  \, p_{12}\, \sqrt{s} \, \sigma(s) \, .
\end{align}
The first integral to be performed is over $E_-$, yielding: 
\begin{align}
\int \frac{E_+ + E-}{2} dE_- = 2 E_+ p_{12} \sqrt{\frac{E_+^2}{s}-1} \left(1 + \frac{m_2^2-m_1^2}{s}\right) \, .
\end{align}
Then, we perform the integral over $E_+$ to find:
\begin{align}
\int e^{-\frac{E_+}{T}} dE_+ \left(\int \frac{E_+ + E-}{2} dE_- \right) = \frac{2 \, p_{12} }{\sqrt{s}}\,  T  \, \left(s +m_2^2 -m_1^2\right) K_2\left(\frac{\sqrt{s}}{T}\right) \, ,
\end{align}
where $K_2$ is a modified Bessel function of the second kind. 

This allows us to write the energy transfer rate in terms of a single integral over $s$:
\begin{align}\label{eq:aux_1}
\frac{\delta \rho}{\delta t} =  \frac{ g_1g_2 }{16 \pi^4 }\, \int_{s_{\rm min}}^{\infty} d s  \,p_{12}^2\, 
  \left(s +m_2^2 -m_1^2\right) \, \sigma(s)  \left[ T' \, K_2\left(\frac{\sqrt{s}}{T'}\right) -  T \, K_2\left(\frac{\sqrt{s}}{T}\right)  \right] \, ,
\end{align}
where $s_{\rm min} = \text{min}[(m_1+m_2)^2, \, (m_3+m_4)^2]$. For the particular case in which $m_1 = m_2 = m_3 = m_4 = 0$, the previous expression reads
\begin{align}\label{eq:aux_2}
\frac{\delta \rho}{\delta t} =  \frac{ g_1g_2 }{64 \pi^4 }\, \int_{0}^{\infty} d s  \, s^2 \, \sigma(s)  \left[ T' \, K_2\left(\frac{\sqrt{s}}{T'}\right) -  T \, K_2\left(\frac{\sqrt{s}}{T}\right)  \right] \, .
\end{align}

The effect of quantum statistics in the collision depends upon the mass of the species involved in them. The larger $m/T$ is the more they resemble the MB formulas presented here. Table~\ref{tab:suppression} outlines the effect on the rates by assuming that all the relevant species are massless and have a very small temperature difference of $\Delta T \equiv (T_1 -T_2)/T_1 = 0.01$.

\begin{table}[t]
\begin{center}
\begin{tabular}{cc|ccc|ccc}
\hline\hline
\multicolumn{2}{c|}{Rates}  & \multicolumn{3}{c|}{$[\delta\rho/\delta t]^{\rm stats}\big/[\delta\rho/\delta t]^{\rm MB}$} & \multicolumn{3}{c}{$[\delta n/\delta t]^{\rm stats}\big/[\delta n/\delta t]^{\rm MB}$} \\\hline
 Initial State &  Final State	& $|\mathcal{M}_4|^2$  & $|\mathcal{M}_2|^2$ & $|\mathcal{M}_0|^2$  &	$|\mathcal{M}_4|^2$  & $|\mathcal{M}_2|^2$ & $|\mathcal{M}_0|^2$   	 \\\hline

 F + F			    &	 F + F   &	0.88 & 0.79 & 0.64 & 0.85  & 0.74 & 0.57	 \\
 F + F			    &	 B + B   &	1.02 & 1.06 & 1.24 & 1.03   & 1.08 & 1.35	 \\
 B + B			    &	 B + B   &	1.20 & 1.54 & 3.40 & 1.28   & 1.83 & 6.44	 \\ \hline\hline
\end{tabular}
\end{center}\vspace{-0.3cm}
\caption{Statistical factors corrections to the annihilation rates as compared to the Maxwell-Boltzmann case for a four-point interaction and for $m_i = 0$. $\mathcal{M}$ is a matrix element, $|\mathcal{M}_n|^2 \propto p^n$, where $p^n$ represents any combination of 4-momenta scalar products with dimension of energy $n$. For example, $(p_1\cdot p_2)(p_3\cdot p_4)$ corresponds to $n= 4$. These spin-statistics correction factors are calculated assuming that $(T_2-T_1)/T_1 = 0.01$ and negligible chemical potentials. }\label{tab:suppression}
\end{table}

\subsubsection*{An example: Neutrino-electron interactions}\label{app:delta_rho_delta_t_1234_example}
We can consider the case of the Standard Model neutrino-electron interactions. In that case, the neutrino-electron annihilation cross section is simply given by
\begin{align}
\sigma_{\nu_\alpha \bar{\nu}_\alpha \to e^+ e^-}(s) = \frac{2}{3}\frac{G_F^2}{\pi^2} (g_L^2+g_R^2) \,s\,,
\end{align}
and by using Equation~\eqref{eq:aux_1} we find:
\begin{align}\label{eq:aux_3}
\left. \frac{\delta \rho}{\delta t}\right|_{\nu_\alpha,\,\nu_\alpha \bar{\nu}_\alpha \to e^+ e^-} =  \frac{64 G_F^2 \left(g_L^2+g_R^2\right) }{\pi ^5} \, (T_\gamma^9-T_{\nu}^9) \,,
\end{align}
where we have used the fact that $g_{\nu_\alpha} = g_{\bar{\nu}_\alpha} = 1$. Equation~\eqref{eq:aux_3} exactly matches Equation A.8 of~\cite{Escudero:2018mvt} and in order to find the joint energy transfer for neutrinos and antineutrinos (as considered in Section~\ref{sec:SM_neutrinodec}) one should simply multiply this expression by 2. 
 
\subsubsection*{$\delta n/\delta t$ for $1+ 2 \leftrightarrow 3 +4 $ the case: $T_1 = T_2 = T$ and $T_3=T_4 = T'$}
Again, by working in terms of $E_+$, $E_-$ and $s$ we find:
\begin{align}
\frac{\delta n}{\delta t} = \frac{g_1g_2}{8 \pi ^4} \int_{s_{\rm min}}^{\infty} d s  \,p_{12}^2\,\sqrt{s}\, \sigma(s)\, \left[T' K_1\left(\frac{\sqrt{s}}{T'}\right)-T K_1\left(\frac{\sqrt{s}}{T}\right)\right]\,.
\end{align}
Since $\frac{\delta n}{\delta t} = \left< \sigma v \right> (n^2-n_{\rm eq}^2) $, this expression clearly matches the usual annihilation cross section formula:
\begin{align}
\left< \sigma v\right> &= \frac{1}{8 T m_1^2 m_2^2 K_2(m_1/T) K_2(m_2/T)} \int_{s_{\rm min}}^{\infty} ds \, s^{3/2} K_1\left(\frac{\sqrt{s}}{T} \right) \lambda \left[1,\frac{m_1^2}{s},\frac{m_2^2}{s}\right] \sigma(s)\,,
 \end{align}
where $\lambda(x,y,z) = x^2+y^2+z^2-2xy-2xz-2yz $ is the Kallen function.

Note that in the limit in which $m_1 = m_2 = m$ and $T\ll m$, $\frac{\delta n}{\delta t} \simeq \frac{1}{m}\frac{\delta \rho}{\delta t}$ for annihilation interactions. 

\subsubsection*{Scatterings}\label{app:MB_scatterings}
The number density transfer rate for scattering interactions vanishes by definition since scattering processes conserve particle number. The energy transfer rate does not necessarily vanish, although if annihilation or decay interactions are active then scattering interactions represent a subdominant contribution to the energy transfer since the typical energy exchanged in an scattering interaction is the momentum transferred $\sim q$ while the energy transferred in an annihilation or decay is $\sim E$.

\begin{table}[t]
\begin{center}
\begin{tabular}{ccccc}
\hline\hline
$\delta\rho_1/\delta t$ & $S|\mathcal{M}|^2$  & $f_{\rm stat}^{FF}$ & $f_{\rm stat}^{FB}$ & $f_{\rm stat}^{BB}$  \\\hline
$\frac{6}{8\pi^5M_*^4} T_1^4\,T_2^4 \, (T_2-T_1)$			    &	 ${(p_1\cdot p_2)(p_3\cdot p_4)}/{M_*^4}$   &0.83 & 1.05 & 1.38 	 \\
$\frac{1}{8\pi^5M_*^4} T_1^4\,T_2^4 \, (T_2-T_1)$			    &	 ${(p_1\cdot p_4)(p_2\cdot p_3)}/{M_*^4}$   & 0.85 & 1.04 & 1.28 	 \\
$\frac{3}{8\pi^5M_*^4} T_1^4\,T_2^4 \, (T_2-T_1)$			    &	 ${(p_1\cdot p_3)(p_2\cdot p_4)}/{M_*^4}$   & 0.81 & 1.05 & 1.43 	 \\
$\frac{3}{64\pi^5M_*^2} T_1^3\,T_2^3 \, (T_2-T_1)$			    &	 ${(p_1\cdot p_2)}/{M_*^2}$     	 	  & 0.72 & 1.14 & 2.02  	 \\
$\frac{2}{64\pi^5M_*^2} T_1^3\,T_2^3 \, (T_2-T_1)$			    &	 ${(p_1\cdot p_3)}/{M_*^2}$     		  &0.71 & 1.15 & 2.12 	 \\
$\frac{1}{64\pi^5M_*^2} T_1^3\,T_2^3 \, (T_2-T_1)$			    &	 ${(p_1\cdot p_4)}/{M_*^2}$     	 	  & 0.74 & 1.14 & 1.81 	 \\
$\frac{1}{64\pi^5M_*^2} T_1^3\,T_2^3 \, (T_2-T_1)$			    &	 ${(p_2\cdot p_3)}/{M_*^2}$     	 	  &0.74 & 1.14 & 1.81  	 \\
$\frac{2}{64\pi^5M_*^2} T_1^3\,T_2^3 \, (T_2-T_1)$			    &	 ${(p_2\cdot p_4)}/{M_*^2}$     	 	  & 0.71 &1.15 & 2.12  	 \\
$\frac{3}{64\pi^5M_*^2} T_1^3\,T_2^3 \, (T_2-T_1)$			    &	 ${(p_3\cdot p_4)}/{M_*^2}$     	 	  & 0.72 & 1.14 &  	2.02 \\
$\frac{\lambda^2}{128\pi^5} T_1^2\,T_2^2 \, (T_2-T_1)$			    &	 $\lambda^2$     	 	  &0.56 & 1.56 & 5.95 	 \\
\hline\hline
\end{tabular}
\end{center}\vspace{-0.3cm}
\caption{Scattering energy exchange rates for scattering processes $12 \to 12$ where $m_1 = m_2 = 0$ and $g_1 = 1$. $\lambda$ is a generic coupling constant and $M_\star$ is a mass scale. $\mathcal{M}$ is the matrix element of the process and $S$ is the symmetry factor, see~\cite{Dolgov:2002wy} for more details. The third, fourth and fifth columns correspond to the spin-statistics corrections to these rates for the scattering of various particles that have negligible chemical potentials and $(T_2-T_1)/T_1 = 0.01$. }\label{tab:scatterings}
\end{table}

The problem with scattering rates is that the collision term for such interactions is typically rather complicated, see e.g.~\cite{Cadamuro:2010cz,Ibe:2019gpv}. This is a mere consequence of the fact that for scattering interactions the collision term is proportional to $e^{-E_1/T_1} e^{-E_2/T_2} - e^{-E_3/T_1} e^{-E_4/T_2}$ and cannot be further simplified as in the case of annihilation interactions, particularly for massive particles. Therefore, to actually calculate the exact collision term or energy transfer rate, one should resort to the rather involved methods of~\cite{Hannestad:1995rs} in order to reduce the phase space to 2 dimensions (see also~\cite{Fradette:2018hhl,Kreisch:2019yzn}). We also point the reader to~\cite{Kawasaki:2000en} where the phase space is reduced to 1 dimension when all species are massless. We, however, warn the reader that typos exist in that reference, see footnote 3 of~\cite{Escudero:2018mvt}. Given the difficulty of the calculation and that no generic expression exists, here, by using the methods of~\cite{Hannestad:1995rs} we outline the energy transfer rate for scattering processes involving massless species interacting via a four-point interaction. They are reported in Table~\ref{tab:scatterings} together with spin-statistics corrections to them.

As an example of such formulae, one can consider the scattering interactions between neutrinos of different flavor. The matrix element for the process $\nu_\alpha \nu_\beta \to \nu_\alpha \nu_\beta$ where $\alpha \neq \beta$ is $S|\mathcal{M}|^2 = 2^5 G_F^2(p_1p_2)(p_3p_4)$~\cite{Dolgov:2002wy}. By using the first row of Table~\ref{tab:scatterings} we can relate $1/M_\star^4 \to 2^5 G_F^2 $. For the case of $\nu_e$ and summing over $\mu$ and $\tau$ flavors, we find:
\begin{align}\label{eq:scatt_nue}
\left. \frac{\delta \rho_{\nu_e}}{\delta t} \right|_{\nu_e \,\nu_\mu  \leftrightarrow \nu_e \nu_\mu+ \nu_e \,\nu_\tau  \leftrightarrow \nu_e \nu_\tau} =  48\,\frac{G_F^2}{\pi^5}  \, T_{\nu_e}^4\, T_{\nu_\mu}^4 \,\left[T_{\nu_\mu}- T_{\nu_e} \right]\,,
\end{align}
which matches expression A.13 of~\cite{Escudero:2018mvt}.

\end{document}